\documentclass[reprint,amsmath,amssymb,aps,prd,nofootinbib]{revtex4-1}
\usepackage{braket}

\usepackage[dvipsnames,table]{xcolor}
\usepackage{mdframed}

\usepackage{amssymb,amsmath,amsfonts}
\usepackage{mathrsfs}
\usepackage{dsfont} 
\usepackage{graphicx,bm}
\usepackage{multirow}
\usepackage{color}
\usepackage{enumerate}
\usepackage{placeins}
\usepackage{dcolumn}
\usepackage{array}
\usepackage{bbold} 
\usepackage{lipsum}  

\usepackage{subfigure}
\usepackage{subfig} 

\newcommand{\Tr}{\text{Tr}}
\newcommand{\n}{\nonumber}

\newcommand{\eref}[1]{Eq.~\eqref{#1}}
\newcommand{\secref}[1]{Sec.~\ref{#1}}
\newcommand{\apref}[1]{Appendix~\ref{#1}}
\newcommand{\tabref}[1]{Table~\ref{#1}}
\newcommand{\diag}{\mathop{}\!\mathrm{diag}}

\providecommand{\e}[1]{\ensuremath{{\scriptscriptstyle E\negthinspace #1}}}

\newcommand{\mbeq}{\overset{!}{=}}

\begin{document}

\title{Phenomenology of isospin-symmetry breaking with vector mesons}
\author{Péter Kovács}
\email{kovacs.peter@wigner.hu}
\affiliation{Institute for Particle and Nuclear Physics, Wigner Research Centre for Physics, 1121 Budapest, Hungary}
\author{György Wolf}
\email{wolf.gyorgy@wigner.hu}
\affiliation{Institute for Particle and Nuclear Physics, Wigner Research Centre for Physics, 1121 Budapest, Hungary}
\author{Nora Weickgenannt}
\affiliation{Institut de Physique Th\'{e}orique, Universit\'{e} Paris Saclay, CEA, CNRS, F-91191 Gif-sur-Yvette, France}
\affiliation{Institute for Theoretical Physics, Goethe University, Max-von-Laue-Str.\ 1, D-60438 Frankfurt am Main, Germany}
\author{Dirk H.\ Rischke}
\email{drischke@th.physik.uni-frankfurt.de}
\affiliation{Institute for Theoretical Physics, Goethe University, Max-von-Laue-Str.\ 1, D-60438 Frankfurt am Main, Germany}
\affiliation{Helmholtz Research Academy Hesse for FAIR, Campus Riedberg, Max-von-Laue-Str.\ 12, D-60438 Frankfurt am Main, Germany}

\begin{abstract}

We study the effect of isospin-symmetry breaking in the framework of the extended Linear Sigma Model (eLSM) in vacuum. 
In this model, several particles mix with each other at tree level, due to the three non-zero scalar condensates (non-strange, strange, isospin).
We resolve these mixings with the help of various field transformations. 
We compute all possible meson mixings and decay widths at tree level and perform a $\chi^2$ fit to PDG data.
A very good fit is found if we exclude the (very small $\sim 130$~keV) $\omega\to \pi\pi$ decay. 
We also investigate the violation of Dashen's theorem.

\end{abstract}

\maketitle

\allowdisplaybreaks

\section{Introduction}
\label{Sec:intro}

Understanding the meson mass spectrum is a fundamental task in particle physics. 
In principle, the QCD Lagrangian contains all relevant information, but the strong coupling becomes large in the low-energy regime, such that QCD is not solvable by perturbative methods. 
In this regime, quarks and gluons are confined in hadrons, which become the relevant degrees of freedom.
Therefore, a possible solution is to use effective models for the hadronic degrees of freedom, which obey the same global symmetries as QCD \cite{Gasiorowicz:1969kn}, but do not contain gauge bosons.
Here, all interactions are expressed by vertices of the hadronic fields. 
One of these effective models is the extended Linear Sigma Model (eLSM) for three flavors, which was discussed assuming isospin symmetry at zero temperature and baryochemical potential  in Ref.\ \cite{Parganlija:2012fy}, and later studied at finite temperature and/or finite baryochemical potential in Refs.\ \cite{Kovacs:2016juc, Kovacs:2021kas, Kovacs:2021ger}.
It successfully describes the meson masses and decay widths at tree level and agrees well with lattice-QCD data at finite temperature \cite{Kovacs:2016juc}. 
However, isospin symmetry is broken in Nature, i.e., the masses of up and down quarks are different (see, e.g., Ref.\ \cite{Weinberg:1977hb}). 
Consequently, the masses of charged and neutral mesons of the same flavor are slightly different. 
Note that the difference is not only due to the strong but also due to the electromagnetic interaction.    

Isospin-symmetry breaking was investigated from different aspects in the literature, like in connection with charge-symmetry breaking \cite{Miller:1990iz, Machleidt:2001}. The latter can be seen in charge-conjugate systems as, e.g., in proton-proton and neutron-neutron binary systems, and is caused by different mechanisms like $\rho-\omega$ mixing \cite{Stephenson:1991yz}, the nucleon mass-difference effect in one-pion exchange interactions \cite{Ge:1986zz}, or isospin-violating meson-nucleon coupling constants \cite{Gardner:1995ya}. 
The effect of $\rho-\omega$ mixing on the nuclear symmetry energy was investigated in Ref.\ \cite{Jiang:2008ztv}, and its relation to low-energy pion-nucleon scattering was studied in Ref.\ \cite{Gibbs:1995dm}. 
Another important topic is the mechanism of production of light scalar mesons, such as $f_0(980)$, which is explained by the $f_0(980)-a_0^0(980)$ mixing \cite{Aceti:2012dj,Aceti:2015zva,ACHASOV201789,Achasov_2019}. 
This mixing, more precisely the triple mixing of the $f_0^{L}$ and $f_0^H$ scalar-isoscalar and $a_0^0$ scalar-isovector states, naturally arises in effective theories including isospin-symmetry breaking in the scalar sector. 
Similarly, in the pseudoscalar sector, there is the $\pi^0-\eta-\eta^{\prime}$ mixing, which was investigated in Ref.\ \cite{Kroll:2005sd}. 
The effects of isospin-symmetry breaking were also thoroughly examined in chiral perturbation theory (ChPT) \cite{Neufeld:1994eg,Maltman:1997ja,Bernstein:1998ip,Amoros:2001cp,Nehme:2001wf,Pallante:2001he,Cirigliano:2003gt}. 

Lattice-QCD calculations have recently also reached such a precision that isospin-symmetry breaking effects become important, e.g., in the case of the precise determination of leptonic and semileptonic decay rates \cite{Boyle:2022lsi}. 
These kinds of investigations can shed some light on the observed deviation from unitarity in the first row of the CKM matrix (see, e.g., Ref.\ \cite{Spadaro:2009zzb}). 
Different mass splittings were also examined on the lattice \cite{deDivitiis:2011eh,deDivitiis:2013xla,Boyle:2017gzv,Giusti:2017dmp}.  

In this paper, we model phenomenologically the violation of isospin symmetry in the eLSM to describe these mass differences and also differences of charged and neutral decay widths. 
It should be mentioned that a similar effective model was already investigated in Ref.\ \cite{Kapoor:1975rp} in some detail. 
Here, however, we pursue a more thorough analysis with current experimental data taken from Ref.\ \cite{Workman:2022ynf}. 
We resolve all the various mixings arising between different meson nonets and within each nonet. 
As it was already mentioned, the charged and neutral masses differ not only due to the $u-d$ quark mass difference, but also because of electromagnetic interactions. 
We take these electromagnetic contributions effectively into account through additive terms to the masses of the charged fields in the different nonets. 

Dashen's theorem \cite{Dashen:1969eg,Gao:1996yan} states that, in the chiral limit, i.e., when the quark masses are zero, the following holds in the pseudoscalar sector,
\begin{align}
    (m_{\pi^{\pm}}-m_{\pi^0})|_{\text{EM}} &= (m_{K^{\pm}}-m_{K^0})|_{\text{EM}}\;, \nonumber \\
    m_{\pi^0}|_{\text{EM}}  &= 0\;, \quad m_{K^0}|_{\text{EM}} = 0\;.
\end{align}
However, there are also corrections to Dashen's theorem. 
In Ref.\ \cite{Baur:1995ig}, using ChPT the authors find only moderate deviations from Dashen's theorem, while in more recent works the deviation seems much more significant \cite{aoki2017review,Miramontes:2022mex}. 
We perform fits to Particle Data Group (PDG) data studying a scenario where Dashen's theorem is valid, as well as various scenarios where the latter is violated.

This paper is organized as follows. 
In \secref{Sec:model} the eLSM is introduced, its tree-level mixing terms are presented, and a collection of transformations is shown to resolve these various mixings. 
In Secs.\ \ref{Sec:phys_mass} and \ref{Sec:dec_widths} the physical masses and decay widths are summarized. 
Section~\ref{Sec:results} is dedicated to the description of the fitting procedure and to the acquired results. Conclusions are given in \secref{Sec:conclusion}. 
Appendix~\ref{App:mass_matrix} lists the explicit expressions for the squared-mass matrix elements, Appendix~\ref{App:fpi0_weak_dec} a derivation of the $f_{\pi^0}$ decay constant from the PCAC relation, Appendix~\ref{App:expl_decays} collects the detailed formulas for the tree-level decay widths, while Appendix~\ref{App:fit_res} contains tables with the detailed results of the fits.    

\section{The model}
\label{Sec:model}

\subsection{Lagrangian with broken isospin}
\label{SSec:lagrangian}

According to Ref.\ \cite{Parganlija:2012fy}, the eLSM Lagrangian is given by
\begin{widetext}
\begin{align}
\mathcal{L} &= \Tr[(D_\mu \Phi)^\dagger(D^\mu \Phi)]-m_0^2\Tr(\Phi
^\dagger\Phi)-\lambda _1[\Tr(\Phi ^\dagger \Phi )]^2-\lambda_2
\Tr(\Phi^\dagger\Phi)^2+\Tr[H(\Phi +\Phi^\dagger)]+c_1(\det \Phi
-\det\Phi ^\dagger)^2\n\\ & -\frac{1}{4}
\Tr(L^{\mu\nu}L_{\mu\nu}+R^{\mu\nu}R_{\mu\nu})+\Tr\left[\left(\frac{m_1^2}{2}+\Delta\right)(L^\mu
  L_\mu+R^\mu R_\mu)\right]
+i\frac{g_2}{2}(\Tr\{L_{\mu\nu}[L^\mu,L^\nu]\}+\Tr\{R_{\mu\nu}[R^\mu,R^\nu]\})\n\\ &
+\frac{h_1}{2}\Tr(\Phi^\dagger\Phi)\Tr(L^\mu L_\mu+R^\mu
R_\mu)+h_2\Tr(\Phi^\dagger L^\mu L_\mu\Phi+R^\mu \Phi^\dagger \Phi
R_\mu)+2h_3\Tr(L_\mu\Phi R^\mu\Phi^\dagger)\n\\ & +g_3[\Tr(L_\mu L_\nu
  L^\mu L^\nu)+\Tr(R_\mu R_\nu R^\mu R^\nu))]+g_4[\Tr(L_\mu L^\mu
  L_\nu L^\nu)+\Tr(R_\mu R^\mu R_\nu R^\nu)]\n\\ & +g_5 \Tr(L_\mu
L^\mu)\Tr(R_\nu R^\nu)+g_6[\Tr(L_\mu L^\mu)\Tr(L_\nu L^\nu)+\Tr(R_\mu
  R^\mu)\Tr(R_\nu R^\nu)]\;,
\label{Lagrangian}
\end{align}
where 
\begin{align*}
D^\mu \Phi&\equiv \partial^\mu\Phi-ig_1(L^\mu\Phi-\Phi R^\mu)-ie A^{\mu}_e[T_3,\Phi]\;,\\
L^{\mu\nu}&\equiv \partial^\mu L^\nu-\partial^\nu L^\mu-ie A^{\mu}_e[T_3,L^{\nu}]+ie A^{\nu}_e[T_3,L^{\mu}]\;,\\
R^{\mu\nu}&\equiv \partial^\mu L^\nu-\partial^\nu L^\mu-ie A^{\mu}_e[T_3,R^{\nu}]+ie A^{\nu}_e[T_3,R^{\mu}]\;.
\end{align*}
 The scalar nonet $\Phi$ is given by
 \begin{equation}
 \Phi\equiv \Phi_{S} + \Phi_{PS}
 = \sum_{a=0}^{8}(S_{a}+iP_{a})T_{a}=\frac{1}{\sqrt{2}}\left(
 \begin{array}{@{}ccc@{}}%
 \frac{(\sigma_{N}+a_{0}^{0})+i(\eta_{N}+\pi^{0})}{\sqrt{2}} & a_{0}^{+}
 +i\pi^{+} & K_{0}^{\star+}+iK^{+}\\
 a_{0}^{-}+i\pi^{-} & \frac{(\sigma_{N}-a_{0}^{0})+i(\eta_{N}-\pi^{0})}
 {\sqrt{2}} & K_{0}^{\star0}+iK^{0}\\
 K_{0}^{\star-}+iK^{-} & {\bar{K}_{0}^{\star0}}+i{\bar{K}^{0}} & \sigma_{S}+i\eta_{S}
 \end{array}
 \right)\;, \label{eq:matrix_field_Phi}
 \end{equation}
while the left- and right-handed vector nonets $L^\mu$ and $R^\mu$ are defined as
\begin{subequations}
\begin{align}
L^{\mu}  &  \equiv V^{\mu} + A^{\mu} \equiv
\sum_{a=0}^{8}(V_{a}^{\mu}+A_{a}^{\mu})T_{a} = \frac{1}{\sqrt{2}
}\left(
\begin{array}{@{}ccc@{}}
\frac{\omega_{N}+\rho^{0}}{\sqrt{2}}+\frac{f_{1N}+a_{1}^{0}}{\sqrt{2}} &
\rho^{+}+a_{1}^{+} & K^{\star+}+K_{1}^{+}\\
\rho^{-}+a_{1}^{-} & \frac{\omega_{N}-\rho^{0}}{\sqrt{2}}+\frac{f_{1N}
	-a_{1}^{0}}{\sqrt{2}} & K^{\star0}+K_{1}^{0}\\
K^{\star-}+K_{1}^{-} & {\bar{K}}^{\star0}+{\bar{K}}_{1}^{0} & \omega_{S}+f_{1S}
\end{array}
\right)  ^{\mu}\;,\label{eq:matrix_field_L}\\
R^{\mu}  & \equiv V^{\mu} - A^{\mu} \equiv
\sum_{a=0}^{8}(V_{a}^{\mu}-A_{a}^{\mu})T_{a} = \frac{1}{\sqrt{2}}
\left(
\begin{array}{@{}ccc@{}}
\frac{\omega_{N}+\rho^{0}}{\sqrt{2}}-\frac{f_{1N}+a_{1}^{0}}{\sqrt{2}} &
\rho^{+}-a_{1}^{+} & K^{\star+}-K_{1}^{+}\\
\rho^{-}-a_{1}^{-} & \frac{\omega_{N}-\rho^{0}}{\sqrt{2}}-\frac{f_{1N}
	-a_{1}^{0}}{\sqrt{2}} & K^{\star0}-K_{1}^{0}\\
K^{\star-}-K_{1}^{-} & {\bar{K}}^{\star0}-{\bar{K}}_{1}^{0} & \omega_{S}-f_{1S}
\end{array}
\right)  ^{\mu}\;. \label{eq:matrix_field_R}
\end{align}
\end{subequations}
\end{widetext}
The fields $H$ and $\Delta$ are defined as
\begin{subequations}
\begin{align}
H&= \sum_{i=0,3,8}\zeta_i T_i = \frac{1}{2}\diag(\zeta_{N}+\zeta_{3},\zeta_{N}-\zeta_{3},\sqrt{2}\zeta_{S})\;,\\
\Delta&= \sum_{i=0,3,8} \Delta_i T_i =\diag(\delta_u,\delta_d,\delta_s)\;,
\end{align}
\end{subequations}
with $T_a$, $a\in\{0,...,8\}$, being the generators of U(3).  
It is worth to note that in the matrices above and throughout the article, the $N-S$ (nonstrange--strange) basis is used instead of the $0-8$ basis, which for a generic field $\xi_a\in(S_a,P_a,V_a^{\mu},A_a^{\mu}, H_a, \Delta_a)$ is defined as
\begin{equation}
\xi_N = \frac{1}{\sqrt{3}} \left( \sqrt{2}\; \xi_0+ \xi_8\right)\;, \quad 
\xi_S = \frac{1}{\sqrt{3}} \left( \xi_0-\sqrt{2}\;\xi_8 \right)\;.   
\label{eq:nsbase}
\end{equation}
If the fields $\zeta_{N/S/3}$ are non-vanishing, chiral symmetry is explicitly broken.
In particular, for $\zeta_3\neq 0$ -- and also for $\delta_3\equiv\delta_u-\delta_d\neq 0$ -- the isospin symmetry is violated, which is the situation in Nature.
In this case, all scalar-isoscalar fields $\sigma_N$, $\sigma_S$, and $a_0^0$ can have nonzero vacuum expectation values denoted as $\phi_{N/S}\equiv\braket{\sigma_{N/S}}$ and $\phi_3\equiv\braket{a_0^0}$. 
The condensates $\phi_{N}$, $\phi_{S}$, $\phi_{3}$ can be considered as order parameters for the chiral phase transition at finite temperature.  
Their values at zero temperature and at tree level are determined by minimizing the classical potential $V_{\text{cl}}(\phi_N,\phi_S,\phi_3)$, which can be read off the Lagrangian (\ref{Lagrangian}) after shifting the scalar-isoscalar fields by their expectation values,
\begin{subequations}
\label{Eq:SSB}
\begin{eqnarray}
\sigma_N &\rightarrow& \phi_N + \sigma_N\;, \\
\sigma_S &\rightarrow& \phi_S + \sigma_S\;, \\
a_0^0 &\rightarrow& \phi_3+ a_0^0\;.
\end{eqnarray}
\end{subequations}
The explicit form of the classical potential reads
\begin{widetext}
\begin{eqnarray}
V_{\text{cl}}(\phi_N,\phi_S,\phi_3) &=& \frac{m_0^2}{2} \left[\phi_N^2+\phi_S^2 + (\phi_3)^2\right] + \frac{\lambda_1}{4} \left[\phi_N^2+\phi_S^2 + (\phi_3)^2\right]^2\n \\ &+& \frac{\lambda_2}{4} \left[\frac{\phi_N^4}{2}+3\phi_N^2(\phi_3)^2 + \frac{(\phi_3)^4}{2} + \phi_S^4\right] - \zeta_N\phi_N -\zeta_S\phi_S-\zeta_3\phi_3\;.
\end{eqnarray}
From the stationary points of $V_{\text{cl}}$, i.e., from the conditions $\partial V_{\text{cl}}/ \partial \phi_{N/S/3} = 0$, the fields $\zeta_{N/S/3}$ are derived as  
\begin{subequations}
\begin{eqnarray}
\zeta_N &=& \phi_N \left\{ m_0^2+\lambda_1 \left[\phi_N^2+\phi_S^2+(\phi_3)^2\right]+\frac{\lambda_2}{2} \left[\phi_N^2+3(\phi_3)^2\right] \right\}\;,\label{Eq:zeta_N}\\
\zeta_S &=& \phi_S \left\{ m_0^2+\lambda_1 \left[\phi_S^2+\phi_N^2+(\phi_3)^2\right]+\lambda_2\phi_S^2 \right\} \;, \label{Eq:zeta_S}\\
\zeta_3 &=& \phi_3 \left\{ m_0^2+\lambda_1\left[\phi_N^2+\phi_S^2+ (\phi_3)^2\right]+\frac{\lambda_2}{2}\left[3\phi_N^2+(\phi_3)^2\right]\right\} \;. \label{Eq:zeta_3}
\end{eqnarray}
\end{subequations}
\end{widetext}
As can be seen, $\zeta_{N/S/3}\propto \phi_{N/S/3}$, i.e., if a field $\zeta_{N/S/3}$ is zero, there is a solution where the corresponding condensate $\phi_{N/S/3} = 0$. 
However, there is another solution where $\zeta_{N/S/3}=0$ and $\phi_{N/S/3}\ne 0$ and this solution corresponds to the physical point.

\subsection{Tree-level masses and mixing terms}
\label{SSec:mixing}

After spontaneous symmetry breaking there will be various mixing terms, namely those between different nonets and those within a given nonet. 
The latter ones are the off-diagonal elements of the squared-mass matrices and are given in \apref{App:mass_matrix}.
The former kind of mixing terms relate certain fields of the axial-vector/vector nonets with those of the pseudoscalar/scalar nonets and read
\begin{widetext}
\begin{eqnarray}
\mathcal{L}_{Nmix} && = - g_1 i\phi_3 ({\rho^-}^\mu \partial_\mu a_0^+ - {\rho^+}^\mu \partial_\mu a_0^-) - i\frac{g_1}{2} \left(\phi_N +
    \phi_3 -\sqrt2\phi_S\right) ({K^{*-}}^\mu\partial_\mu K_0^{\star+} - {K^{\star+}}^\mu\partial_\mu K_0^{\star-}) \n \\
  && -i\frac{g_1}{2} \left(\phi_N - \phi_3 -\sqrt2\phi_S\right) ({\bar K}^{\star 0\mu} \partial_\mu K^0 - {K^{\star 0}}^\mu \partial_\mu {\bar K}^0)
  - g_1\phi_N({a_1^-}^\mu\partial_\mu\pi^+ + {a_1^+}^\mu\partial_\mu\pi^-) \n \\
  && -\frac{g_1}{2}\left(\phi_N+\phi_3+\sqrt2\phi_S\right)\left({K_1^-}^\mu\partial_\mu K^+ +{K_1^+}^\mu\partial_\mu K^-\right) -
  \frac{g_1}{2} \left(\phi_N-\phi_3+\sqrt2\phi_S\right) ({\bar K}_1^{0\mu}\partial_\mu K^0 + {K_1^0}^\mu\partial_\mu \bar{K^0}) \n \\ 
  && -g_1 \sqrt2\phi_Sf_{1S}^\mu\partial_\mu\eta_S - g_1\Big[f_{1N}^\mu(\phi_N\partial_\mu \eta_N + \phi_3\partial_\mu\pi^0) + {a_1^0}^\mu
  (\phi_3\partial_\mu\eta_N + \phi_N\partial_\mu\pi^0)\Big] \;. 
\label{L_n_mix}
\end{eqnarray}
\end{widetext}
In order to calculate the tree-level meson masses, these mixing terms must be eliminated, i.e., the mass matrices have to be diagonalized. 
First we deal with mixings between nonets, then continue with the two-state mixings in the $N-3$ sectors of the vector and axial-vector nonets, and finally we resolve the three-state mixings in the $N-3-S$ sectors of the scalar and pseudoscalar nonets.

\subsubsection{Mixings between different nonets}
\label{SSSec:nonetmix}

In order to eliminate the mixings between different nonets as listed in \eref{L_n_mix}, the (axial-)vector fields have to be shifted by appropriately chosen derivative terms of the (pseudo)scalar fields. 
Such a shift spoils the canonical normalization of the (pseudo)scalar fields. 
Consequently, the (pseudo)scalar fields must be rescaled by adequate wave-function renormalization factors $Z_i$, which will subsequently result in the appearance of factors $Z_i^2$ in the expressions of the (pseudo)scalar squared masses. 
The situation is slightly more complicated in the $N-3-S$ sector of the axial-vector -- pseudoscalar mixing. 
The transformations are found to be,
\begin{subequations}\label{Eq:nmix_rho_pm}
   \begin{eqnarray}
    \rho_{\mu}^{\pm} &\rightarrow& \tilde{\rho}_{\mu}^{\pm} + Z_{a_0^{\pm}}w_{\rho^{\pm}}\partial_\mu \tilde{a}_0^{\pm}\;,\\
     a_0^\pm &\rightarrow& Z_{a_0^\pm}\tilde{a}_0^\pm\;,
   \end{eqnarray}
\end{subequations}
\begin{subequations}\label{Eq:nmix_Kstar_pm}
   \begin{eqnarray}
    K_{\mu}^{\star\pm} &\rightarrow&  \tilde{K}_{\mu}^{\star\pm} + Z_{K_0^{\star\pm}}w_{K^{\star\pm}} \partial_\mu \tilde{K}_0^{\star \pm }\;,\\
    K_0^{\star \pm} &\rightarrow& Z_{K_0^{\star\pm}} \tilde{K}_0^{\star \pm}\;,
   \end{eqnarray}
\end{subequations}
\begin{subequations}\label{Eq:nmix_Kstar_0}
   \begin{eqnarray}
    K_{\mu}^{\star 0,\bar{0}} &\rightarrow&  \tilde{K}_{\mu}^{\star 0,\bar{0}} + Z_{K_0^{\star 0}}w_{K^{\star 0, \bar{0}}}\partial_\mu \tilde{K}_0^{\star 0, \bar{0}}\;,\\
    K_0^{\star 0,\bar{0}} &\rightarrow& Z_{K_0^{\star 0}} \tilde{K}_0^{\star 0,\bar{0}}\;,
   \end{eqnarray}
\end{subequations}
\begin{subequations}\label{Eq:nmix_a1_pm}
   \begin{eqnarray}
    {a}_{1 \mu}^{\pm} &\rightarrow& {\tilde{a}}_{1 \mu}^{\pm} + Z_{\pi^{\pm}}w_{a_1^{\pm}} \partial_\mu \tilde{\pi}^{\pm}\;,\\
    \pi^\pm &\rightarrow& Z_{\pi^\pm}\tilde{\pi}^\pm\;,
   \end{eqnarray}
\end{subequations}
\begin{subequations}\label{Eq:nmix_K1_pm}
   \begin{eqnarray}
    K_{1 \mu}^{\pm} &\rightarrow&  \tilde{K}_{1 \mu}^{\pm} + Z_{K^{\pm}}w_{K_1^{\pm}} \partial_\mu \tilde{K}^{\pm}\;,\\
    K^{\pm} &\rightarrow& Z_{K^{\pm}} \tilde{K}^{\pm}\;,
   \end{eqnarray}
\end{subequations}
\begin{subequations}\label{Eq:nmix_K1_0}
   \begin{eqnarray}
    K_{1 \mu}^{ 0,\bar{0}} &\rightarrow&  \tilde{K}_{1 \mu}^{0,\bar{0}} + Z_{K^{0}}w_{K_1^{0}}\partial_\mu \tilde{K}^{0,\bar{0}}\;,\\
    K^{0,\bar{0}} &\rightarrow& Z_{K^{0}} \tilde{K}^{0,\bar{0}}\;,
   \end{eqnarray}
\end{subequations}
while the nonet mixing in the axial-vector -- pseudoscalar $N-3-S$ sector can be resolved by
\begin{subequations}\label{Eq:transf_axial_ps_N3}
   \begin{eqnarray}
    \left(
    \begin{array}{@{}c@{}}
    f_{1N} \\
    a_1^0
    \end{array}
    \right)^{\mu}
    &\rightarrow & 
    \left(
    \begin{array}{@{}c@{}}
    \tilde{f}_{1N} \\
    \tilde{a}_1^{\,0}
    \end{array}
    \right)^{\mu} + 
    \mathds{W} \left(
    \begin{array}{@{}c@{}}
    \partial^{\mu}\tilde\eta_N \\
    \partial^{\mu}\tilde\pi^0
    \end{array}
    \right)\;,\label{Eq:transf_axial_f1_a1} \\
    f_{1S}^\mu &\rightarrow& f_{1}^{H,\,\mu} + w_{f_{1S}}\partial^\mu\tilde\eta_S\;,\label{Eq:transf_axial_f1S} \\ 
    \eta_{N}&\rightarrow& \tilde\eta_{N}\;, \quad \pi^0\rightarrow \tilde\pi^0 \;, \quad  \eta_{S}\rightarrow \tilde\eta_{S} \;, \label{Eq:transf_axial_eta_pi}
\end{eqnarray}
\end{subequations}
where
\begin{subequations}
\begin{eqnarray}
  w_{\rho^{\pm}} &=& \pm i\frac{g_1\phi_3}{m_{\rho^{\pm}}^2}\;, \\
  w_{K^{\star\pm}} &=& \pm i\frac{g_1(\phi_N + \phi_3 -\sqrt2\phi_S)}{2m_{K^{\star\pm}}^2}\;,\\
  w_{K^{\star 0,\bar{0}}} &=& \pm i\frac{g_1(\phi_N - \phi_3 -\sqrt2\phi_S)}{2m_{K^{\star 0}}^2}\;, \\
  w_{a_1^{\pm}} &=& \frac{g_1\phi_N}{m_{a_1^{\pm}}^2}\;,\\
    w_{K_1^{\pm}} &=& \frac{g_1(\phi_N + \phi_3 + \sqrt2\phi_S)}{2m_{K_1^{\pm}}^2}\;, \\ 
  w_{K_1^{0}} &=& \frac{g_1(\phi_N - \phi_3 + \sqrt2\phi_S)}{2m_{K_1^{0}}^2}\;, \\
    w_{f_{1S}} &=& \frac{\sqrt2 g_1\phi_S}{m_{f_{1S}}^2}\;,
\end{eqnarray}
\end{subequations}
\begin{subequations}
   \begin{eqnarray}
     \mathds{W} &\equiv& \left(
  	\begin{array}{@{}cc@{}}
  	w_{\eta}^{f} & w_{\pi}^{f}\\
  	w_{\eta}^{a} & w_{\pi}^{a} 
  	\end{array}
  	\right)\;, \label{Eq:Wdef}\\
    w_{\eta}^{f} &=& \frac{g_1}{\det \mathds{M}^2_{A}} (\phi_N m^2_{a_1^{0}} -\phi_3 m^2_{f_{1N}a_1^0}) \;,\\
    w_{\pi}^{f} &=& \frac{g_1}{\det \mathds{M}^2_{A}}(-\phi_N m^2_{f_{1N}a_1^0} + \phi_3 m^2_{a_1^{0}}) \;,\\
    w_{\eta}^{a} &=& \frac{g_1}{\det \mathds{M}^2_{A}}(-\phi_N m^2_{f_{1N}a_1^0} + \phi_3 m^2_{f_{1N}}) \;,\\
    w_{\pi}^{a} &=& \frac{g_1}{\det \mathds{M}^2_{A}}(\phi_N m^2_{f_{1N}} - \phi_3 m^2_{f_{1N}a_1^0})\;,
   \end{eqnarray}
   \end{subequations}
   \begin{subequations}
\begin{eqnarray}
Z_{a_0^\pm}&=&\frac{m_{\rho^\pm}}{\sqrt{m_{\rho^\pm}^2-g_1^2(\phi_3)^2}}\;, \\ 
Z_{K_0^{\star\pm}} &=& \frac{2m_{K^{\star\pm}}}{\sqrt{4m_{K^{\star\pm}}^2-g_1^2(\phi_N+\phi_3-\sqrt2\phi_S)^2}}\;,\\ 
Z_{K_0^{\star 0}} &=& \frac{2m_{K^{\star 0}}}{\sqrt{4m_{K^{\star 0}}^2 - g_1^2(\phi_N - \phi_3 - \sqrt2 \phi_S)^2}}\;, \\ 
Z_{\pi^\pm}&=&\frac{m_{a_1^\pm}}{\sqrt{m_{a_1^\pm}^2-g_1^2\phi_N^2}}\;, \\
Z_{K^\pm}&=&\frac{2m_{K_1^\pm}}{\sqrt{4m_{K_1^\pm}^2-g_1^2(\phi_N+\phi_3+\sqrt2\phi_S)^2}}\;,\\
Z_{K^0}&=&\frac{2m_{K_1^0}}{\sqrt{4m_{K_1^0}^2-g_1^2(\phi_N-\phi_3+\sqrt2\phi_S)^2}}\;,
\end{eqnarray} 
\end{subequations}
and $\mathds{M}^2_{A}$ is the squared-mass matrix in the $N-3$ sector of the axial-vector nonet.

On the right-hand sides of Eqs.~\eqref{Eq:nmix_rho_pm} -- \eqref{Eq:transf_axial_ps_N3} the transformed fields are denoted by a tilde.  
After substituting  Eqs.~\eqref{Eq:nmix_rho_pm} -- \eqref{Eq:transf_axial_ps_N3} into the quadratic part of the Lagrangian the tildes are dropped, except for the $\tilde{f}_{1N}$, $\tilde{a}_1^0$, $\tilde{\eta}_N$, $\tilde{\pi}^0$, $\tilde{\eta}_S$ fields, where additional transformations are needed (see the next sections).  



\subsubsection{Two\,-\,state mixings in the $N-3$ sector of (axial-)vectors}
\label{SSSec:AVmixing}
The vector and axial-vector mass matrices in the $N-3$ sector are 	
\begin{subequations}
\begin{eqnarray}
\mathds{M}^2_{V} &=& \left(
\begin{array}{@{}cc@{}}
m^2_{\omega_{N}} & m^2_{\omega_{N}\rho^0} \\
m^2_{\omega_{N}\rho^0} & m^2_{\rho^{0}} 
\end{array}
\right)\;, \\
\mathds{M}^2_{A} &=& \left(
\begin{array}{@{}cc@{}}
m^2_{f_{1N}} & m^2_{f_{1N}a_1^0} \\
m^2_{f_{1N}a_1^0} & m^2_{a_1^{0}} 
\end{array}
\right)\;,
\end{eqnarray}
\end{subequations}
where the explicit form of the matrix elements $m^2_{\omega_{N}}$, $m^2_{\rho^{0}}$,
$m^2_{\omega_{N}\rho^0}$, $m^2_{f_{1N}}$, $m^2_{a_1^{0}}$,
$m^2_{f_{1N}a_1^0}$ are given by Eqs.~\eqref{Eq:omN} -- \eqref{Eq:omNrho0} and \eqref{Eq:f1N} -- \eqref{Eq:f1Na10}, respectively. 
The matrices $\mathds{M}^2_{V/A}$ can be diagonalized by orthogonal transformations,
\begin{subequations}
\begin{eqnarray}
{{\tilde{\mathds{M}}}^2}_{V/A} &=& \mathds{O}_{V/A}\, \mathds{M}^2_{V/A}\,\mathds{O}_{V/A}^T\;,\\ 
\mathds{O}_{V/A} &=&\left(
\begin{array}{@{}cc@{}}
\cos \vartheta_{V/A} & \sin \vartheta_{V/A} \\
-\sin \vartheta_{V/A} & \cos \vartheta_{V/A}
\end{array}
\right)\;.
\end{eqnarray}
\end{subequations}
Consequently, the resulting eigenvalues and mixing angles are 
\begin{subequations}
\begin{eqnarray}
m^2_{\omega/\rho^0} &=& \frac{1}{2}\Bigg(m^2_{\omega_{N}} + m^2_{\rho^{0}}\nonumber\\ 
&\pm&\sqrt{\Big(m^2_{\omega_{N}} - m^2_{\rho^{0}}\Big)^2 + 4 m^4_{\omega_{N}\rho^0}}\Bigg) \;,\\
m^2_{f_1^L/a_1^0} &=& \frac{1}{2}\Bigg(m^2_{f_{1N}} + m^2_{a_1^{0}}\nonumber\\ 
&\pm&\sqrt{\Big(m^2_{f_{1N}} - m^2_{a_1^{0}}\Big)^2 + 4 m^4_{f_{1N}a_1^0}}\Bigg)\;,\\
\tan (2\vartheta_{V}) &=& \frac{m^2_{\omega_{N}\rho^0}}{m^2_{\omega_{N}} -  m^2_{\rho^{0}}}\;,\\ 
\tan (2\vartheta_{A}) & = & \frac{m^2_{f_{1N}a_1^0}}{m^2_{f_{1N}} -  m^2_{a_1^{0}}}\;.
\end{eqnarray}
\end{subequations}
While the field transformations that should be performed in the Lagrangian are,
\begin{subequations}
\begin{eqnarray}
\label{Eq:ax_vec_twomix}
	\left(
	\begin{array}{@{}c@{}}
	\omega_{N} \\
	\rho^0
	\end{array}
	\right)^{\mu} &\rightarrow& 
	\mathds{O}_V^T \left(
	\begin{array}{@{}c@{}}
	\omega \\
	\rho^0
	\end{array}
	\right)^{\mu}\;,\\
	\left(
	\begin{array}{@{}c@{}}
		\tilde{f}_{1N} \\
		\tilde{a}_1^0
	\end{array}
	\right)^{\mu}
	&\rightarrow& 
	\mathds{O}_A^T\left(
	\begin{array}{@{}c@{}}
		f_{1}^{L} \\
		a_1^{\,0}
	\end{array}
	\right)^{\mu}\;,
\end{eqnarray}
\end{subequations}
here, if we combine the second transformation with \eref{Eq:transf_axial_f1_a1} we end up with the following transformation for the original $f_{1N}$ and $a_1^0$ axial-vector fields of the Lagrangian,
\begin{equation}
    	\left(
	\begin{array}{@{}c@{}}
		f_{1N} \\
		a_1^0
	\end{array}
	\right)^{\mu}
	\rightarrow 
	\mathds{O}_A^T\left(
	\begin{array}{@{}c@{}}
		f_{1}^{L} \\
		a_1^{\,0}
	\end{array}
	\right)^{\mu} + 
	\left(
	\begin{array}{@{}cc@{}}
		w_{\eta}^{f} & w_{\pi}^{f}\\
		w_{\eta}^{a} & w_{\pi}^{a} 
	\end{array}
	\right) \left(
	\begin{array}{@{}c@{}}
		\partial^{\mu}\tilde\eta_N \\
		\partial^{\mu}\tilde\pi^0
	\end{array}
	\right)\;.\label{Eq:transf_f1_a1_final}
\end{equation}
After the diagonalization the $\omega$, $\rho^0$ vector and the $f_1^L$, $a_1^0$ axial-vector fields correspond to the physical $\omega(782)$, $\rho^0(770)$, $f_1(1280)$, and $a_1^0(1186)$ states, respectively.

\subsubsection{Three\,-\,state mixings within the scalar nonet}
\label{SSSec:Scalar_N3Smix}

There is a three-state mixing in the $N-3-S$ sector of the scalar nonet among the $\sigma_N$, $a_0^0$, and $\sigma_S$ fields of the Lagrangian. 
This mixing can be resolved by a $3$\,-\,dimensional orthogonal transformation $\mathds{O}_S$ resulting in the squared-mass eigenvalues $\lambda_{f_0^L}$, $\lambda_{a_0^0}$, and $\lambda_{f_0^H}$ and eigenstates $f_0^L$, $a_0^0$, and $f_0^H$, respectively. 
The symmetric scalar squared-mass mixing matrix is
\begin{equation}
\mathds{M}^2_{S} = \left(
\begin{array}{@{}ccc@{}}
  m^2_{\sigma_{N}} & m^2_{\sigma_{N} a_0^0} & m^2_{\sigma_{N} \sigma_{S}}\\
  m^2_{\sigma_{N} a_0^0} & m^2_{a_0^0} & m^2_{a_0^0 \sigma_{S}}\\
  m^2_{\sigma_{N} \sigma_{S}} & m^2_{a_0^0 \sigma_{S}} & m^2_{\sigma_{S}}
\end{array}
\right)\;,
\end{equation}
where the explicit form of the matrix elements are given by Eqs.~\eqref{mS_NN} -- \eqref{mS_3S}. 
The diagonalization can be written as
\begin{subequations}
\begin{eqnarray}
  \mathds{O}_{S}\, \mathds{M}^2_{S}\,\mathds{O}_{S}^T &=& \tilde{\mathds{M}}^2_{S} \equiv \diag(\lambda_{S_l}, \lambda_{S_m}, \lambda_{S_h})\;,\\
  \text{requiring}&& \lambda_{S_l} \leq \lambda_{S_m} < \lambda_{S_h}\;, 
\end{eqnarray}
\end{subequations}
where $\mathds{O}_{S}$ and the eigenvalues are to be calculated numerically. 
The particle assignment of this sector is not as straightforward as for the others due to the fact that there are two $a_0$'s and five $f_0$'s below $2$~GeV according to PDG data \cite{Workman:2022ynf}. 
In this sector the squared-mass eigenvalues and the field transformations are given by,
\begin{subequations}
\begin{eqnarray}
m^2_{f_0^L} &=& \lambda_{S_l}\;,\quad m^2_{a_0^0} = \lambda_{S_m}\;,\quad m^2_{f_0^H} = \lambda_{S_h} 
\;,\\
&& \left(
\begin{array}{@{}c@{}}
     \sigma_N \\
     a_0^0 \\
     \sigma_S
\end{array}
\right) \longrightarrow \mathds{O}_S^T \left(
\begin{array}{@{}c@{}}
     f_0^L \\
     a_0^0 \\
     f_0^H
\end{array}
\right)\;.
\end{eqnarray}
\end{subequations}
Here we tried all possible assignments and chose the one that resulted in the lowest $\chi^2$, but due to the large error in this sector, this part of the fit is not very restrictive (see also \secref{Sec:results} for the specific assignment).   

\subsubsection{Three\,-\,state mixings within the pseudoscalar nonet}
\label{SSSec:PseudoScalar_N3Smix}
Even after the transformations \eqref{Eq:transf_axial_f1S}, \eqref{Eq:transf_axial_eta_pi}, and \eqref{Eq:transf_f1_a1_final} have been applied, there is still a three-state mixing within the $N-3-S$ sector of the pseudoscalar nonet, which concerns both the kinetic and the mass terms. 
The affected fields are $\mathbf{x}^T = (\tilde\eta_N, \tilde\pi^0, \tilde\eta_S)$. 
The relevant part of the Lagrangian reads
\begin{equation}
  \mathcal{L}_{P_{N3S}} = \frac12 \partial^{\mu} \mathbf{x}^T \mathds{D}_{P} \partial_{\mu}\mathbf{x} - \frac12 \mathbf{x}^T \mathds{M}_{P}^2 \mathbf{x}\;,
\end{equation}
where
\begin{subequations}
\begin{eqnarray}
  \mathds{D}_{P} &=& \left(\begin{array}{@{}c|c@{}}
  \mathds{1}-g_1\mathds{W^{T}}\mathds{N} & \mathbf{0} \\\hline
  \mathbf{0} & 1-\frac{2g_1^2\phi_S^2}{m_{f_{1S}}^2}
  \end{array}\right) \;, \\
 \mathds{N} &=& \left(\begin{array}{@{}cc@{}} 
   \phi_N & \phi_3 \\
   \phi_3 & \phi_N
 \end{array}\right)\;, \\
\mathds{M}^2_{P} &=& \left(
\begin{array}{@{}ccc@{}}
  m^2_{\eta_{N}} & m^2_{\eta_{N} \pi^0} & m^2_{\eta_{N} \eta_{S}}\\
  m^2_{\eta_{N} \pi^0} & m^2_{\pi^0} & m^2_{\pi^0 \eta_{S}}\\
  m^2_{\eta_{N} \eta_{S}} & m^2_{\pi^0 \eta_{S}} & m^2_{\eta_{S}}
\end{array}
\right)\;,
\end{eqnarray}
\end{subequations}
and $\mathds{W}$ is defined in \eref{Eq:Wdef}, while the explicit expressions for the elements of the squared-mass matrix $\mathds{M}^2_{P}$ are given by Eqs.~\eqref{mP_NN} -- \eqref{mP_3S}.
The matrix $\mathds{D}_{P}$ is symmetric and can be diagonalized by a rotation in the $N-3$ plane, 
\begin{equation}
\mathds{O}_{D} = \left(
  \begin{array}{@{}ccc@{}}
    \cos \vartheta_{D} & \sin \vartheta_{D} & 0 \\
    -\sin \vartheta_{D} & \cos \vartheta_{D} & 0 \\
    0 & 0 & 1 \\
  \end{array}
  \right)\;.
\end{equation}
Consequently, $\mathcal{L}_{P_{N3S}}$ can be written as
\begin{equation}
  \mathcal{L}_{P_{N3S}} = \frac12 \partial^{\mu} \mathbf{y}^T \tilde{\mathds{D}}_{P} \partial_{\mu}\mathbf{y} - \frac12 \mathbf{y}^T \mathds{M}_{P}^{\prime \,2} \mathbf{y}\;,\quad \mathbf{y} \equiv \mathds{O}_{D} \mathbf{x}\; ,
\end{equation}
with
\begin{subequations}
\begin{align}
  \tilde{\mathds{D}}_{P} &\equiv \mathds{O}_{D} \mathds{D}_{P} \mathds{O}_{D}^T \nonumber\\
  &= \diag\left(\lambda_{D_1}, \lambda_{D_2}, 1-\frac{2g_1^2\phi_S^2}{m_{f_{1S}}^2}\right)\;,\\
  \lambda_{D_{1,2}} &=  \frac{\left( \Tr\mathds{D}_P^{2\times 2} \pm \sqrt{ {(\Tr\mathds{D}_P^{2\times 2}})^2 - 4 \det{\mathds{D}_P^{2\times 2}}} \right)}{2}\;, \\
  \mathds{M}_{P}^{\prime \,2} &\equiv \mathds{O}_{D} \mathds{M}_{P}^2 \mathds{O}_{D}^T\;. 
\end{align}
\end{subequations}
Here, $\mathds{D}_{P}^{2\times 2} \equiv \mathds{1}-g_1\mathds{W^{T}}\mathds{N}$ denotes the upper left $(2\times 2)$ block of $\mathds{D}_{P}$, which is assumed to be positive definite. 
Accordingly, we can define
\begin{subequations}
\begin{align}
  Z_{P_N} &= \frac{1}{\sqrt{\lambda_{D_1}}}\;,\quad Z_{P_3} = \frac{1}{\sqrt{\lambda_{D_2}}}\;,\\
  Z_{\eta_S} &= \frac{m_{f_{1S}}}{\sqrt{m_{f_{1S}}^2-2g_1^2\phi_S^2}}\;, \\
  \mathds{Z}_{P} &= \diag\left(Z_{P_N}, Z_{P_3}, Z_{\eta_S}\right)\;, \quad \mathbf{y}^{\prime} = \mathds{Z}_{P}^{-1} \mathbf{y}\;,
\end{align}
\end{subequations}
which subsequently leads to 
\begin{subequations}
\begin{align}
  \mathcal{L}_{P_{N3S}} &=
  \frac12 \partial^{\mu} {\mathbf{y}^{\prime}}^T \partial_{\mu}\mathbf{y}^{\prime} - \frac12 {\mathbf{y}^{\prime}}^T \tilde{\mathds{M}}_{P}^2 \mathbf{y}^{\prime}\;,\label{Eq:lagr_pi_eta}\\
  \tilde{\mathds{M}}_{P}^2 &\equiv \mathds{Z}_{P} \mathds{M}_{P}^{\prime \, 2} \mathds{Z}_{P} \;.
\end{align}
\end{subequations}
Now the kinetic part has become canonical, and the symmetric matrix $\tilde{\mathds{M}}_{P}^2$ can be diagonalized by an orthogonal transformation $\mathds{O}_M$,
\begin{subequations}
\begin{eqnarray}
  \mathds{O}_{M}\, \tilde{\mathds{M}}_{P}^2\,\mathds{O}_{M}^T &=& \diag\left(\lambda_{P_l}, \lambda_{P_m}\;, \lambda_{P_h}\right),\label{Eq:M_P_diag}\\
  \text{requiring} &&\lambda_{P_l} < \lambda_{P_m} < \lambda_{P_h}\;, 
\end{eqnarray}
\end{subequations}
where $\mathds{O}_{M}$ and the eigenvalues are to be calculated numerically. 
The field transformations are given by,
\begin{equation}
\left(
\begin{array}{@{}c@{}}
     \eta_N \\
     \pi^0 \\
     \eta_S
\end{array}
\right) \longrightarrow
\left(
\begin{array}{@{}c@{}}
     \tilde\eta_N \\
     \tilde\pi^0 \\
     \tilde\eta_S
\end{array}
\right) = 
\mathds{O}_{D}^T \mathds{Z}_{P} \mathds{O}_{M}^T \left(
\begin{array}{@{}c@{}}
     \pi^0 \\
     \eta \\
     \eta'
\end{array}
\right)\equiv \mathds{O}_{P} \left(
\begin{array}{@{}c@{}}
     \pi^0 \\
     \eta \\
     \eta^{\prime}
\end{array}
\right)\;.
\label{Eq:pseudoN3S_transf}
\end{equation}
The particle assignment of this sector is straightforward,
\begin{equation}
  m^2_{\pi} = \lambda_{P_l}\;, \quad  m^2_{\eta} = \lambda_{P_m}\;, \quad m^2_{\eta'} = \lambda_{P_h}\;,
\end{equation}
and the resulting field vector contains the physical $\pi$, $\eta$, $\eta'$ fields,
\begin{equation}
\label{Eq:phys_pi_eta_fields}
  \mathds{O}_{M} \mathbf{y}^{\prime} = \mathbf{y}^{\text{ph}} \equiv (\pi^0, \eta, \eta')^T\;.
\end{equation}

It is worth to note that $\mathds{O}_{P}$ is not an orthogonal transformation. 
Using Eqs.~\eqref{Eq:transf_axial_ps_N3}, \eqref{Eq:transf_f1_a1_final}, and \eqref{Eq:pseudoN3S_transf}, finally the transformations of the $f_{1N}^{\mu}$, $a_1^{0\mu}$, and $f_{1S}^\mu$ fields into physical fields are,
\begin{subequations}
\begin{eqnarray}
\left(
\begin{array}{@{}c@{}}
f_{1N} \\
a_1^0
\end{array}
\right)^{\mu}
& \rightarrow & 
\mathds{O}_A^T\left(
\begin{array}{@{}c@{}}
f_{1}^{L} \\
a_1^{\,0}
\end{array}
\right)^{\mu} \\
&+& \left(
\begin{array}{@{}cc@{}}
w_{\eta}^{f} & w_{\pi}^{f}\\
w_{\eta}^{a} & w_{\pi}^{a} 
\end{array}
\right) 
\left(
\begin{array}{@{}ccc@{}}
{\mathds{O}_{P}}_{11} & {\mathds{O}_{P}}_{12} & {\mathds{O}_{P}}_{13}\\
{\mathds{O}_{P}}_{21} & {\mathds{O}_{P}}_{22} & {\mathds{O}_{P}}_{23}
\end{array}
\right)
\left(
\begin{array}{@{}c@{}}
\partial^{\mu}\pi^0 \\
\partial^{\mu}\eta \\
\partial^{\mu}\eta^{\prime}
\end{array}
\right)	\;,\nonumber\\
f_{1S}^\mu &\rightarrow& f_{1}^{H,\,\mu} \\
&+& w_{f_{1S}}\left({\mathds{O}_{P}}_{31} \partial^\mu\pi^0 + {\mathds{O}_{P}}_{32}\partial^\mu \eta + {\mathds{O}_{P}}_{33} \partial^\mu\eta^{\prime}\right)\;.\nonumber
\end{eqnarray}
\end{subequations}

\section{Tree-level physical masses}
\label{Sec:phys_mass}

After taking care of all the mixings, the squared-mass eigenvalues for the pseudoscalars, scalars, vectors, and axial-vectors, respectively, are given by
\begin{subequations}
\begin{align}
M_{\pi^{\pm}}^2 &= Z_{\pi^{\pm}}^2 m_{\pi^{\pm}}^2 + m^2_{\text{em},P}\;,\\
M_{K^{\pm}}^2 &= Z_{K^{\pm}}^2 m_{K^{\pm}}^2 + m^2_{\text{em},P} + m^2_{\text{em},P_K}\;,\\ 
M_{K^{0}}^2 &= Z_{K^{0}}^2 m_{K^{0}}^2\;,\\
M_{\eta}^2 &= m^2_{\eta}\;, \\  
M_{\pi^0}^2 &= m^2_{\pi^0}\;, \\ 
M_{\eta^{\prime}}^2 &= m^2_{\eta^{\prime}}\;, 
\end{align}
\end{subequations}
\begin{subequations}
\begin{align}
M_{a_0^{\pm}}^2 &= Z_{a_0^{\pm}}^2 m_{a_0^{\pm}}^2 + m^2_{\text{em},S}\;,\\
M_{K_0^{\star\pm}}^2 &= Z_{K_0^{\star\pm}}^2 m_{K_0^{\star\pm}}^2 + m^2_{\text{em},S} + m^2_{\text{em},S_K}\;,\\ 
M_{K_0^{\star 0}}^2 &= Z_{K_0^{\star 0}}^2 m_{K_0^{\star 0}}^2\;,\\
M_{f_0^L}^2 &= m^2_{f_0^L}\;,\\  
M_{a_0^0}^2 &= m^2_{a_0^0}\;, \\ 
M_{f_0^H}^2 &= m^2_{f_0^H}\;,
\end{align}
\end{subequations}
\begin{subequations}
\begin{align}
M_{\rho^{\pm}}^2 &= m_{\rho^{\pm}}^2 + m^2_{\text{em},V}\;, \\
M_{K^{\star\pm}}^2 &= m_{K^{\star\pm}}^2 + m^2_{\text{em},V} + m^2_{\text{em},V_K}\;,\\ 
M_{K^{\star 0}}^2 &= m_{K^{\star 0}}^2\;, \\
M_{\omega/\rho^0}^2 &=\frac{1}{2}\Bigg(m^2_{\omega_{N}} + m^2_{\rho^{0}}\\ 
&\pm \sqrt{\Big(m^2_{\omega_{N}} - m^2_{\rho^{0}}\Big)^2 + 4 m^2_{\omega_{N}\rho^0}}\Bigg)\;, \nonumber\\
M_{\phi}^2 &= m_{\omega_S}^2\;,
\end{align}
\end{subequations}
\begin{subequations}
\begin{align}
M_{a_1^{\pm}}^2 &= m_{a_1^{\pm}}^2 + m^2_{\text{em},A}\;, \\
M_{K_1^{\pm}}^2 &= m_{K_1^{\pm}}^2 + m^2_{\text{em},A} + m^2_{\text{em},A_K}\;, \\ 
M_{K_1^0}^2 &= m_{K_1^0}^2 \\
M_{f_{1}^L/a_1^0}^2 &= \frac{1}{2}\Bigg(m^2_{f_{1N}} + m^2_{a_1^{0}}\nonumber\\ 
&\pm \sqrt{\Big(m^2_{f_{1N}} - m^2_{a_1^{0}}\Big)^2 + 4 m^2_{f_{1N}a_1^0}}\Bigg)\;,\\
M_{f_1^H}^2 &= m_{f_{1S}}^2\;, \label{Eq:m_f1H}
\end{align}
\end{subequations}
where the explicit expressions for the tree-level masses $m_{x}^2$  are given in \apref{App:mass_matrix} (squared-mass matrix elements), \secref{SSSec:Scalar_N3Smix} (scalar $N-3-S$ squared-mass eigenvalues), and \secref{SSSec:PseudoScalar_N3Smix} (pseudoscalar $N-3-S$ squared-mass eigenvalues). 
Moreover, we introduced electromagnetic mass terms in each nonet, namely, $m^2_{\text{em},S}$, $m^2_{\text{em},P}$, $m^2_{\text{em},V}$, $m^2_{\text{em},A}$, and $m^2_{\text{em},S_K}$, $m^2_{\text{em},P_K}$, $ m^2_{\text{em},V_K}$, $m^2_{\text{em},A_K}$, which are of the same order as the contribution from isospin-symmetry breaking. 
We will consider three different cases. In the first, we consider Dashen's theorem to be valid (see the introduction), so $m^2_{\text{em},S_K} = m^2_{\text{em},P_K} = m^2_{\text{em},V_K} = m^2_{\text{em},A_K} = 0$. 
This means an electromagnetic mass contribution for each charged particle in each sector. 
The second case is when Dashen's theorem is violated, but the additional electromagnetic contribution for the kaonic particles is the same in all sectors, $m^2_{\text{em},S_K} = m^2_{\text{em},P_K} = m^2_{\text{em},V_K} = m^2_{\text{em},A_K} \equiv  m^2_{\text{em},K}$. 
The third -- and most general -- case is where these contributions can be different in each sector. 

\section{Tree-level decay widths}
\label{Sec:dec_widths}	
		
We start from the usual tree-level expression for two-body decay, 
\begin{equation}
	\Gamma_{A\to BC} = \frac{k}{8\pi M_A^2}\left| \mathcal{M}_{A\to BC}\right|^2\;, 
\end{equation}
where $k\equiv \sqrt{\vec{k}^2} $ is the absolute value of the three-momentum of the produced particles $B, C$ in the rest frame of the decaying particle $A$ and $\mathcal{M}_{A\to BC}$ is the tree-level matrix element of the process.	

In a straightforward, but lengthy calculation we computed tree-level decay widths for the following processes:\\
(i) \textit{Vector-meson decays.}
\begin{subequations}
\begin{align}
    \rho^{0} &\to \pi^{+}\pi^{-}\;,  \\
     \rho^{-}&\to \pi^{-}\pi^{0}\;, \quad \rho^{-} \to \pi^{-} \eta\;,\\
    \omega &\to \pi^+\pi^- \;,\\
    \bar K^{\star 0} &\to \pi^{0,+}K^{\bar{0},-}\;, \\ 
    K^{\star -}& \to  \pi^{0,-}K^{-,\bar 0} \;,\\
    \Phi &\to  K^0 \bar K^0\;, \quad \Phi\to K^+ K^- \;.
\end{align}
\end{subequations}
(ii) \textit{Axial-vector-meson decays.}
\begin{subequations}
\begin{align}
    a_{1}^0 &\to \rho^{+}\pi^{-}\;, \\
    a_{1}^{-} &\to \pi^{-}\gamma \;,\quad a_{1}^-\to\rho^{-,0}\pi^{0,-}\;,\\
    f_1^H &\to K^{\star \pm,0,\bar 0}K^{\mp,\bar 0,0}\;.
\end{align}
\end{subequations}
(iii) \textit{Scalar-meson decays.}
\begin{subequations}
\begin{align}
\bar K_0^{\star 0} &\to \pi^{0,+}K^{\bar{0},-}\;,  \\
K_0^{\star -}& \to  \pi^{0,-}K^{-, \bar{0}}\;,\\
a_0^{0} &\to \pi^{0}\eta\;,\quad a_0^{0}\to \pi^{0}\eta^{\prime}\;, \quad a_0^0 \to K^{0,+}K^{\bar{0},-}\;,\\
a_0^{-} &\to \pi^{-}\eta\;, \quad a_0^{-}\to \pi^{-}\eta^{\prime}\;,\quad a_0^{-}\to K^{0}K^{-} \;,\\
f_0^{L/H}& \to \pi^{0,+}\pi^{0,-}\;, \quad f_0^{L/H} \to K^{0,+}K^{\bar{0},-}\;.
\end{align}
\end{subequations}
It should be noted that above we only listed the negatively-charged particle decays (in case of charged particles), because the charge-conjugated decays have the same decay width. 
Similarly, in the case of the kaonic particles, we gave only the decays of the conjugate particles, like $\bar K^{\star 0}$ and $\bar K_0^{\star 0}$, since their charge-conjugated partners ($K^{\star 0}$ and $K_0^{\star 0}$) have the same decay width in the given channel. The explicit expressions for all decay widths can be found in \apref{App:expl_decays}.

\section{Fit and results}
\label{Sec:results}

As discussed in detail in Ref.\ \cite{Parganlija:2012fy} there are $13$ unknown parameters in the Lagrangian in the case of isospin symmetry, namely $m_0^2$, $m_1^2$, $c_1$, $\delta_S$, $g_1$, $g_2$, $\zeta_N$, $\zeta_S$, $\lambda_1$, $\lambda_2$, $h_1$, $h_2$, and $h_3$.\footnote{In Ref.\ \cite{Parganlija:2012fy} we have used $h_{0N}$ and $h_{0S}$ to denote the explicit symmetry-breaking parameters instead of $\zeta_N$, $\zeta_S$.} 
In addition to the parameters, the condensates $\phi_{N}$, $\phi_{S}$ are also unknown. 
However, the external fields $\zeta_N$ and $\zeta_S$ can be calculated using the field equations at $T=0$ once $\phi_{N}$ and $\phi_{S}$ are known, thus instead of these fields, the condensates can be used in the fitting procedure. 
Consequently, the unknown parameters to be determined in the case of isospin symmetry are: $m_0^2$, $m_1^2$, $c_1$, $\delta_S$, $g_1$, $g_2$, $\phi_N$, $\phi_S$, $\lambda_1$, $\lambda_2$, $h_1$, $h_2$, and $h_3$.  
After determining these parameters, the fields $\zeta_N$, $\zeta_S$ can be calculated via Eqs.~\eqref{Eq:zeta_N}, \eqref{Eq:zeta_S}. 

In the case of isospin-symmetry violation there are several changes in the parameter set:
\begin{itemize}
    \item The fields $\delta_u$, $\delta_d$, and $\delta_s$, as well as $m_1^2$ appear in all  the meson masses (see \apref{App:mass_matrix}) in the following three combinations:
    \begin{align*}
        \tilde{m}_1^2 &= m_1^2 + \delta_u + \delta_d\;, \\
        \tilde{\delta}_s &= \delta_s - \frac{1}{2}(\delta_u + \delta_d) \;,\\
        \delta_3 &= \delta_u - \delta_d  \;.   
    \end{align*}
Thus, instead of $m_1^2$ and $\delta_S$ ($\delta_N = \delta_u+\delta_d$ can be incorporated into $m_1^2$, similarly as in Ref.\ \cite{Parganlija:2012fy}), we can fit $\tilde{m}_1^2$ and $\tilde{\delta}_s$. 
All in all, instead of the parameters $\delta_u$, $\delta_d$, $\delta_s$, and $m_1^2$ we have to fit only $\tilde{m}_1^2$, $\tilde{\delta}_s$, and $\delta_3$. 
    \item There is a new condensate $\phi_3$.
    \item There are four, five, or eight electromagnetic mass contributions as it is described below Eq.\ \eqref{Eq:m_f1H}. 
    In the first case these are $m^2_{\text{em},S}$, $m^2_{\text{em},P}$, $m^2_{\text{em},V}$, and $m^2_{\text{em},A}$ (Dashen theorem-respecting scenario or DS), in the second case we have $m^2_{\text{em},S}$, $m^2_{\text{em},P}$, $m^2_{\text{em},V}$, $m^2_{\text{em},A}$, and $m^2_{\text{em},K}$ (Dashen theorem-violating scenario I or DVS-I), while in the third case these are $m^2_{\text{em},S}$, $m^2_{\text{em},P}$, $m^2_{\text{em},V}$, $m^2_{\text{em},A}$, $m^2_{\text{em},S_K}$, $m^2_{\text{em},P_K}$, $m^2_{\text{em},V_K}$, and $m^2_{\text{em},A_K}$ (Dashen theorem-violating scenario II or DVS-II). 
    It is worth noting that during the fit an upper bound of $10$~MeV was introduced, since a higher value for the electromagnetic correction to the mass is physically unrealistic.
    \item Finally, there are two additional mass terms, $\delta m^2_V$ in the vector and $\delta m^2_A$ in the axial-vector sector, in order to generate a splitting between the $\rho^0$ and $\omega$ and similarly between the $a_1^0$ and $f_1^L$ masses.
\end{itemize}
Altogether there are 21, 22, or 25 unknown parameters depending on the handling of the electromagnetic mass contributions (see above) in the isospin-symmetry broken case -- compared to the 13 parameters in the isospin-symmetric case -- namely $m_0^2$, $m_1^2$, $c_1$, $\delta_S$, $\delta_3$, $g_1$, $g_2$, $\phi_N$, $\phi_S$, $\phi_3$, $\lambda_1$, $\lambda_2$, $h_1$, $h_2$, $h_3$, $m^2_{\text{em},S}$, $m^2_{\text{em},P}$, $m^2_{\text{em},V}$, $m^2_{\text{em},A}$, $\delta m^2_V$, $\delta m^2_A$, and optionally $m^2_{\text{em},K}$, or $m^2_{\text{em},S_K}$, $m^2_{\text{em},P_K}$, $m^2_{\text{em},V_K}$, and $m^2_{\text{em},A_K}$.

In order to determine these parameters, we calculated physical quantities at tree level and used a multiparametric $\chi^2$ minimization method (MINUIT \cite{James:1975dr}) similar to Ref.\ \cite{Parganlija:2012fy}. 
The values for the physical quantities are taken from  experiment, that is, from the PDG \cite{Workman:2022ynf}. 
More precisely, we take only the mean value from the PDG, and in most cases we use an artificially increased error instead of the experimental value if the latter is smaller than a prescribed percentage (see below). 
We do this because some of the masses are known with very high precision, e.g., the mass of the $\eta$ meson is known with $0.003\%$ precision, and we do not expect such a phenomenological model to describe a mass with that high accuracy. 
Our expected accuracy is around several percent. 
Since the effect of isospin-symmetry breaking is about $3\%$ for instance for the mass of the pion and below $1\%$ for all the other quantities, we cannot simply fit the neutral and charged quantities separately (our increased error would be greater than the effect). 
Thus, we chose to fit instead the isospin-averaged neutral and charged quantities and their differences.
We use a small artificial minimal error of 5\% for the isospin-averaged masses of those particles which can be modeled very precisely within our model, i.e., the pseudoscalars and vector mesons. 
For the corresponding mass differences we use a somewhat larger minimal error of 20\%.
The same minimal error is also used for the axial-vectors and the scalar $K_0^{\star}$ and $a_0$, while the minimal error is $50\%$ for the masses of the $f_0^{L/H}$ mesons, since the latter cannot be described very precisely within our model, because it does not contain the other three isoscalar-scalar states. 
For the sake of completeness we list here all the values used for our fit (it should be noted that the decay widths of the scalar $f_0^{L/H}$ fields are not used in the fit): 
\begin{enumerate}[(i)]
    \item \emph{Weak-decay constants.} For the pion and kaon decay constants we use \cite{Workman:2022ynf},
    \begin{subequations}
    \begin{align}
        f_{\pi} &= 92.06\pm 4.60\,\text{MeV},\\ 
        f_K &= 110.10\pm 5.51\,\text{MeV}, 
         \end{align}
         \end{subequations}
        where we have also applied the $5\%$ minimal-error prescription. 
        The decay constants are related to the condensates through the PCAC (partially conserved axial current) relations, which in our model leads to,
        \begin{subequations}
        \begin{align}
        f_{\pi^{\pm}} &= \frac{\phi_N}{Z_{\pi^{\pm}}}\;,\\
        f_{\pi^{0}} &= \left({\mathds{O}^{-1}_{P}}_{11}+{\mathds{O}^{-1}_{P}}_{12}\right)(\phi_N + \phi_3) \label{Eq:f_pi_0} \nonumber\\
        &+{\mathds{O}^{-1}_{P}}_{13}\phi_S\;,\\
        f_{K^{\pm}} &= \frac{\phi_N + \phi_3 + \sqrt{2}\phi_S}{2 Z_{K^{\pm}}}\;, \label{Eq:f_K_pm}\\
        f_{K^0} &= \frac{\phi_N - \phi_3 + \sqrt{2}\phi_S}{2 Z_{K^{0}}}\;, \label{Eq:f_K_0}
        \end{align}
        \end{subequations}
        where the somewhat unusual expression for $f_{\pi^0}$ is due to the mixing in the $N-3-S$ sector.
        Its explicit form is derived in \apref{App:fpi0_weak_dec}.        
        We have fitted only the charged weak-decay constants, since from the combination of measurements and theory only this can be determined (see Ref.\ \cite{Workman:2022ynf} part 72. Leptonic Decays of Charged Pseudoscalar Mesons),
        \begin{subequations}
        \begin{align}
            f_{\pi} &\mbeq f_{\pi^\pm}\;,\\
            f_{K} &\mbeq f_{K^\pm}\;.
        \end{align}
        \end{subequations}
        It is worth noting that $f_{\pi^{0}}$ and $f_{K^0}$ will be predictions. 
    \item \emph{Pseudoscalar masses.} For the charged particles, we use the isospin averages and differences,
    \begin{subequations}
    \begin{align}
        \bar M_{\pi} &= \frac{M_{\pi^0} + 2 M_{\pi^{\pm}}}{3} = 138.04\pm 6.90\,\text{MeV}\;,\\ 
        \Delta M_{\pi} &= M_{\pi^{0}} - M_{\pi^{\pm}} = -4.59\pm 0.92 \,\text{MeV}\;,\\ 
        \bar M_{K} &= \frac{M_{K^0} + M_{K^{\pm}}}{2} = 495.64 \pm 24.78\,\text{MeV}\;,\\
        \Delta M_{K} &= M_{K^{0}} - M_{K^{\pm}} = 3.93 \pm 0.79\,\text{MeV}\;, \\ 
        M_{\eta} &= 547.86 \pm 27.39\,\text{MeV}\;,\\ 
        M_{\eta^{\prime}} &= 957.78 \pm 47.89\,\text{MeV}\;. 
    \end{align}
    \end{subequations}
    \item \emph{Scalar-meson masses.} For the charged particles, we use only the isospin averages, since there are no data for the differences,
    \begin{subequations}
    \begin{align}
        \bar M_{a_0} &= \frac{M_{a_0^0} + 2 M_{a_0^{\pm}}}{3} = 1474 \pm 294.8 \,\text{MeV}\;,\\
        \bar M_{K_0^{\star}} &= \frac{M_{K_0^{\star 0}} + M_{K_0^{\star \pm}}}{2} = 1425 \pm 285\,\text{MeV}\;,\\
        M_{f_0^L} &= 1350 \pm 675\,\text{MeV}\;, \\
        M_{f_0^H} &= 1733 \pm 867\,\text{MeV}\;. 
    \end{align}
    \end{subequations}
    Here, $a_0$ is assigned to $a_0(1450)$ and $K_0^{\star}$ to $K_0^{\star}(1430)$.\footnote{For the sake of completeness, we also checked the other options, when $a_0$ and $K_0^{\star}$ are assigned to $a_0(980)$ and $K_0^{\star}(700)$.} 
    For the two $f_0$ any two combination from the five states $f_0(500)$, $f_0(980)$, $f_0(1370)$, $f_0(1500)$ and $f_0(1710)$ has been checked previously -- i.e., in the isospin-symmetric case --- and the assignment $f_0^L = f_0(1370)$, $f_0^H = f_0(1710)$ was found to be favored \cite{Parganlija:2012fy}. 
    Thus, we started our minimization procedure with this assignment.    
    \item \emph{Vector-meson masses.} For the charged particles, we use the isospin averages and differences,
    \begin{subequations}
     \begin{align}
        \bar M_{\rho} &= \frac{M_{\rho^0} + 2 M_{\rho^{\pm}}}{3} = 775.16 \pm 38.76 \,\text{MeV}\;,\\ 
        \Delta M_{\rho} &= M_{\rho^{0}} - M_{\rho^{\pm}} = 0.15 \pm 0.57 \,\text{MeV}\;,\\
        \bar M_{K^{\star}} &= \frac{M_{K^{\star 0}} + M_{K^{\star \pm}}}{2} =  895.51 \pm 44.78 \,\text{MeV}\;,\\
        \Delta M_{K^{\star}} &= M_{K^{\star 0}} - M_{K^{\star \pm}} = 0.08 \pm 0.94 \,\text{MeV}\;, \\ 
        M_{\omega} &=  782.66 \pm 38.13\,\text{MeV}\;,\\ 
        M_{\phi} &=  1019.46 \pm 50.97 \,\text{MeV}\;. 
    \end{align}
    \end{subequations}
    \item \emph{Axial-vector masses.} For the charged particles, we use only the isospin averages, since there are no data for the differences,
    \begin{subequations}
    \begin{align}
        \bar M_{a_1} &= \frac{M_{a_1^0} + 2 M_{a_1^{\pm}}}{3} =  1230 \pm 246 \,\text{MeV}\;,\\
        \bar M_{K_1} &= \frac{M_{K_1^{0}} + M_{K_1^{\pm}}}{2} =  1253 \pm 250.6 \,\text{MeV}\;,\\
        M_{f_1^L} &=  1281.9 \pm  256.38 \,\text{MeV}\;, \\
        M_{f_1^H} &=  1426.3 \pm 285.26 \,\text{MeV}\;. 
    \end{align}
    \end{subequations}
    \item \emph{Vector-meson decays.} We use
    \begin{subequations}
    \begin{align}
     \bar \Gamma_{\rho \to \pi\pi} &= \frac{\Gamma_{\rho^{0}\to \pi^{+}\pi^{-}} + 2\Gamma_{\rho^{\pm}\to \pi^{\pm}\pi^{0}}}{3} \nonumber \\
     &= 148.533 \pm 7.426 \,\text{MeV}\;,  \\
     \Delta\Gamma_{\rho \to \pi\pi} &=  \Gamma_{\rho^{0}\to \pi^{+}\pi^{-}} -\Gamma_{\rho^{\pm}\to \pi^{\pm}\pi^{0}} \nonumber\\
     &= -1.7 \pm 1.6 \,\text{MeV}\;, \\
     \bar \Gamma_{K^{\star}\to K \pi} &= \frac{\Gamma_{\bar K^{\star 0}\to \pi^{0,+}K^{\bar{0},-}} + \Gamma_{K^{\star -}\to \pi^{0,-}K^{-,\bar 0}}}{2} \nonumber\\
     &= 46.75 \pm 2.34 \,\text{MeV} \;, \\
     \Delta \Gamma_{K^{\star}\to K \pi} &= \Gamma_{\bar K^{\star 0}\to \pi^{0,+}K^{\bar{0},-}} - \Gamma_{K^{\star -}\to \pi^{0,-}K^{-,\bar 0}} \nonumber \\ 
     &= 1.1 \pm 1.8 \,\text{MeV} \;, \\
     \Gamma_{\omega\to \pi^{+}\pi^{-}} &= 0.133 \pm 0.026 \,\text{MeV} \;, \\
     \bar \Gamma_{\phi \to K K} &= \frac{\Gamma_{\phi \to K^0 \bar K^0} + \Gamma_{\phi \to K^+ K^-}}{2} \nonumber\\
     &= 1.763 \pm  0.088 \,\text{MeV}\;,  \\
     \Delta \Gamma_{\phi \to K K} &=  \Gamma_{\phi \to K^0 \bar K^0} -\Gamma_{\phi \to K^+ K^-} \nonumber\\
     &= -0.646 \pm 0.129 \,\text{MeV}\;.
    \end{align}
    \end{subequations}
    \item \emph{Axial-vector-meson decays.} We use
    \begin{subequations}
    \begin{align}
        \bar \Gamma_{a_1\to \rho \pi} &= \frac{\Gamma_{a_{1}^0\to\rho^{\pm}\pi^{\mp}} + 2 \Gamma_{a_{1}^\pm\to\rho^{\pm,0}\pi^{0,\pm}}}{3} \nonumber\\
        &= 425 \pm 175 \,\text{MeV}\;,\\
        \Gamma_{a_1^{\pm}\to \pi^{\pm} \gamma} &= 0.64 \pm 0.246 \,\text{MeV}\;, \\
        \Gamma_{f_1^H\to K^{\star \pm,0,\bar 0}K^{\mp,\bar 0,0}} &= 43.60 \pm 8.72 \,\text{MeV}\;. 
    \end{align}
    \end{subequations}
    The $f_1^H\to K^{\star} K$ decay width is determined as described in Ref.\ \cite{Parganlija:2012fy}.    
    \item \emph{Scalar-meson decays.} We use
    \begin{subequations}
    \begin{align}
        \Gamma_{a_0} &= \bar \Gamma_{a_0\to K\bar K} + \bar \Gamma_{a_0\to \pi \eta} + \bar \Gamma_{a_0\to \pi \eta^{\prime}} \nonumber \\
        &= 265 \pm 53 \,\text{MeV}\;,\\
        \bar \Gamma_{K_0^{\star}\to K \pi} &= \frac{\Gamma_{\bar K_0^{\star 0}\to \pi^{0,+}K^{\bar{0},-}} + \Gamma_{K_0^{\star -}\to \pi^{0,-}K^{-, \bar{0}}}}{2}  \nonumber \\
        &= 270 \pm 80\,\text{MeV} \;, \\
        \Gamma_{f_0^L \to \pi^{0,+} \pi^{0,-}} &= 250 \pm 125 \,\text{MeV}\; ,\\ 
        \Gamma_{f_0^{L}\to K^{0,+}K^{\bar{0},-}} &= 150 \pm 75\,\text{MeV}\; ,\\ 
        \Gamma_{f_0^H \to \pi^{0,+} \pi^{0,-}} &= 20.2 \pm 10.1 \,\text{MeV}\; ,\\ 
        \Gamma_{f_0^{H}\to K^{0,+}K^{\bar{0},-}} &= 87.7 \pm 43.9 \,\text{MeV}\; ,
    \end{align}
        where 
    \begin{align}
        \bar \Gamma_{a_0\to K\bar K} &= \frac{\Gamma_{a_0^0\to K^{0,+}K^{\bar{0},-}} + 2 \Gamma_{a_0^{\pm}\to K^{0}K^{\pm}}}{3}\;, \\
       \bar \Gamma_{a_0\to \pi \eta} &= \frac{\Gamma_{a_0^0\to \pi^{0}\eta} + 2 \Gamma_{a_0^{\pm}\to \pi^{\pm} \eta}}{3}\;, \\
       \bar \Gamma_{a_0\to \pi \eta^{\prime}} &= \frac{\Gamma_{a_0^0\to \pi^{0}\eta^{\prime}} + 2 \Gamma_{a_0^{\pm}\to \pi^{\pm} \eta^{\prime}}}{3}\;.
    \end{align}
    \end{subequations}
The decay widths for $f_0^L = f_0(1370)$ are not taken from the PDG \cite{Workman:2022ynf}, but are instead estimates based on Refs.\ \cite{Workman:2022ynf,Bugg:2007ja,Polychronakos:1978ur,Wicklund:1980qd} (see also Ref.\ \cite{Parganlija:2012fy})
\end{enumerate}
 At first the best solution taken from Ref.\ \cite{Parganlija:2012fy} was used to check whether we reproduce the isospin-symmetric solution with the new numerical fitting algorithm, where we do not fit the isospin-symmetry breaking quantities. 
 With that parameter set we got a solution with a reasonable $\chi^2 = 8.9$.





Then we initialized new parameter sets from $1.2\times 10^7$ random points in the parameter space, set the error level for the isospin-symmetry breaking quantities to $20\%$, and used the DVS-I. 
The best solution leads to $\chi^2 = 29.8$, where the largest contribution comes from $\Gamma_{\omega \to \pi\pi}$, which is about $25$, while all other quantities give a rather small $\chi^2$ contribution.
(The detailed results of this fit and the corresponding parameter set are shown in Table~\ref{Tab:fit_res} and \tabref{Tab:par_sets} in Appendix~\ref{App:fit_res}). 
This may be due to the fact that $\Gamma_{\omega \to \pi\pi}$ is the only quantity that is directly proportional to $\phi_3^2$ (cf.\ Eq.~\eqref{Eq:decay_omega}), so it is very sensitive to the value of $\phi_3$, which -- due to the fitting of other quantities -- tends to be very small, $\sim \mathcal{O}(10^{-6})-\mathcal{O}(10^{-5})$~GeV, compared to $\phi_{N/S}$. 

\begin{figure}[htbp]
  \centering
  \includegraphics[width=0.5\textwidth]{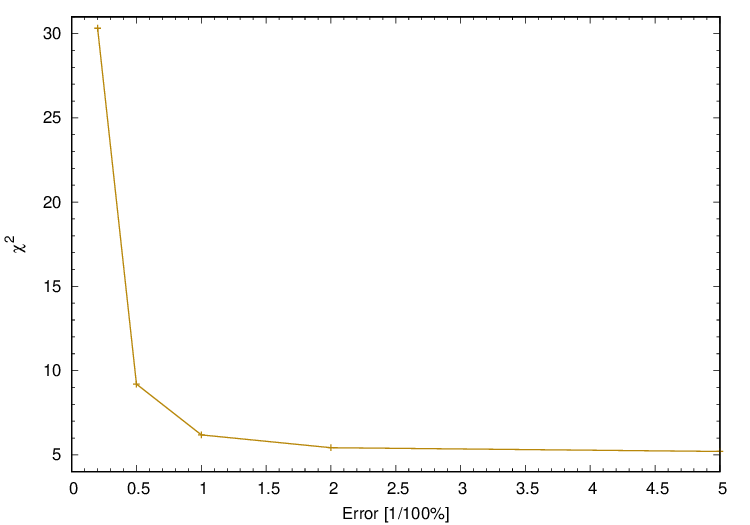}
  \caption{\label{fig:chi2} $\chi^2$ as a function of prescribed error starting from $1.2\times 10^7$ random points in the parameter space.}
\end{figure}

Next, the error level for the isospin-symmetry breaking quantities was increased in several steps from $20\%$ to $500\%$ and the resulting $\chi^2$ values are shown in Fig.~\ref{fig:chi2} for the best solution in each case. 
If the error level for the isospin-symmetry breaking quantities is increased to infinity, i.e., their $\chi^2$ contributions are neglected, an isospin-symmetric solution is obtained. 
In this way a solution with $\chi^2 = 3.2$ can be achieved. 
It is somehow surprising that with this limiting procedure a better solution can be obtained than if we start from $1.2\times 10^7$ random points in the isospin-symmetric case, which leads to a solution with $\chi^2 > 7$.  

In every fit the $\Gamma_{\omega\to \pi\pi}$ decay width produced the largest contribution to $\chi^2$. Therefore, we also made fits in which the $\omega$ decay is omitted. 
In the case of DVS-I, this leads to a solution with $\chi^2 = 6.5$ and $\chi^2_{\text{red}} \equiv \chi^2/N_{\text{d.o.f.}} = 0.6$ (see Table~\ref{Tab:fit_res} and \tabref{Tab:par_sets} in Appendix~\ref{App:fit_res}). 
It is worth noting that in this case the number of degrees of freedom $N_{\text{d.o.f.}}$, which is the difference between the number of fitted quantities and the number of fit parameters, was $11$. 

We also investigated all possible scenarios (described in Sec.~\ref{Sec:phys_mass} below \eqref{Eq:m_f1H}), namely DS, DVS-I, and DVS-II), both with or without fitting the $\omega \to \pi\pi$ decay width. 
The resulting $\chi^2$ and $\chi^2_{\text{red}}$ values are displayed in \tabref{Tab:chi2_scens}.
\begin{table}[h!]
    \centering
    \vspace{0.3cm}
    \begin{tabular}{|c||c|c|c|c|}
          \hline
            &  \multicolumn{2}{c}{with $\omega$ decay} & \multicolumn{2}{|c|}{without $\omega$ decay} \\
            \hline
          Case & $\chi^2$ & $\chi^2_{\text{red}}$ & $\chi^2$ & $\chi^2_{\text{red}}$ \\
          \hline\hline
          DS    & $31.6$ & $2.4$ & $6.5$ & $0.6$  \\
          \hline
          DVS I & $30.3$  & $2.5$ & $6.2$ &  $0.6$\\
          \hline
          DVS II & $31.3$ & $3.5$ & $6.0$ & $0.8$ \\
          \hline
    \end{tabular}
    \caption{$\chi^2$ and $\chi^2_{\text{red}}$ values for the different scenarios (see \secref{Sec:results}), with or without fitting the $\omega$ decay width.}
    \label{Tab:chi2_scens}
\end{table}
According to the reduced $\chi^2_{\text{red}}$ values, the quality of the fit is very similar for the DS and DVS-I, while the DVS-II is worse both with or without fitting the $\omega \to \pi \pi$ decay width. 
In the case without the $\omega \to \pi\pi$ fit, the $\chi^2_{\text{red}}$ is below $1$ in every case, which testifies for very high-quality fits. 
Looking only at the $\chi^2_{\text{red}}$ values, the DS or DVS-I are favored over the DVS-II. 
According to Refs.\ \cite{aoki2017review,Miramontes:2022mex}, Dashen's theorem is significantly violated, favoring DVS-I, but this is not reflected in our fit. 
On the other hand, we can exclude DVS-II, since in this scenario -- besides the larger $\chi^2_{\text{red}}$ values -- the predictive power of our model decreases significantly. 
This is because in each sector the isospin averages and differences can be set with the help of the electromagnetic mass contributions, and the only constraints on isospin-symmetry violation come from the decay widths. 
The detailed results of the fits and the parameter values for the DS and DVS-I with and without the $\omega$ decay are shown in \tabref{Tab:fit_res} and \tabref{Tab:par_sets} of Appendix~\ref{App:fit_res}), respectively. 

We also checked if instead of omitting the $\omega\to\pi^+\pi^-$ decay, either the $\rho^0\to\pi^+\pi^-$ or the $\rho^{\pm}\to\pi^{\mp}\pi^0$ decay is omitted, because $\omega$ mixes with $\rho^0$. 
In both cases, the $\chi^2$ gets much worse, $22.7$ and $23.2$, respectively, while the $\chi^2_{\text{red}}$ value is $2.1$ in both cases. 
So even if there is $\rho^0-\omega$ mixing, omitting the $\rho^0$ decay will not improve the fit as much as omitting the $\omega$ decay.

According to Eqs.\ \eqref{Eq:f_pi_0} and \eqref{Eq:f_K_0} we can calculate the neutral-pion and neutral-kaon decay constants that cannot be measured directly. 
In the case of the DVS-I without the $\omega$ decay the mean values are:
\begin{subequations}
\begin{align}
    f_{\pi^0} &= 109.00 \,\text{MeV}\;, \\
    f_{K^0} &= 109.55\, \text{MeV}\;. 
\end{align}
\end{subequations}
In our case the difference $\Delta f_K = f_K^0 - f_K^+$ is very small ($\approx 5$~keV) compared to a calculation \cite{Chen:2006hr} based on QCD sum rules, where it was found to be $\Delta f_K = 1.5$~MeV.

\FloatBarrier

\section{Conclusion}
\label{Sec:conclusion}

In this paper we have investigated isospin-symmetry breaking effects in the framework of the eLSM at tree level. 
We have shown that the various particle mixings caused by the three non-zero condensates are quite complicated and can only be resolved by several field transformations. 
We have calculated the physical masses of the mesons at tree level. 
All possible tree-level decay widths are also determined, which leads to long expressions due to the many field transformations. 
We took existing experimental data on the considered mesons and determined the unknown parameters of the model by a $\chi^2$ fitting procedure. 
In our approach, we also investigated the fulfillment of Dashen's theorem. 
For this purpose, we introduced an additional electromagnetic mass contribution in the kaonic sectors and analyzed the possible best fits. 
We found that there is no strong violation of Dashen's theorem at tree level in our approach, i.e., fits respecting or violating Dashen's theorem are of equally good quality. 
We found that the decay width of $\omega\to \pi\pi$ gave the largest contribution to the $\chi^2$, showing that this decay cannot be described at tree level in this model due to the smallness of $\phi_3$ required by the other isospin-symmetry breaking channels.  


\section*{Acknowledgments}

The research was supported by the Hungarian National Research, Development and Innovation Fund under Project numbers FK 131982 and K 138277. We also acknowledge support for the computational resources provided by the Wigner Scientific Computational Laboratory (WSCLAB). N.W.\ acknowledges support by the German National Academy of Sciences Leopoldina through the Leopoldina fellowship program with funding code LPDS 2022-11.

\appendix

\begin{widetext}
\section{Elements of the mass squared matrices}
\label{App:mass_matrix}

The elements of the squared meson mass matrices at tree level after isospin-symmetry breaking -- before the field shifts and orthogonal transformations in the $N-3-S$ sectors -- are given as follows.
\begin{enumerate}
\item[(i)] \textit{Pseudoscalars:}
\begin{subequations}
\begin{eqnarray}
m_{\pi^{\pm}}^2 &=& m_0^2+\lambda_1(\phi_N^2+\phi_3^2+\phi_S^2)+\frac{\lambda_2}{2}(\phi_N^2+3\phi_3^2)\;,\\
m_{K^{\pm}}^2 &=& m_0^2+\lambda_1(\phi_N^2+\phi_3^2+\phi_S^2)+\lambda_2 \left[-\frac{1}{\sqrt2}(\phi_N + \phi_3)\phi_S +
    \frac{(\phi_N+\phi_3)^2}{2}  + \phi_S^2\right]\;,\\
m_{K^0}^2 &=& m_0^2 + \lambda_1(\phi_N^2 + \phi_3^2+\phi_S^2) + \lambda_2\left[-\frac{1}{\sqrt2}(\phi_N-\phi_3) \phi_S +
    \frac{(\phi_N-\phi_3)^2}{2} + \phi_S^2\right]\;,\\
m_{\eta_N}^2 &=& m_0^2+\lambda_1(\phi_N^2+\phi_3^2+\phi_S^2)+\frac{\lambda_2}{2}(\phi_N^2+\phi_3^2)+c_1\phi_N^2 \phi_S^2\;,\label{mP_NN}\\
m_{{\pi}^0}^2 &=& m_0^2+\lambda_1(\phi_N^2+\phi_3^2+\phi_S^2)+\frac{\lambda_2}{2}(\phi_N^2+\phi_3^2)
+c_1\phi_S^2 \phi_3^2\;,\label{mP_33}\\
m_{\eta_S}^2 &=& m_0^2+\lambda_1(\phi_N^2+\phi_3^2+\phi_S^2)+\lambda_2\phi_S^2+\frac{c_1}{4}
(\phi_N^2 -\phi_3^2)^2\;,\label{mP_SS}\\
m_{\eta_N \pi^0}^2 &=& m_{\pi^0 \eta_N}^2 = (\lambda_2 - c_1\phi_S^2)\phi_N\phi_3,\label{mP_N3}\\
m_{\eta_N \eta_{S}}^2 &=& m_{\eta_S \eta_{N}}^2 = \frac{c_1}{2}\phi_N\phi_S(\phi_N^2 - \phi_3^2)\;,\label{mP_NS}\\
m_{\pi^0 \eta_{S}}^2 &=& m_{\eta_{S} \pi^0}^2 =-\frac{c_1}{2}\phi_3\phi_S(\phi_N^2 - \phi_3^2)\;.\label{mP_3S}
\end{eqnarray}
\end{subequations}
\item[(ii)] \textit{Scalars:}
\begin{subequations}
\begin{eqnarray}
m_{a_0^{\pm}}^2 &=& m_0^2+\lambda_1(\phi_N^2+\phi_3^2+\phi_S^2)+\frac{\lambda_2}{2}(3\phi_N^2+\phi_3^2)\;,\\
m_{K_0^{\star\pm}}^2 &=& m_0^2 + \lambda_1 (\phi_N^2+\phi_3^2+\phi_S^2) + \lambda_2 \left[\frac{1}{\sqrt2} (\phi_N+\phi_3) \phi_S +
    \frac{(\phi_N+\phi_3)^2}{2} + \phi_S^2\right]\;,\\
m_{K_0^{\star 0}}^2 &=& m_0^2 + \lambda_1 (\phi_N^2+\phi_3^2+\phi_S^2) + \lambda_2 \left[\frac{1}{\sqrt2} (\phi_N-\phi_3) \phi_S +
    \frac{(\phi_N-\phi_3)^2}{2} + \phi_S^2\right]\;,\\
m_{\sigma_N}^2 &=& m_0^2+\lambda_1(3\phi_N^2+\phi_3^2+\phi_S^2)+\frac{3\lambda_2}{2}(\phi_N^2+\phi_3^2)\;,\label{mS_NN}\\
m_{a_0^0}^2 &=& m_0^2+\lambda_1(\phi_N^2+3\phi_3^2+\phi_S^2)+\frac{3\lambda_2}{2}(\phi_N^2+\phi_3^2)\;,\label{mS_33}\\
m_{\sigma_S}^2 &=& m_0^2+\lambda_1(\phi_N^2+\phi_3^2+3\phi_S^2)+3\lambda_2\phi_S^2\;,\label{mS_SS}\\
m_{\sigma_N a_0^0}^2 &=& m_{a_0^0 \sigma_N}^2 = (2\lambda_1 + 3 \lambda_2)\phi_N\phi_3\;,\label{mS_N3}\\
m_{\sigma_N \sigma_{S}}^2 &=& m_{\sigma_S \sigma_N}^2 = 2\lambda_1\phi_N\phi_S\;,\label{mS_NS}\\
m_{a_0^0 \sigma_{S}}^2 &=& m_{\sigma_{S} a_0^0}^2 = 2\lambda_1\phi_3\phi_S\;.\label{mS_3S}
\end{eqnarray}
\end{subequations}
\item[(iii)] \textit{Vectors:}
\begin{subequations}
\begin{eqnarray}
m_{\rho^{\pm}}^2 &=& \tilde{m}_1^2 + \frac{\phi_N^2}{2}(h_1+h_2+h_3)+\frac{\phi_3^2}{2}(h_1+h_2-h_3+2g_1^2)+\frac{h_1}{2}\phi_S^2\;,\\
m_{K^{\star\pm}}^2 &=& \tilde{m}_1^2 + \tilde{\delta}_s + \frac{1}{2}\delta_3 +\frac{h_1}{2} (\phi_N^2+\phi_3^2) + \frac{1}{4}(\phi_N+\phi_3)^2(h_2+g_1^2) +\frac{\phi_S^2}{2}(h_1+h_2+g_1^2)\\
& &+\frac{1}{\sqrt2}(\phi_N+\phi_3)\phi_S(h_3-g_1^2)\;,\\
m_{K^{*0}}^2 &=& \tilde{m}_1^2 + \tilde{\delta}_s - \frac{1}{2}\delta_3 +\frac{h_1}{2}(\phi_N^2+\phi_3^2) +\frac{1}{4}(\phi_N-\phi_3)^2(h_2+g_1^2)+
\frac{\phi_S^2}{2}(h_1+h_2+g_1^2)\\
& &+\frac{1}{\sqrt2}(\phi_N-\phi_3)\phi_S(h_3-g_1^2)\;,\\
m_{\omega_N}^2 &=& \tilde{m}_1^2 + \frac{\phi_N^2}{2}(h_1+h_2+h_3) +\frac{\phi_3^2}{2}(h_1+h_2+h_3) + \frac{h_1}{2}\phi_S^2 + \delta m^2_V\;,\\ \label{Eq:omN}
m_{\rho^0}^2 &=& \tilde{m}_1^2 + \frac{\phi_N^2}{2}(h_1+h_2+h_3) +\frac{\phi_3^2}{2}(h_1+h_2+h_3) + \frac{h_1}{2}\phi_S^2 - \delta m^2_V\;,\\ \label{Eq:rho0}
m_{\omega_N \rho^0}^2 &=& (h_2 + h_3)\phi_N\phi_3 + \delta_3\;,\\ \label{Eq:omNrho0}
m_{\omega_S}^2 &=& \tilde{m}_1^2 + 2\tilde{\delta}_s + \frac{h_1}{2}(\phi_N^2 + \phi_3^2) + \phi_S^2\left(\frac{h_1}{2}+h_2+h_3\right)\;,\\
m_{\omega_N \omega_{S}}^2&=& m_{\rho^0 \omega_{S}}^2 = 0\;.
\end{eqnarray}
\end{subequations}
\item[(iv)] \textit{Axial-vectors:}
\begin{subequations}
\begin{eqnarray}
m_{a_1^{\pm}}^2 &=& \tilde{m}_1^2+\frac{\phi_N^2}{2}(h_1+h_2-h_3+2g_1^2)+\frac{\phi_3^2}{2}(h_1+h_2+h_3)+\frac{h_1}{2}\phi_S^2\;,\\
m_{K_1^{\pm}}^2 &=& \tilde{m}_1^2 + \tilde{\delta}_s + \frac{1}{2}\delta_3 + \frac{h_1}{2}(\phi_N^2+\phi_3^2) + \frac{1}{4}(\phi_N+\phi_3)^2(h_2+g_1^2) +
\frac{\phi_S^2}{2}(h_1+h_2+g_1^2)\\
& &-\frac{1}{\sqrt2}(\phi_N+\phi_3)\phi_S(h_3-g_1^2)\;,\\
m_{K_1^0}^2 &=& \tilde{m}_1^2 + \tilde{\delta}_s - \frac{1}{2}\delta_3+\frac{h_1}{2}(\phi_N^2+\phi_3^2)+\frac{1}{4}(\phi_N-\phi_3)^2(h_2+g_1^2)+
\frac{\phi_S^2}{2}(h_1+h_2+g_1^2)\\
& &-\frac{1}{\sqrt2}(\phi_N-\phi_3)\phi_S(h_3-g_1^2)\;,\\
m_{f_{1N}}^2 &=& \tilde{m}_1^2 + \frac{1}{2}(\phi_N^2+\phi_3^2)(h_1+h_2-h_3+2g_1^2) + \frac{h_1}{2}\phi_S^2 + \delta m^2_A\;,\\ \label{Eq:f1N}
m_{a_1^0}^2 &=& \tilde{m}_1^2 + \frac{1}{2}(\phi_N^2+\phi_3^2)(h_1+h_2-h_3+2g_1^2) + \frac{h_1}{2}\phi_S^2 - \delta m^2_A\;,\\ \label{Eq:a10}
m_{f_{1N} a_1^0}^2 &=& (h_2 - h_3 +2g_1^2)\phi_N\phi_3 + \delta_3\;,\\ \label{Eq:f1Na10}
m_{f_{1S}}^2 &=& \tilde{m}_1^2+2\tilde{\delta}_s+\frac{h_1}{2}(\phi_N^2 + \phi_3^2) + \phi_S^2\left(\frac{h_1}{2}+h_2-h_3+2g_1^2\right)\;,\\
m_{f_{1N} f_{1S}}^2 &=& m_{a_1^0 f_{1S}}^2 = 0\;,
\end{eqnarray}
where 
\begin{equation}
  \tilde{m}_1^2 = m_1^2 + \delta_u + \delta_d\;,\quad \tilde{\delta}_s = \delta_s - (\delta_u + \delta_d)/2\;,\quad \delta_3 = \delta_u - \delta_d\;.
\end{equation} 
\end{subequations}
\end{enumerate}

\section{The $f_{\pi^0}$ weak-decay constant}
\label{App:fpi0_weak_dec}

We start from the PCAC hypotheses for the neutral pion,
\begin{equation}
    \langle 0 | \mathcal{J}^{\mu}_{\pi^0}(0)|\pi^0\rangle = ip^{\mu}f_{\pi^0}\;,\label{Eq:PCAC_hyp}
\end{equation}
where $\mathcal{J}^{\mu}_{\pi^0}$ is the Noether current pertaining to axial transformations. 
We need to calculate the current only in the $N-3-S$ sector, since $\pi_3$ only mixes with $\eta_N$ and $\eta_S$ to form the physical $\pi^0$,
\begin{equation}
    \mathcal{J}^{\mu}_i\omega^{A}_i = \frac{\delta \mathcal{L}_{y}}{\delta (\partial_{\mu} y^{\text{ph}}_k)}\delta y^{\text{ph}}_k\;,\quad i,k \in (N,3,S)\;,
\end{equation}
where $\omega_i^A$ are the arbitrary infinitesimal parameters of the axial transformations and $y^{\text{ph}}_k = (\pi^0, \eta, \eta')$ are the physical isoscalar fields in the pseudoscalar sector (see Eq.~\eqref{Eq:phys_pi_eta_fields}), while $\delta y^{\text{ph}}_k$ are the infinitesimal axial transformations of the considered fields that also need to be determined. 
The relevant part of the Lagrangian based on Eqs.~\eqref{Eq:lagr_pi_eta} and \eqref{Eq:phys_pi_eta_fields} is
\begin{equation}
    \mathcal{L}_{y} = \frac{1}{2}\partial^{\mu}{y^{\text{ph}}_k}\partial_{\mu}{y^{\text{ph}}_k} - \frac{1}{2}{y^{\text{ph}}_k}{\diag{(\tilde{\mathds{M}}_{P}^2})}_{kl}{y^{\text{ph}}_l}\;,
\end{equation}
where $\diag{(\tilde{\mathds{M}}_{P}^2)}$ is given by Eq.~\eqref{Eq:M_P_diag}. 
According to Eq.~\eqref{Eq:pseudoN3S_transf}, $y^{\text{ph}}_k = (\mathds{O}_{M} \mathds{Z}_{P}^{-1} \mathds{O}_{D})_{kl} x_l \equiv (\mathds{O}_{P}^{-1})_{kl} x_l$ with $x_l = (\tilde\eta_N,\tilde\pi_0,\tilde\eta_S)$. 
Thus,
\begin{equation}
    \mathcal{J}^{\mu}_i\omega^{A}_i = \partial^{\mu} y^{\text{ph}}_k(\mathds{O}_{P}^{-1})_{kl}\delta x_l\;,\quad k,l \in (N,3,S)\;.\label{Eq:J_omega}
\end{equation}
While $\delta x_l$ can be determined as follows, let $\Phi = \sum_{l\in N,3,S} \phi_l T_l$, $\phi_l = \sigma_l + ix_l$ be the scalar-pseudoscalar $N,3,S$ fields and $T_l =\lambda_l/2$ the generators of $U(3)$ in the $N,3,S$ directions. 
The effect of an infinitesimal axial transformation on $\Phi$ can be written as
\begin{equation}
    \delta \Phi = T_l \delta \phi_l = (\mathds{1} + i\omega_i^A T_i)T_k\phi_k(\mathds{1} + i\omega_i^A T_i) - T_k\phi_k = i\omega_i^A \left\{T_i,T_k\right\}\phi_k \equiv i\omega_i^A d_{ikl}\phi_k T_l\;, \label{Eq:delta_Phi_axial}
\end{equation}
where $d_{ikl}$ are the totally symmetric structure constants of $U(3)$. 
Since the $i$ and $k$ indices are in $(N,3,S)$ it can be shown that $d_{ikl}$ is nonzero only if $l \in (N,3,S)$. 
Calculating the $d_{ikl}$ structure constants in the $N-3-S$ basis, the nonzero elements are $d_{NNN} =1$, $d_{N33} = d_{3N3} = d_{33N} = 1$, and $d_{SSS} = \sqrt{2}$. 
Thus, Eq.~\eqref{Eq:delta_Phi_axial} can be written as,
\begin{eqnarray}
    \frac{1}{2}( i\delta x_l + \delta \sigma_l)\lambda_l &=& \frac{1}{2}\left[\left(i(\omega_N^A \sigma_N + \omega_3^A \sigma_3) - (\omega_N^A \tilde\eta_N + \omega_3^A \pi_3)\right)\lambda_N + \left(i(\omega_N^A \sigma_3 + \omega_3^A \sigma_N) - (\omega_N^A \pi_3 + \omega_3^A \tilde\eta_N)\right)\lambda_3 \right.\nonumber \\
     &+& \left.\left(i\omega_S^A \sigma_S  - \omega_S^A \tilde\eta_S\right)\lambda_S\right], 
\end{eqnarray}
where $\sigma_3 \equiv a_0^0$, $\pi_3 = \tilde\pi^0$, and the $\delta x_l$ can be read off as
\begin{subequations}
\begin{eqnarray}
  \delta x_1 &\equiv& \delta \tilde \eta_N = \omega_N^A \sigma_N + \omega_3^A \sigma_3\; , \\
  \delta x_2 &\equiv& \delta \tilde \pi^0 = \omega_N^A \sigma_3 + \omega_3^A \sigma_N\; , \\
  \delta x_3 &\equiv& \delta \tilde \eta_S = \omega_S^A \sigma_S\; .
\end{eqnarray}
\end{subequations}
Since the $\omega_i^A$'s are independent, we can set $\omega_N^A \ne 0$, $\omega_3^A = \omega_S^A = 0$, which -- after substitution into Eq.~\eqref{Eq:J_omega} and considering only the physical $\pi^0$ -- leads to
\begin{equation}
    \mathcal{J}_N^{\mu}\omega_N^A = \partial^{\mu} \pi^0\left({\mathds{O}_{P}^{-1}}_{11}\sigma_N + {\mathds{O}_{P}^{-1}}_{12}\sigma_3\right)\omega_N^A\;.
\end{equation}
Similarly, $\omega_3^A \ne 0, \omega_{N,S}^A=0$ and $\omega_S^A \ne 0, \omega_{N,3}^A =0$ gives $J_3^{\mu}$ and $J_S^{\mu}$, respectively. 
Thus, the relevant parts of the currents are,
\begin{subequations}
\begin{eqnarray}
  \mathcal{J}_N^{\mu} &=& \partial^{\mu} \pi^0\left({\mathds{O}_{P}^{-1}}_{11}\sigma_N + {\mathds{O}_{P}^{-1}}_{12}\sigma_3\right)\;, \\
  \mathcal{J}_3^{\mu} &=& \partial^{\mu} \pi^0\left({\mathds{O}_{P}^{-1}}_{11}\sigma_3 + {\mathds{O}_{P}^{-1}}_{12}\sigma_N\right)\;, \\
  \mathcal{J}_S^{\mu} &=& \partial^{\mu} \pi^0\left({\mathds{O}_{P}^{-1}}_{13}\sigma_S \right)\;,
\end{eqnarray}
\end{subequations}
while the total current is
\begin{eqnarray}
\mathcal{J}^{\mu}_{\text{tot}} &=& \mathcal{J}_N^{\mu} + \mathcal{J}_3^{\mu} + \mathcal{J}_S^{\mu} \nonumber\\
&=& \partial^{\mu} \pi^0\left[\left({\mathds{O}_{P}^{-1}}_{11} + {\mathds{O}_{P}^{-1}}_{12}\right)(\sigma_N + \sigma_3) + {\mathds{O}_{P}^{-1}}_{13}\sigma_S \right]\;.
\end{eqnarray}
In the PCAC relation~\eqref{Eq:PCAC_hyp} one can only get a nonzero contribution if the scalar fields assume nonvanishing expectation values (Eq.~\eqref{Eq:SSB}).
Thus, $f_{\pi^0}$ is given by
\begin{equation}
    f_{\pi^0} = \left({\mathds{O}_{P}^{-1}}_{11} + {\mathds{O}_{P}^{-1}}_{12}\right)(\phi_N + \phi_3) + {\mathds{O}_{P}^{-1}}_{13}\phi_S\;.
\end{equation}
It should be noted that due to the $\mathds{O}_{P}$ transformation this quantity can be negative. 
However, according to Ref.\ \cite{Workman:2022ynf} (part 72. {\it Leptonic Decays of Charged Pseudoscalar Mesons}) the measurements of leptonic decays of charged pseudoscalar mesons (like $\pi^+$, $K^+$, $D_s^+$, $B^+$) can only determine the combination $|f_{P}||V_{q_1,q_2}|$, where $|V_{q_1,q_2}|$ is a CKM matrix element. 
Thus, we can consider simply $|f_{\pi^0}|$, which is a prediction in our case since only the charged decay widths are measured. 


\section{Explicit forms of the tree-level decay widths}
\label{App:expl_decays}

In this appendix we explicitly list all tree-level expressions for the decay widths, grouped as vector-meson, axial-vector-meson, and scalar-meson decays.

\subsection{Vector-meson decays}
\label{SSec:vector_dec}

\subsubsection*{$\rho\to \pi\pi$}
\label{SSSec:rhopipi}
The relevant part of the Lagrangian reads

\begin{eqnarray}
\mathcal{L}_{\rho\pi\pi} &=& i B^{\rho}_1\rho^{0}_{\mu}\pi^-\partial^{\mu}\pi^+ + i C_1^{\rho}(\partial_{\mu}\rho^{0}_{\nu})(\partial^{\mu}\pi^-)\partial^{\nu}\pi^+ +\nonumber\\
&+& i \rho^{+}_{\mu}\left[ D^{\rho}_1 \pi^0\partial^{\mu}\pi^- + D^{\rho}_2 \pi^- \partial^{\mu}\pi^0\right] + i (\partial_{\mu}\rho^{+}_{\nu})\left[ F^{\rho}_1 (\partial^{\mu}\pi^0)\partial^{\nu}\pi^- + F^{\rho}_2 (\partial^{\mu}\pi^-) \partial^{\nu}\pi^0 \right]+\text{h.c.}\;,
\end{eqnarray}
where
\begin{subequations}
\begin{eqnarray}
B^{\rho}_1 &=& Z_{\pi^\pm}^2\left\{\left[g_1 + \phi_N \left(h_3-g_1^2\right) w_{a_1^\pm}\right]\cos\vartheta_V - \frac{2}{3}\phi_3(h_2 + h_3)w_{a_1^\pm}\sin\vartheta_V\right\}\;,\\
C^{\rho}_1 &=& -g_2 Z_{\pi^\pm}^2 w_{a_1^\pm}^2 \cos\vartheta_V \\
D^{\rho}_1 &=& Z_{\pi^\pm}\left\{\left[g_1 + \phi_N \left(h_3-g_1^2\right) w_{a_1^\pm}\right]{\mathds{O}_{P}}_{21} - \phi_3 \big(h_3-g_1^2\big) w_{a_1^\pm} {\mathds{O}_{P}}_{11} \right\}\;,\\
D^{\rho}_2 &=& -Z_{\pi^\pm}\left[g_1 {\mathds{O}_{P}}_{21}+ \phi_N \big(h_3-g_1^2\big) \big(w_{\eta}^a{\mathds{O}_{P}}_{11} + w_{\pi}^a{\mathds{O}_{P}}_{21} \big) - \phi_3\big(h_2 - h_3 + 2g_1^2\big)\big(w_{\eta}^f{\mathds{O}_{P}}_{11} + w_{\pi}^f{\mathds{O}_{P}}_{21}\big)\right]\;,\\
F^{\rho}_1 &=& -g_2 Z_{\pi^\pm} w_{a_1^\pm} \big(w_{\eta}^a{\mathds{O}_{P}}_{11} + w_{\pi}^a{\mathds{O}_{P}}_{21} \big),\quad F^{\rho}_2 = -F^{\rho}_1\;.
\end{eqnarray}
\end{subequations}
The explicit forms of the tree-level decay widths for the neutral and charged $\rho$ read
\begin{subequations}
\begin{eqnarray}
\Gamma_{\rho^{0}\to \pi^{+}\pi^{-}} &=& \frac{k_{\rho^0}^3}{6\pi M_{\rho^0}^2} \left| B^{\rho}_1  + \frac12 C^{\rho}_1 M_{\rho^{0}}^2\right|^2 \;,\\
\Gamma_{\rho^{\pm}\to \pi^{\pm}\pi^{0}} &=& \frac{k_{\rho^\pm}^3}{24\pi M_{\rho^\pm}^2} \left| D^{\rho}_1 - D^{\rho}_2 + F^{\rho}_1 M_{\rho^{\pm}}^2\right|^2\;,
\end{eqnarray}
\end{subequations}
where
\begin{subequations}
\begin{eqnarray}
k_{\rho^0} &=& \frac12 \sqrt{M_{\rho^0}^2-4M_{\pi^\pm}^2}\;,\\
k_{\rho^{\pm}} &=& \frac{\sqrt{(M_{\rho^\pm}^2 - M_{\pi^\pm}^2 - M_{\pi^0}^2)^2-4M_{\pi^\pm}^2 M_{\pi^0}^2}}{2M_{\rho^\pm}}\;.
\end{eqnarray}
\end{subequations}

\subsubsection*{$\rho\to \pi\eta$}
\label{SSSec:rhopieta}
This $G$ parity-violating process will be a prediction. 
The relevant part of the Lagrangian reads
\begin{eqnarray}
\mathcal{L}_{\rho\pi\eta} &=& i \rho^{+}_{\mu}\left[ G^{\rho}_1 \eta\partial^{\mu}\pi^- + G^{\rho}_2 \pi^- \partial^{\mu}\eta\right] + i (\partial_{\mu}\rho^{+}_{\nu})\left[ H^{\rho}_1 (\partial^{\mu}\eta)\partial^{\nu}\pi^- + H^{\rho}_2 (\partial^{\mu}\pi^-) \partial^{\nu}\eta \right]+\text{h.c.}\;,
\end{eqnarray}
where
\begin{subequations}
\begin{eqnarray}
G^{\rho}_1 &=& Z_{\pi^\pm}\left\{\left[g_1 + \phi_N \left(h_3-g_1^2\right) w_{a_1^\pm}\right]{\mathds{O}_{P}}_{22} - \phi_3 \big(h_3-g_1^2\big) w_{a_1^\pm} {\mathds{O}_{P}}_{12} \right\}\;,\\
G^{\rho}_2 &=& -Z_{\pi^\pm}\left[g_1 {\mathds{O}_{P}}_{22}+ \phi_N \big(h_3-g_1^2\big) \big(w_{\eta}^a{\mathds{O}_{P}}_{12} + w_{\pi}^a{\mathds{O}_{P}}_{22} \big) - \phi_3\big(h_2 - h_3 + 2g_1^2\big)\big(w_{\eta}^f{\mathds{O}_{P}}_{12} + w_{\pi}^f{\mathds{O}_{P}}_{22}\big)\right]\;,\\
H^{\rho}_1 &=& -g_2 Z_{\pi^\pm} w_{a_1^\pm} \big(w_{\eta}^a{\mathds{O}_{P}}_{12} + w_{\pi}^a{\mathds{O}_{P}}_{22} \big)\;,\quad H^{\rho}_2 = -H^{\rho}_1\;,
\end{eqnarray}
\end{subequations}
and the decay width of the process is
\begin{eqnarray}
\Gamma_{\rho^{\pm}\to \pi^{\pm}\eta} &=& \frac{k_{\rho^\pm}^3}{24\pi M_{\rho^\pm}^2} \left| G^{\rho}_1 - G^{\rho}_2 + H^{\rho}_1 M_{\rho^{\pm}}^2\right|^2\;,
\end{eqnarray}
where
\begin{equation}
k_{\rho^{\pm}} = \frac{1}{2M_{\rho^\pm}}\sqrt{(M_{\rho^\pm}^2 - M_{\pi^\pm}^2 - M_{\eta}^2)^2-4M_{\pi^\pm}^2 M_{\eta}^2}\;.
\end{equation}
\subsubsection*{$\omega \to \pi\pi$}	
The relevant part of the Lagrangian reads,
\begin{equation}
\mathcal{L}_{\omega\pi\pi} = i B^{\omega}_1\omega_{\mu}\pi^-\partial^{\mu}\pi^+ +\text{h.c.}\;,
\end{equation}
where
\begin{equation}
B^{\omega}_1 = Z_{\pi^\pm}^2 w_{a_1^\pm} \phi_3 (h_2+h_3)\cos \vartheta_{V}\;,
\end{equation}
while the decay width reads
\begin{equation}
\label{Eq:decay_omega}
\Gamma_{\omega\to \pi^{+}\pi^{-}} = \frac{k_{\omega}^3}{6\pi M_{\omega}^2} \left| B^{\omega}_1 \right|^2\;,\quad k_{\omega} = \frac12 \sqrt{M_{\omega}^2-4M_{\pi^\pm}^2}\;.
\end{equation}

\subsubsection*{$K^{\star}\to K\pi$}	
The relevant part of the Lagrangian reads
\begin{align}
\mathcal{L}_{K^{\star}K\pi} &= i K^{\star 0}_{\mu}\left[ B^{K^{\star}}_1 \pi^0\partial^{\mu}\bar{K^0} + B^{K^{\star}}_2 \bar{K^0} \partial^{\mu}\pi^0 + B^{K^{\star}}_3 \pi^+\partial^{\mu}K^- + B^{K^{\star}}_4 K^- \partial^{\mu}\pi^+\right]\nonumber\\
&+ i (\partial_{\mu}K^{\star 0}_{\nu})\left[ C^{K^{\star}}_1 (\partial^{\mu}\pi^0)\partial^{\nu}\bar{K^0} + C^{K^{\star}}_2 (\partial^{\mu}\bar{K^0}) \partial^{\nu}\pi^0 + C^{K^{\star}}_3 (\partial^{\mu}\pi^+)\partial^{\nu}K^- + C^{K^{\star}}_4 (\partial^{\mu}K^-) \partial^{\nu}\pi^+\right]\nonumber\\
&+ i K^{\star +}_{\mu}\left[ D^{K^{\star}}_1 \pi^0\partial^{\mu}K^- + D^{K^{\star}}_2 K^- \partial^{\mu}\pi^0 + D^{K^{\star}}_3 \pi^-\partial^{\mu}\bar{K^0} + D^{K^{\star}}_4 \bar{K^0} \partial^{\mu}\pi^-\right]\nonumber\\
&+ i (\partial_{\mu}K^{\star +}_{\nu})\left[ F^{K^{\star}}_1 (\partial^{\mu}\pi^0)\partial^{\nu}K^- + F^{K^{\star}}_2 (\partial^{\mu}K^-) \partial^{\nu}\pi^0 + F^{K^{\star}}_3 (\partial^{\mu}\pi^-)\partial^{\nu}\bar{K^0} + F^{K^{\star}}_4 (\partial^{\mu}\bar{K^0}) \partial^{\nu}\pi^-\right]+\text{h.c.}\;,
\end{align}
where
\begin{subequations}
\begin{eqnarray}
B^{K^{\star}}_1 &=& Z_{K^0}\left\{\frac{g_1}{2} \left({\mathds{O}_{P}}_{11} - {\mathds{O}_{P}}_{21} -\sqrt2 {\mathds{O}_{P}}_{31}\right) + \frac{h_3-g_1^2}{\sqrt{2}} w_{K_1^0} \left[(\phi_3-\phi_N){\mathds{O}_{P}}_{31} +\phi_S\left({\mathds{O}_{P}}_{11} - {\mathds{O}_{P}}_{21}\right)\right]\right\}\;,\\
B^{K^{\star}}_2 &=& -Z_{K^0}\Bigg\{\frac{g_1}{2} \left({\mathds{O}_{P}}_{11} - {\mathds{O}_{P}}_{21} -\sqrt2 {\mathds{O}_{P}}_{31}\right) + \frac{\phi_N-\phi_3}{4}\left[W_m (h_2 - 2h_3 + 3 g_1^2) - \sqrt{2} W_S (h_2 + g_1^2)\right] \nonumber\\
&-& \frac{\sqrt{2}\phi_S}{4}\left[W_m (h_2 + g_1^2) - \sqrt2 W_S (h_2 - 2h_3 + 3 g_1^2)\right]\Bigg\}\;,\\
B^{K^{\star}}_3 &=& \frac{1}{\sqrt{2}}Z_{\pi^{\pm}}Z_{K^{\pm}} \left\{g_1 + w_{K_1^\pm}\left[-\phi_3(h_2 + g_1^2) + \sqrt{2}\phi_S (h_3 - g_1^2)\right]\right\}\;,\\
B^{K^{\star}}_4 &=& \frac{1}{\sqrt{2}}Z_{\pi^{\pm}}Z_{K^{\pm}} \left\{-g_1 + \frac{1}{2} w_{a_1^\pm}\left[\phi_N(h_2 - 2h_3 + 3g_1^2) + \phi_3(h_2 + 2h_3 - g_1^2) - \sqrt{2}\phi_S (h_2 + g_1^2)\right]\right\}\;,\\
C^{K^{\star}}_1 &=& \frac{g_2}{2}Z_{K^0}w_{K_1^0}\left(W_m + \sqrt{2} W_S\right)\;,\quad C^{K^{\star}}_2 = -C^{K^{\star}}_1 \;,\\
C^{K^{\star}}_3 &=& -\frac{g_2}{\sqrt{2}}Z_{\pi^{\pm}}Z_{K^{\pm}}w_{a_1^\pm}w_{K_1^\pm},\quad C^{K^{\star}}_4 = -C^{K^{\star}}_3\;,\\
D^{K^{\star}}_1 &=& Z_{K^\pm}\left\{\frac{g_1}{2} \left({\mathds{O}_{P}}_{11} + {\mathds{O}_{P}}_{21} -\sqrt2 {\mathds{O}_{P}}_{31}\right) + \frac{h_3-g_1^2}{\sqrt{2}} w_{K_1^\pm} \left[-(\phi_N+\phi_3){\mathds{O}_{P}}_{31} + \phi_S\left({\mathds{O}_{P}}_{11} + {\mathds{O}_{P}}_{21}\right)\right]\right\}\;,\\
D^{K^{\star}}_2 &=& -Z_{K^\pm}\Bigg\{\frac{g_1}{2} \left({\mathds{O}_{P}}_{11} + {\mathds{O}_{P}}_{21} -\sqrt2 {\mathds{O}_{P}}_{31}\right) - \frac{\phi_N+\phi_3}{4}\left[W_m (h_2 - 2h_3 + 3 g_1^2) + \sqrt{2} W_S (h_2 + g_1^2)\right] \nonumber\\
&+& \frac{\sqrt{2}\phi_S}{4}\left[W_m (h_2 + g_1^2) + \sqrt2 W_S (h_2 - 2h_3 + 3 g_1^2)\right]\Bigg\}\;,\\
D^{K^{\star}}_3 &=& \frac{1}{\sqrt{2}}Z_{\pi^{\pm}}Z_{K^0} \left\{g_1 + w_{K_1^0}\left[\phi_3(h_2 + g_1^2) + \sqrt{2}\phi_S (h_3 - g_1^2)\right]\right\}\;,\\
D^{K^{\star}}_4 &=& \frac{1}{\sqrt{2}}Z_{\pi^{\pm}}Z_{K^0} \left\{-g_1 + \frac{1}{2} w_{a_1^\pm}\left[\phi_N(h_2 - 2h_3 + 3g_1^2) - \phi_3(h_2 + 2h_3 - g_1^2) - \sqrt{2}\phi_S (h_2 + g_1^2)\right]\right\}\;,\\
F^{K^{\star}}_1 &=& -\frac{g_2}{2}Z_{K^\pm}w_{K_1^\pm}\left(W_p - \sqrt{2} W_S\right),\quad F^{K^{\star}}_2 = -F^{K^{\star}}_1 \;,\\
F^{K^{\star}}_3 &=& -\frac{g_2}{\sqrt{2}}Z_{\pi^{\pm}}Z_{K^0}w_{a_1^\pm}w_{K_1^0}\;,\quad F^{K^{\star}}_4 = -F^{K^{\star}}_3\;,
\end{eqnarray}
\end{subequations}
and
\begin{subequations}
\begin{eqnarray}
W_m &=& (w_{\eta}^a - w_{\eta}^f) {\mathds{O}_{P}}_{11} + (w_{\pi}^a - w_{\pi}^f) {\mathds{O}_{P}}_{21} \;,\\
W_p &=& (w_{\eta}^a + w_{\eta}^f) {\mathds{O}_{P}}_{11} + (w_{\pi}^a + w_{\pi}^f) {\mathds{O}_{P}}_{21} \;,\\
W_S &=& w_{f_{1S}}{\mathds{O}_{P}}_{31}\;.
\end{eqnarray}
\end{subequations}
The explicit forms of the tree-level decay widths for the neutral and charged $K^\star$ -- both of them containing two subchannels -- read
\begin{subequations}
\begin{align}
\Gamma_{\bar K^{\star 0}\to \pi^{0,+}K^{\bar{0},-}} &= \Gamma_{\bar K^{\star 0}\to \pi^{0}\bar K^{0}} + \Gamma_{\bar K^{\star 0}\to \pi^{+}K^{-}}\nonumber\\
= \frac{1}{24\pi M_{K^{\star 0}}^2} & \left\lbrace k_{\bar K^{\star 0}\to \pi^{0}\bar K^{0}}^3 \left| B^{K^{\star}}_1 - B^{K^{\star}}_2 + C^{K^{\star}}_1 M_{K^{\star 0}}^2\right|^2 + k_{\bar K^{\star 0}\to \pi^{+}K^{-}}^3 \left| B^{K^{\star}}_3 - B^{K^{\star}}_4 + C^{K^{\star}}_3 M_{K^{\star 0}}^2\right|^2 \right\rbrace\;, \\
\Gamma_{K^{\star -}\to \pi^{0,-}K^{-,\bar 0}} &= \Gamma_{K^{\star -}\to \pi^{0}K^{-}} + \Gamma_{K^{\star -}\to \pi^{-}\bar K^{0}} \nonumber \\
= \frac{1}{24\pi M_{K^{\star \pm}}^2} &\left\lbrace k_{K^{\star -}\to \pi^{0}K^{-}}^3 \left| D^{K^{\star}}_1 - D^{K^{\star}}_2 + F^{K^{\star}}_1 M_{K^{\star \pm}}^2\right|^2 + k_{K^{\star -}\to \pi^{-}\bar K^{0}}^3 \left| D^{K^{\star}}_3 - D^{K^{\star}}_4 + F^{K^{\star}}_3 M_{K^{\star 0}}^2\right|^2 \right\rbrace\;,  
\end{align}
\end{subequations}
where
\begin{equation}
 k_{A\to BC} = \frac{1}{2M_A}\sqrt{(M_A^2 - M_B^2 - M_C^2)^2-4M_B^2 M_C^2}\;.
\label{Eq:decay_momABC}
\end{equation}

\subsubsection*{$\Phi \to KK $}	

The relevant part of the Lagrangian reads
\begin{equation}
\mathcal{L}_{\Phi KK} = i \Phi_{\mu}\left[ B^{\Phi}_1 K^0\partial^{\mu}\bar{K^0} + B^{\Phi}_2 K^+\partial^{\mu}K^-\right] + i (\partial_{\mu}\Phi_{\nu})\left[ C^{\Phi}_1 (\partial^{\mu}K^0)\partial^{\nu}\bar{K^0} + C^{\Phi}_2 (\partial^{\mu}K^+)\partial^{\nu}K^-\right] + \text{h.c.}\;,
\end{equation}
where
\begin{subequations}
\begin{align}
B^{\Phi}_1 &= \frac{1}{\sqrt{2}}Z_{K^0}^2 \left\{ g_1 - \frac12 w_{K_1^0}\left[(\phi_N - \phi_3 )(h_2 + g_1^2) - \sqrt{2}\phi_S (h_2 + 2h_3 - g_1^2)\right]\right\}\;,\\
B^{\Phi}_2 &= \frac{1}{\sqrt{2}}Z_{K^{\pm}}^2 \left\{g_1 - \frac12 w_{K_1^\pm}\left[(\phi_N + \phi_3)(h_2 + g_1^2) - \sqrt{2}\phi_S (h_2 + 2h_3 - g_1^2)\right]\right\}\;,\\
C^{\Phi}_1 &= -\frac{g_2}{\sqrt2}Z_{K^0}^2 w_{K_1^0}^2\;, \\
C^{\Phi}_2 &= -\frac{g_2}{\sqrt2}Z_{K^\pm}^2 w_{K_1^\pm}^2\;.
\end{align}
\end{subequations}
The explicit forms of the tree\,-\,level decay widths for the neutral and charged part of the $\Phi \to KK$ process read
\begin{subequations}
\begin{align}
\Gamma_{\Phi \to K^0 \bar K^0} &= \frac{k_{\Phi\to K^0\bar K^0}^3}{6\pi M_{\Phi}^2} \left| B^{\Phi}_1  + \frac12 C^{\Phi}_1 M_{\Phi}^2\right|^2 \;,\\
\Gamma_{\Phi\to K^+ K^-} &= \frac{k_{\Phi\to K^+ K^-}^3}{6\pi M_{\Phi}^2} \left| B^{\Phi}_2  + \frac12 C^{\Phi}_2 M_{\Phi}^2\right|^2\;,  
\end{align}
\end{subequations}
where
\begin{subequations}
\begin{align}
k_{\Phi\to K^0\bar K^0} &= \frac12 \sqrt{M_{\Phi}^2-4M_{K^0}^2}\;,\\
k_{\Phi\to K^+ K^-} &= \frac12 \sqrt{M_{\Phi}^2-4M_{K^\pm}^2}\;.
\end{align}
\end{subequations}

\subsection{Axial-vector-meson decays}
\label{SSec:axialvector_dec}

\subsubsection*{$a_1\to \rho\pi$}
\label{SSSec:a1rhopi}
The relevant part of the Lagrangian reads
\begin{align}
\mathcal{L}_{a_1\rho\pi} &= i B_1^{a_1} {a_1^{0}}_{\mu}{\rho^+}^{\mu}\pi^-
+ iC_1^{a_1} {a_1^{0}}_{\mu} (\partial^{\mu}{\rho^{+}}^{\nu}) \partial_{\nu}\pi^- - iC_1^{a_1} {a_1^{0}}_{\nu} (\partial^{\mu}{\rho^{+}}^{\nu}) \partial_{\mu}\pi^- - i C_1^{a_1}(\partial_{\mu}{a_1^{0}}_{\nu})\left[{\rho^+}^{\mu}\partial^{\nu}\pi^- - {\rho^+}^{\nu}\partial^{\mu}\pi^- \right]\nonumber\\
&+  i {a_1^{+}}_{\mu} \left[ D^{a_1}_1{\rho^-}^{\mu}\pi^0 + D^{a_1}_2{\rho^0}^{\mu}\pi^-\right] + i {a_1^{+}}_{\mu} \left[ E_1^{a_1} (\partial^{\mu}{\rho^{-}}^{\nu}) \partial_{\nu}\pi^0 + E_2^{a_1} (\partial^{\mu}{\rho^{0}}^{\nu}) \partial_{\nu}\pi^- \right] - 
i {a_1^{+}}_{\nu} \left[ E_1^{a_1} (\partial^{\mu}{\rho^{-}}^{\nu}) \partial_{\mu}\pi^0 \right. \nonumber\\ 
&+ \left. E_2^{a_1} (\partial^{\mu}{\rho^{0}}^{\nu}) \partial_{\mu}\pi^- \right] - i (\partial_{\mu}{a_1^{+}}_{\nu})\left[ E_1^{a_1}({\rho^-}^{\mu}\partial^{\nu}\pi^0 - {\rho^-}^{\nu}\partial^{\mu}\pi^0) + E_2^{a_1}({\rho^0}^{\mu}\partial^{\nu}\pi^- - {\rho^0}^{\nu}\partial^{\mu}\pi^-)  \right]+\text{h.c.}\;,
\end{align}
where
\begin{subequations}
\begin{align}
B_1^{a_1} &= -Z_{\pi^\pm}\phi_N \big(h_3-g_1^2\big)\cos\vartheta_{A}\;,\\
C_1^{a_1} &= -g_2 Z_{\pi^\pm} w_{a_1^\pm} \cos\vartheta_{A}\;,\\
D_1^{a_1} &= -\big(h_3-g_1^2\big) \left(\phi_N {\mathds{O}_{P}}_{21} - \phi_3 {\mathds{O}_{P}}_{11}\right)\;,\\
D_2^{a_1} &= Z_{\pi^\pm}\left[\phi_N \big(h_3-g_1^2\big) \cos\vartheta_{V}-\phi_3 (h_2 + h_3) \sin\vartheta_{V} \right]\;,\\
E_1^{a_1} &= -g_2 \left( w_{\eta}^a {\mathds{O}_{P}}_{11} + w_{\pi}^a {\mathds{O}_{P}}_{21}\right)\;,\\
E_2^{a_1} &= g_2 Z_{\pi^\pm} w_{a_1^\pm} \cos\vartheta_{V}\;.
\end{align}
\end{subequations}
The decay width for the neutral $a_1^0$ reads
\begin{equation}
\Gamma_{a_{1}^0\to\rho^{\pm}\pi^{\mp}}
=\frac{k_{a_{1}^0}}{12 M_{a_{1}^0}^2\pi} \left(  |V^{1}_{\mu\nu}|^{2}-\frac{|V^1_{\mu\nu
	}k_{a_1^0}^{\nu}|^{2}}{M_{a_1^0}^{2}}-\frac{|V^1_{\mu\nu}k_{\rho^{\pm}}^{\mu}|^{2}%
}{M_{\rho^{\pm}}^{2}}+\frac{|V^1_{\mu\nu}k_{a_{1}^0}^{\mu}k_{\rho^\pm}^{\nu}|^{2}}{M_{a_1^0}^{2}M_{\rho^{\pm}}^{2}}\right) \; , \label{eq:decay_a1_rho_pi}%
\end{equation}
with (omitting the $a_1$ superscript from $B_1, C_1$)
\begin{subequations}
\begin{align}
V^1_{\mu\nu}  &= i\left\{  B_1 g_{\mu\nu}+ C_1 \left[k_{\pi^\mp}\cdot(k_{a_1^{0}} + k_{\rho^{\pm}})g_{\mu\nu} - {k_{\rho^{\pm}}}_\mu
{k_{\pi^{\mp}}}_\nu - {k_{\pi^{\mp}}}_\mu {k_{a_1^0}}_\nu\right]\right\} \; ,\\
|V^1_{\mu\nu}|^{2} &= \left\{  4 B_1^{2}+ C_1^{2} \left[
\frac{5}{2}(M_{a_1^0}^{2}-M_{\rho^{\pm}}^{2})^{2}+\frac12 M_{\pi^{\pm}}^{2} (2M_{a_1^0}^{2} + 2M_{\rho^{\pm}}^{2} - M_{\pi^\pm}^{2})\right] + 6 B_1 C_1(M_{a_{1}^0}^{2} - M_{\rho^{\pm}}^{2})\right\}\;, \\
|V^1_{\mu\nu}k_{a_1^0}^{\nu}|^{2} &= B_1^{2} M_{a_1^0}^2 -  k_{a_{1}^0}^2 C_1 M_{a_1^{0}}^{2} \left[C_1 M_{\rho^\pm}^{2} - 2 B_1\right]\;,\\
|V^1_{\mu\nu}k_{\rho^{\pm}}^{\mu}|^{2} &= B_1^{2} M_{\rho^\pm}^2 - k_{a_{1}^0}^2 C_1 M_{a_1^{0}}^{2} \left[C_1 M_{a_1^{0}}^{2} + 2 B_1\right] \;,\\
|V^1_{\mu\nu}k_{a_{1}^0}^{\mu}k_{\rho^\pm}^{\nu}|^{2} &= B_1^2 M_{a_1^{0}}^{2} \left(k_{a_{1}^0}^2 + M_{\rho^\pm}^2\right)\;,
\end{align}
\end{subequations}
where $k_{a_{1}^0} \equiv k_{a_{1}^0\to\rho^{\pm}\pi^{\mp}}$ is given in \eref{Eq:decay_momABC}.
The decay width of the charged $a_1^\pm$ consists of two subchannels and is given by
\begin{align}
\Gamma_{a_{1}^\pm\to\rho^{\pm,0}\pi^{0,\pm}}
&= \Gamma_{a_{1}^\pm\to\rho^{\pm}\pi^{0}} + \Gamma_{a_{1}^\pm\to\rho^{0}\pi^{\pm}} \nonumber\\
&= \frac{k_{a_{1}^\pm \to \rho^{\pm}\pi^{0}}}{24 M_{a_{1}^\pm}^2\pi} \left( |V^{2}_{\mu\nu}|^{2}-\frac{|V^2_{\mu\nu
	}k_{a_1^\pm}^{\nu}|^{2}}{M_{a_1^\pm}^{2}}-\frac{|V^2_{\mu\nu}k_{\rho^{\pm}}^{\mu}|^{2}%
}{M_{\rho^{\pm}}^{2}}+\frac{|V^2_{\mu\nu}k_{a_{1}^\pm}^{\mu}k_{\rho^\pm}^{\nu}|^{2}}{M_{a_1^\pm}^{2}M_{\rho^{\pm}}^{2}} \right)\nonumber \\ 
&+ \frac{k_{a_{1}^\pm \to \rho^{0}\pi^{\pm}}}{24 M_{a_{1}^\pm}^2\pi} \left(|V^{3}_{\mu\nu}|^{2}-\frac{|V^3_{\mu\nu
	}k_{a_1^\pm}^{\nu}|^{2}}{M_{a_1^\pm}^{2}}-\frac{|V^3_{\mu\nu}k_{\rho^{0}}^{\mu}|^{2}%
}{M_{\rho^{0}}^{2}}+\frac{|V^3_{\mu\nu}k_{a_{1}^\pm}^{\mu}k_{\rho^0}^{\nu}|^{2}}{M_{a_1^\pm}^{2}M_{\rho^{0}}^{2}}\right) \; ,\label{eq:decay_a1ch_rho_pi}  %
\end{align}
with (omitting the $a_1$ superscript from $D_1, E_1, D_2, E_2$)
\begin{subequations}
\begin{align}
V^2_{\mu\nu}  &= i\left\{  D_1\, g_{\mu\nu}+ E_1 \left[k_{\pi^0}\cdot(k_{a_1^{\pm}} + k_{\rho^{\pm}})g_{\mu\nu} - {k_{\rho^{\pm}}}_\mu
{k_{\pi^{0}}}_\nu - {k_{\pi^{0}}}_\mu {k_{a_1^\pm}}_\nu\right]\right\} \; ,\\
|V^2_{\mu\nu}|^{2} &=   4 D_1^{2}+ E_1^{2} \left[
\frac{5}{2}(M_{a_1^\pm}^{2}-M_{\rho^{\pm}}^{2})^{2}+\frac12 M_{\pi^{0}}^{2} (2M_{a_1^\pm}^{2} + 2M_{\rho^{\pm}}^{2} - M_{\pi^0}^{2})\right] + 6 D_1 E_1(M_{a_{1}^\pm}^{2} - M_{\rho^{\pm}}^{2})\;, \\
|V^2_{\mu\nu}k_{a_1^\pm}^{\nu}|^{2} &= D_1^{2} M_{a_1^\pm}^2 -  k_{a_{1}^\pm\to\rho^{\pm}\pi^{0}}^2 E_1 M_{a_1^{\pm}}^{2} \left(E_1 M_{\rho^\pm}^{2} - 2 D_1\right)\;,\\
|V^2_{\mu\nu}k_{\rho^{\pm}}^{\mu}|^{2} &= D_1^{2} M_{\rho^\pm}^2 -  k_{a_{1}^\pm\to\rho^{\pm}\pi^{0}}^2 E_1 M_{a_1^{\pm}}^{2} \left(E_1 M_{a_1^{\pm}}^{2} + 2 D_1\right)\; ,\\
|V^2_{\mu\nu}k_{a_{1}^\pm}^{\mu}k_{\rho^\pm}^{\nu}|^{2} &= D_1^2 M_{a_1^{\pm}}^{2} \left(k_{a_{1}^\pm\to\rho^{\pm}\pi^{0}}^2 + M_{\rho^\pm}^2\right)\;,\\[10pt]
V^3_{\mu\nu}  &= i\left\{  D_2\, g_{\mu\nu}+ E_2 \left[k_{\pi^\pm}\cdot(k_{a_1^{\pm}} + k_{\rho^{0}})g_{\mu\nu} - {k_{\rho^{0}}}_\mu
{k_{\pi^{\pm}}}_\nu - {k_{\pi^{\pm}}}_\mu {k_{a_1^\pm}}_\nu\right]\right\} \; ,\\
|V^3_{\mu\nu}|^{2} &=  4 D_2^{2}+ E_2^{2} \left[
\frac{5}{2}(M_{a_1^\pm}^{2}-M_{\rho^{0}}^{2})^{2}+\frac12 M_{\pi^{\pm}}^{2} (2M_{a_1^\pm}^{2} + 2M_{\rho^{0}}^{2} - M_{\pi^\pm}^{2})\right] + 6 D_2 E_2(M_{a_{1}^\pm}^{2} - M_{\rho^{0}}^{2})\;, \\
|V^3_{\mu\nu}k_{a_1^\pm}^{\nu}|^{2} &= D_2^{2} M_{a_1^\pm}^2 - k_{a_{1}^\pm\to\rho^{0}\pi^{\pm}}^2 E_2 M_{a_1^{\pm}}^{2} \left(E_2 M_{\rho^{0}}^{2} - 2 D_2\right)\;,\\
|V^3_{\mu\nu}k_{\rho^{0}}^{\mu}|^{2} &= D_2^{2} M_{\rho^{0}}^2 - k_{a_{1}^\pm\to\rho^{0}\pi^{\pm}}^2 E_2 M_{a_1^{\pm}}^{2} \left(E_2 M_{a_1^{\pm}}^{2} + 2 D_2\right) \;,\\
|V^3_{\mu\nu}k_{a_{1}^\pm}^{\mu}k_{\rho^{0}}^{\nu}|^{2} &= D_2^2 M_{a_1^{\pm}}^{2} \left(k_{a_{1}^\pm\to\rho^{0}\pi^{\pm}}^2 + M_{\rho^{0}}^2\right)\;,
\end{align}
\end{subequations}
where $k_{a_{1}^\pm\to\rho^{\pm}\pi^{0}}$ and $k_{a_{1}^\pm\to\rho^{0}\pi^{\pm}}$ are given in \eref{Eq:decay_momABC}.

\subsubsection*{$a_1\to \pi\gamma$}
\label{SSSec:a1pigamma
}
The relevant part of the Lagrangian reads
\begin{equation}
    \mathcal{L}_{a_1\pi\gamma} =  i B^{\gamma} {a_1^{+}}_{\mu}\pi^- A_e^{\mu} + i C^{\gamma} \left(\partial_{\mu}{a_1^{+}}_{\nu} - \partial_{\nu}{a_1^{+}}_{\mu} \right) \left(\partial^{\nu}{\pi^{-}}\right)A_e^{\mu} +\text{h.c.}\;,
\end{equation}
where
\begin{equation}
B^{\gamma} = - e g_1 \phi_N Z_{\pi^\pm}\;,\quad C^{\gamma} = - e Z_{\pi^\pm}w_{a_1^{\pm}} \;. 
\end{equation}
The decay width for the charged $a_1^{\pm}$ has the simple form
\begin{equation}
\Gamma_{a_{1}^{\pm}\rightarrow\pi^{\pm}\gamma}=\frac{e^{2}g_{1}^{2}\phi_{N}^{2}}{96\pi
M_{a_{1}^{\pm}}}Z_{\pi^{\pm}}^{2}\left[  1-\left(  \frac{M_{\pi^{\pm}}}{M_{a_{1}^{\pm}}}\right)
^{2}\right]^{3}\;. \label{eq:a1pigamma}%
\end{equation}

\subsubsection*{$f_1^H\to K^{\star}K$}
\label{SSSec:f1HKstrK}
The relevant part of the Lagrangian reads
\begin{align}
\mathcal{L}_{f_1^H K^{\star}K} &= 
i {f_1^{H}}_{\mu} \left[ B^{f_1}_1{K^{\star +}}^{\mu}K^- + B^{f_1}_2{K^{\star 0}}^{\mu}{\bar K}^0\right] + i {f_1^{H}}_{\mu} \left[ C_1^{f_1} (\partial^{\mu}{K^{\star +}}^{\nu}) \partial_{\nu}K^- + C_2^{f_1} (\partial^{\mu}{K^{\star 0}}^{\nu}) \partial_{\nu}{\bar K}^0 \right] \nonumber \\
&- 
i {f_1^{H}}_{\nu} \left[ C_1^{f_1} (\partial^{\mu}{K^{\star +}}^{\nu}) \partial_{\mu}K^-  
+ C_2^{f_1} (\partial^{\mu}{K^{\star 0}}^{\nu}) \partial_{\mu}{\bar K}^0 \right] \nonumber \\
&- i (\partial_{\mu}{f_1^{H}}_{\nu})\left[ C_1^{f_1}({K^{\star +}}^{\mu} \partial^{\nu}K^- - {K^{\star +}}^{\nu}\partial^{\mu}K^-) + C_2^{f_1}({K^{\star 0}}^{\mu}\partial^{\nu}{\bar K}^0 - {K^{\star 0}}^{\nu}\partial^{\mu}{\bar K}^0)  \right]+\text{h.c.}\;,
\end{align}
where
\begin{subequations}
\begin{align}
B_1^{f_1} &= \frac{\sqrt2}{4} Z_{K^{\pm}} \left[ (\phi_N + \phi_3
)(h_2 + g_1^2) - \sqrt2 \phi_S (h_2 - 2h_3 + 3g_1^2)\right]\;,\\
B_2^{f_1} &= \frac{\sqrt2}{4} Z_{K^{0}} \left[ (\phi_N - \phi_3
)(h_2 + g_1^2) - \sqrt2 \phi_S (h_2 - 2h_3 + 3g_1^2)\right]\;,\\
C_1^{f_1} &= \frac{g_2}{\sqrt2} Z_{K^\pm} w_{K_1^\pm}\;,\\
C_2^{f_1} &= \frac{g_2}{\sqrt2} Z_{K^0} w_{K_1^0}\;.
\end{align}
\end{subequations}
Similarly as in the case of $a_1^\pm$ there are two subchannels, and the decay width is given by
\begin{align}
\Gamma_{f_1^H\to K^{\star \pm,0,\bar 0}K^{\mp,\bar 0,0}}
&= \Gamma_{f_1^H\to K^{\star \pm }K^{\mp}} + \Gamma_{f_1^H\to K^{0,\bar 0} K^{\bar 0,0}} \nonumber\\
&= \frac{k_{f_1^H \to {K^{\star \pm}}K^\mp}}{12 M_{f_1^H}^2\pi} \left( |V^4_{\mu\nu}|^{2}-\frac{|V^4_{\mu\nu
	}k_{f_1^H}^{\nu}|^{2}}{M_{f_1^H}^{2}}-\frac{|V^4_{\mu\nu}k_{{K^{\star \pm}}}^{\mu}|^{2}%
}{M_{{K^{\star \pm}}}^{2}}+\frac{|V^4_{\mu\nu}k_{f_1^H}^{\mu}k_{{K^{\star \pm}}}^{\nu}|^{2}}{M_{f_1^H}^{2}M_{{K^{\star \pm}}}^{2}} \right) \label{eq:decay_f1H_Kstr_K}\nonumber \\ 
&+ \frac{k_{f_1^H \to {K^{\star 0}}{\bar K}^0}}{12 M_{f_1^H}^2\pi} \left(|V^5_{\mu\nu}|^{2}-\frac{|V^5_{\mu\nu
	}k_{f_1^H}^{\nu}|^{2}}{M_{f_1^H}^{2}}-\frac{|V^5_{\mu\nu}k_{{K^{\star 0}}}^{\mu}|^{2}%
}{M_{{K^{\star 0}}}^{2}}+\frac{|V^5_{\mu\nu}k_{f_1^H}^{\mu}k_{{K^{\star 0}}}^{\nu}|^{2}}{M_{f_1^H}^{2}M_{{K^{\star 0}}}^{2}}\right)\; , %
\end{align}
with (omitting the $f_1$ superscript from $B_1, C_1, B_2, C_2$)
\begin{subequations}
\begin{align}
V^4_{\mu\nu}  &= i\left\{  B_1\, g_{\mu\nu}+ C_1 \left[k_{K^\mp}\cdot(k_{f_1^H} + k_{{K^{\star \pm}}})g_{\mu\nu} - {k_{{K^{\star \pm}}}}_\mu
{k_{K^\mp}}_\nu - {k_{K^\mp}}_\mu {k_{f_1^H}}_\nu\right]\right\} \;,\\
|V^4_{\mu\nu}|^{2} &=   4 B_1^{2}+ C_1^{2} \left[
\frac{5}{2}\left(M_{f_1^H}^{2}-M_{{K^{\star \pm}}}^{2}\right)^{2}+\frac12 M_{K^\pm}^{2} \left(2M_{f_1^H}^{2} + 2M_{{K^{\star \pm}}}^{2} - M_{K^\pm}^{2}\right)\right] + 6 B_1 C_1\left(M_{f_1^H}^{2} - M_{{K^{\star \pm}}}^{2}\right)\;, \\
|V^4_{\mu\nu}k_{f_1^H}^{\nu}|^{2} &= B_1^{2} M_{f_1^H}^2 -  k_{f_1^H\to{K^{\star \pm}}K^\mp}^2 C_1 M_{f_1^H}^{2} \left(C_1 M_{{K^{\star \pm}}}^{2} - 2 B_1\right)\;,\\
|V^4_{\mu\nu}k_{{K^{\star \pm}}}^{\mu}|^{2} &= B_1^{2} M_{{K^{\star \pm}}}^2 -  k_{f_1^H\to{K^{\star \pm}}K^\mp}^2 C_1 M_{f_1^H}^{2} \left(C_1 M_{f_1^H}^{2} + 2 B_1\right)\; ,\\
|V^4_{\mu\nu}k_{f_1^H}^{\mu}k_{{K^{\star \pm}}}^{\nu}|^{2} &= B_1^2 M_{f_1^H}^{2} \left(k_{f_1^H\to{K^{\star \pm}}K^\mp}^2 + M_{{K^{\star \pm}}}^2\right)\;,\\[10pt]
V^5_{\mu\nu}  &= i\left\{  B_2\, g_{\mu\nu}+ C_2 \left[k_{{\bar K}^0}\cdot(k_{f_1^H} + k_{{K^{\star 0}}})g_{\mu\nu} - {k_{{K^{\star 0}}}}_\mu
{k_{{\bar K}^0}}_\nu - {k_{{\bar K}^0}}_\mu {k_{f_1^H}}_\nu\right]\right\}\;  ,\\
|V^5_{\mu\nu}|^{2} &=  4 B_2^{2}+ C_2^{2} \left[
\frac{5}{2}\left(M_{f_1^H}^{2}-M_{{K^{\star 0}}}^{2}\right)^{2}+\frac12 M_{K^0}^{2} \left(2M_{f_1^H}^{2} + 2M_{{K^{\star 0}}}^{2} - M_{K^0}^{2}\right) \right] + 6 B_2 C_2\left(M_{f_1^H}^{2} - M_{{K^{\star 0}}}^{2}\right)\;, \\
|V^5_{\mu\nu}k_{f_1^H}^{\nu}|^{2} &= B_2^{2} M_{f_1^H}^2 - k_{f_1^H\to{K^{\star 0}}{\bar K}^0}^2 C_2 M_{f_1^H}^{2} \left(C_2 M_{{K^{\star 0}}}^{2} - 2 B_2\right)\;,\\
|V^5_{\mu\nu}k_{{K^{\star 0}}}^{\mu}|^{2} &= B_2^{2} M_{{K^{\star 0}}}^2 - k_{f_1^H\to{K^{\star 0}}{\bar K}^0}^2 C_2 M_{f_1^H}^{2} \left(C_2 M_{f_1^H}^{2} + 2 B_2\right) \;,\\
|V^5_{\mu\nu}k_{f_1^H}^{\mu}k_{{K^{\star 0}}}^{\nu}|^{2} &= B_2^2 M_{f_1^H}^{2} \left(k_{f_1^H\to{K^{\star 0}}{\bar K}^0}^2 + M_{{K^{\star 0}}}^2\right)\;.
\end{align}
\end{subequations}

\subsection{Scalar-meson decays}
\label{SSec:scalarmeson_dec}

\subsubsection*{$K_0^{\star}\to K\pi$}
\label{SSSec:K0starKpi}
The relevant part of the Lagrangian reads
\begin{align}
\mathcal{L}_{K_0^{\star}K\pi} &= K_0^{\star 0} \left[ B^{K_0^{\star}}_1  \pi^0 \bar{K^0} + B^{K_0^{\star}}_2 (\partial_{\mu}\pi^0)\partial^{\mu}\bar{K^0} + B^{K_0^{\star}}_3 \pi^+ K^- + B^{K_0^{\star}}_4 (\partial_{\mu}\pi^+) \partial^{\mu}K^- \right]\nonumber\\
&+ (\partial_{\mu}K_0^{\star 0})\left[ C^{K_0^{\star}}_1 \pi^0\partial^{\mu}\bar{K^0} + C^{K_0^{\star}}_2 (\partial^{\mu}\pi^0) \bar{K^0} + C^{K_0^{\star}}_3 \pi^+\partial^{\mu}K^- + C^{K_0^{\star}}_4 (\partial^{\mu}\pi^+) K^-\right]\nonumber \\
&+ K_0^{\star +}\left[D^{K_0^{\star}}_1 \pi^0 K^- + D^{K_0^{\star}}_2 (\partial_{\mu}\pi^0)\partial^{\mu}K^- + D^{K_0^{\star}}_3 \pi^- \bar{K^0} + D^{K_0^{\star}}_4 (\partial_{\mu}\pi^-) \partial^{\mu}\bar{K^0}\right]\nonumber\\ 
&+ (\partial_{\mu}K_0^{\star +})\left[ F^{K_0^{\star}}_1 \pi^0\partial^{\mu}K^- + F^{K_0^{\star}}_2 (\partial^{\mu}\pi^0)K^- + F^{K_0^{\star}}_3 \pi^- \partial^{\mu}\bar{K^0} + F^{K_0^{\star}}_4 (\partial^{\mu}\pi^-) \bar{K^0}\right] + \text{h.c.}\;,
\end{align}
where
\begin{subequations}
\begin{align}
  B^{K_0^{\star}}_1 &= \frac{1}{\sqrt{2}} Z_{K_0^{\star 0}} Z_{K^0} {\mathds{O}_{P}}_{21} \phi_S \left[\lambda_2 - c_1\phi_3 (\phi_N + \phi_3) \right]\;,\\
  B^{K_0^{\star}}_2 &= \frac14 Z_{K_0^{\star 0}} Z_{K^0}\left\{2g_1(W_a + w_{K_1^0}{\mathds{O}_{P}}_{21}) - W_a w_{K_1^0}\left[(\phi_N-\phi_3)(h_2 - 2h_3 + 3g_1^2) + \sqrt{2}\phi_S (h_2 + g_1^2)\right]\right\}\;, \\
  B^{K_0^{\star}}_3 &= - Z_{K_0^{\star 0}} Z_{K^{\pm}}Z_{\pi^{\pm}}\left( \phi_S - \sqrt2 \phi_3 \right)\lambda_2\; \\
  B^{K_0^{\star}}_4 &= -\frac{1}{2\sqrt{2}} Z_{K_0^{\star 0}} Z_{K^{\pm}} Z_{\pi^{\pm}} \left\{2g_1(w_{a_1^{\pm}} + w_{K_1^{\pm}}) - w_{a_1^\pm}w_{K_1^{\pm}} \big[\phi_N(h_2 - 2h_3 + 3g_1^2) - \phi_3(h_2 + 2h_3 - g_1^2) \right. \nonumber \\
      &+ \left. \sqrt{2}\phi_S (h_2 + g_1^2)\big]\right\}\;,\\
  C^{K_0^{\star}}_1 &= -\frac12 Z_{K_0^{\star 0}} Z_{K^0} {\mathds{O}_{P}}_{21} \left[g_1(w_{K_1^{0}} + \bar{w}_{K^{\star 0}}) - w_{K_1^{0}}\bar{w}_{K^{\star 0}} \sqrt2 \phi_S (g_1^2 - h_3)\right]\;, \\
  C^{K_0^{\star}}_2 &= -\frac14 Z_{K_0^{\star 0}} Z_{K^0} \left\{2g_1(W_a - \bar{w}_{K^{\star 0}}{\mathds{O}_{P}}_{21}) + W_a\bar{w}_{K^{\star 0}} \left[(\phi_N-\phi_3)(h_2 - 2h_3 + 3g_1^2) - \sqrt{2}\phi_S (h_2 + g_1^2)\right]\right\}\;, \\
  C^{K_0^{\star}}_3 &= \frac{1}{\sqrt2} Z_{K_0^{\star 0}} Z_{K^{\pm}} Z_{\pi^{\pm}} \left\{g_1(w_{K_1^{\pm}} + \bar{w}_{K^{\star 0}}) - w_{K_1^{\pm}}\bar{w}_{K^{\star 0}} \left[ \phi_3 (h_2 + g_1^2) + \sqrt2 \phi_S (g_1^2 - h_3)\right]\right\}\;, \\
  C^{K_0^{\star}}_4 &= \frac{1}{2\sqrt2} Z_{K_0^{\star 0}} Z_{K^{\pm}} Z_{\pi^{\pm}} \left\{2g_1(w_{a_1^{\pm}} - \bar{w}_{K^{\star 0}}) + w_{a_1^{\pm}}\bar{w}_{K^{\star 0}} \big[(\phi_N(h_2 - 2h_3 + 3g_1^2) + \phi_3(h_2 + 2h_3 - 3g_1^2)\right. \nonumber \\
      &- \left. \sqrt{2}\phi_S (h_2 + g_1^2)\big]\right\}\;, \\
  D^{K_0^{\star}}_1 &= -\frac{1}{\sqrt{2}} Z_{K_0^{\star \pm}} Z_{K^{\pm}} {\mathds{O}_{P}}_{21} \phi_S \left[\lambda_2 - c_1\phi_3 (\phi_N - \phi_3) \right] \;,\\
  D^{K_0^{\star}}_2 &= -\frac14 Z_{K_0^{\star \pm}} Z_{K^{\pm}}\left\{2g_1(W_a + w_{K_1^{\pm}}{\mathds{O}_{P}}_{21}) - W_a w_{K_1^{\pm}}\left[(\phi_N + \phi_3)(h_2 - 2h_3 + 3g_1^2) + \sqrt{2}\phi_S (h_2 + g_1^2)\right]\right\}\;, \\
  D^{K_0^{\star}}_3 &= - Z_{K_0^{\star \pm}} Z_{K^{0}}Z_{\pi^{\pm}}\left( \phi_S + \sqrt2 \phi_3 \right)\lambda_2\;, \\
  D^{K_0^{\star}}_4 &= -\frac{1}{2\sqrt{2}} Z_{K_0^{\star \pm}} Z_{K^{0}} Z_{\pi^{\pm}} \left\{2g_1(w_{a_1^{\pm}} + w_{K_1^{0}}) - w_{a_1^\pm}w_{K_1^{0}} \big[\phi_N(h_2 - 2h_3 + 3g_1^2) + \phi_3(h_2 + 2h_3 - g_1^2) \right. \nonumber  \\
      &+ \left. \sqrt{2}\phi_S (h_2 + g_1^2)\big]\right\}\;, \\
  F^{K_0^{\star}}_1 &= \frac12 Z_{K_0^{\star \pm}} Z_{K^{\pm}} {\mathds{O}_{P}}_{21} \left[g_1(w_{K_1^{\pm}} + \bar{w}_{K^{\star +}}) - w_{K_1^{\pm}}\bar{w}_{K^{\star +}} \sqrt2 \phi_S (g_1^2 - h_3)\right] \;,\\
  F^{K_0^{\star}}_2 &= \frac14 Z_{K_0^{\star \pm}} Z_{K^{\pm}} \left\{2g_1(W_a - \bar{w}_{K^{\star +}}{\mathds{O}_{P}}_{21}) + W_a\bar{w}_{K^{\star +}} \left[(\phi_N + \phi_3) (h_2 - 2h_3 + 3g_1^2) - \sqrt{2}\phi_S (h_2 + g_1^2)\right]\right\}\;, \\
  F^{K_0^{\star}}_3 &= \frac{1}{\sqrt2} Z_{K_0^{\star \pm}} Z_{K^{0}} Z_{\pi^{\pm}} \left\{g_1(w_{K_1^{0}} + \bar{w}_{K^{\star +}}) - w_{K_1^{0}}\bar{w}_{K^{\star +}} \left[ -\phi_3 (h_2 + g_1^2) + \sqrt2 \phi_S (g_1^2 - h_3)\right]\right\}\;, \\
  F^{K_0^{\star}}_4 &= \frac{1}{2\sqrt2} Z_{K_0^{\star \pm}} Z_{K^{0}} Z_{\pi^{\pm}} \left\{2g_1(w_{a_1^{\pm}} - \bar{w}_{K^{\star +}}) + w_{a_1^{\pm}}\bar{w}_{K^{\star +}} \big[\phi_N(h_2 - 2h_3 + 3g_1^2) - \phi_3(h_2 + 2h_3 - 3g_1^2)\right. \nonumber \\
      &- \left. \sqrt{2}\phi_S (h_2 + g_1^2)\big]\right\}\;,
\end{align}
\end{subequations}
and
\begin{subequations}
\begin{align}
  W_a &= w_{\eta}^a {\mathds{O}_{P}}_{11} + w_{\pi}^a {\mathds{O}_{P}}_{21} \;,\\
  \bar{w}_{K^{\star 0}} &\equiv iw_{K^{\star 0}} = -\frac{g_1(\phi_N - \phi_3 -\sqrt2\phi_S)}{2m_{K^{\star 0}}^2} \;,\\
  \bar{w}_{K^{\star +}} &\equiv iw_{K^{\star +}} = -\frac{g_1(\phi_N + \phi_3 -\sqrt2\phi_S)}{2m_{K^{\star\pm}}^2}\;.
\end{align}
\end{subequations}
The tree-level decay widths for the neutral and charged $K_0^\star$ are given by
\begin{align}
\Gamma_{\bar K_0^{\star 0}\to \pi^{0,+}K^{\bar{0},-}} &= \Gamma_{\bar K_0^{\star 0}\to \pi^{0}\bar K^{0}} + \Gamma_{\bar K_0^{\star 0}\to \pi^{+}K^{-}}= \nonumber\\
= \frac{1}{8\pi M_{K_0^{\star 0}}^2} & \Bigg\lbrace k_{\bar K_0^{\star 0}\to \pi^{0}\bar K^{0}} \left| B^{K_0^{\star}}_1 + \frac12 \left(C^{K_0^{\star}}_1 + C^{K_0^{\star}}_2 - B^{K_0^{\star}}_2\right) (M_{K_0^{\star 0}}^2 - M_{K^{0}}^2 - M_{\pi^{0}}^2) + C^{K_0^{\star}}_1 M_{K^{0}}^2 + C^{K_0^{\star}}_2 M_{\pi^{0}}^2 \right|^2 \nonumber\\
&+  k_{\bar K_0^{\star 0}\to \pi^{+}K^{-}} \left| B^{K_0^{\star}}_3 + \frac12 \left(C^{K_0^{\star}}_3 + C^{K_0^{\star}}_4 - B^{K_0^{\star}}_4\right) (M_{K_0^{\star 0}}^2 - M_{K^{\pm}}^2 - M_{\pi^{\pm}}^2) + C^{K_0^{\star}}_3 M_{K^{\pm}}^2 + C^{K_0^{\star}}_4 M_{\pi^{\pm}}^2 \right|^2 \Bigg\rbrace \;,\\
\Gamma_{K_0^{\star -}\to \pi^{0,-}K^{-, \bar{0}}} &= \Gamma_{K_0^{\star -}\to \pi^{0} K^{-}} + \Gamma_{K_0^{\star -}\to \pi^{-}\bar K^{0}}= \nonumber\\
= \frac{1}{8\pi M_{K_0^{\star \pm}}^2} & \Bigg\lbrace k_{K_0^{\star -}\to \pi^{0} K^{-}} \left| D^{K_0^{\star}}_1 + \frac12 \left(F^{K_0^{\star}}_1 + F^{K_0^{\star}}_2 - D^{K_0^{\star}}_2\right) (M_{K_0^{\star \pm}}^2 - M_{K^{\pm}}^2 - M_{\pi^{0}}^2) + F^{K_0^{\star}}_1 M_{K^{\pm}}^2 + F^{K_0^{\star}}_2 M_{\pi^{0}}^2 \right|^2 \nonumber\\
 &+  k_{K_0^{\star -}\to \pi^{-}\bar K^{0}} \left| D^{K_0^{\star}}_3 + \frac12 \left(F^{K_0^{\star}}_3 + F^{K_0^{\star}}_4 - D^{K_0^{\star}}_4\right) (M_{K_0^{\star \pm}}^2 - M_{K^{0}}^2 - M_{\pi^{\pm}}^2) + F^{K_0^{\star}}_3 M_{K^{0}}^2 + F^{K_0^{\star}}_4 M_{\pi^{\pm}}^2 \right|^2 \Bigg\rbrace \;, 
\end{align}
where $k_{A\to BC}$ is defined in \eref{Eq:decay_momABC}.

\subsubsection*{$a_0\to K K$}
\label{SSSec:a0KK}
The relevant part of the Lagrangian reads
\begin{align}
\mathcal{L}_{a_0 K K} &= a_0^{0} \left[ B^{a_0}_1  K^0 \bar{K^0} + B^{a_0}_2 (\partial_{\mu}K^0)\partial^{\mu}\bar{K^0} + B^{a_0}_3 K^+ K^- + B^{a_0}_4 (\partial_{\mu}K^+) \partial^{\mu}K^- \right]\nonumber\\
&+ (\partial_{\mu}a_0^{0})\left[ C^{a_0}_1 K^0\partial^{\mu}\bar{K^0} + C^{a_0}_2 (\partial^{\mu}K^0) \bar{K^0} + C^{a_0}_3 K^+\partial^{\mu}K^- + C^{a_0}_4 (\partial^{\mu}K^+) K^-\right]\nonumber\\
&+ a_0^{+}\left[D^{a_0}_1 K^0 K^- + D^{a_0}_2 (\partial_{\mu}K^0)\partial^{\mu}K^- \right] 
+ (\partial_{\mu}a_0^{+})\left[ F^{a_0}_1 K^0\partial^{\mu}K^- + F^{a_0}_2 (\partial^{\mu}K^0)K^- \right] + \text{h.c.}\;,	
\end{align}
where we introduce
\begin{subequations}
\begin{align}
 {\mathds{O}_{S}}^{a_0}_{2} &= {\mathds{O}_{S}}_{22}\;,\\
 {\mathds{O}_{S}}^{a_0}_{NS0} &= \sqrt{2}{\mathds{O}_{S}}_{21}+{\mathds{O}_{S}}_{23}\;,\\
 {\mathds{O}_{S}}^{a_0}_{NS8} &= {\mathds{O}_{S}}_{21}-\sqrt{2}{\mathds{O}_{S}}_{23}\;,
\end{align}
\end{subequations}
\begin{subequations}
\begin{align}
  B^{a_0}_1 &= \frac{1}{2} Z_{K^0}^2 {\mathds{O}_{S}}^{a_0}_{2} \left[\lambda_2(2\phi_N -\sqrt2\phi_S) - 2\phi_3 (2\lambda_1 + \lambda_2) \right]\nonumber\\
  &-\frac{\sqrt{2}}{6}Z_{K^0}^2 {\mathds{O}_{S}}^{a_0}_{NS0}\left[ \phi_N(4\lambda_1+\lambda_2)-\phi_3\lambda_2+\sqrt{2}\phi_S(2\lambda_1+\lambda_2)\right]\nonumber\\
  &-\frac{1}{6}Z_{K^0}^2 {\mathds{O}_{S}}^{a_0}_{NS8}\left[ 4\phi_N(\lambda_1+\lambda_2)-4\phi_3\lambda_2-\sqrt{2}\phi_S(4\lambda_1+5\lambda_2)\right]\;,\\
  B^{a_0}_2 &= \frac{1}{2} Z_{K^0}^2 {\mathds{O}_{S}}^{a_0}_{2}w_{K_1^0} \left\{2g_1 - w_{K_1^0} \left[\phi_N(h_2 + g_1^2) - \phi_3(2h_1 + h_2 + g_1^2) + \sqrt{2}\phi_S (g_1^2 - h_3)\right]\right\} \nonumber\\
  &- \frac{\sqrt{2}}{6}Z_{K^0}^2 {\mathds{O}_{S}}^{a_0}_{NS0} w_{K_1^0} \left\{4g_1 - w_{K_1^0} \left[\phi_N(2h_1+ h_2 -h_3 + 2g_1^2) - \phi_3(h_2 - h_3 + 2g_1^2) \right.\right. \nonumber\\
  &+\left.\left. \sqrt{2}\phi_S (h_1 + h_2 - h_3 + 2g_1^2)\right]\right\} \nonumber\\
  &+ \frac{1}{6}Z_{K^0}^2 {\mathds{O}_{S}}^{a_0}_{NS8} w_{K_1^0} \left\{2g_1 + w_{K_1^0} \left[\phi_N(2h_1+ h_2 + 2h_3 - g_1^2) - \phi_3(h_2 + 2h_3 - g_1^2) \right.\right. \nonumber\\
  &-\left.\left. \sqrt{2}\phi_S (2h_1 + 2h_2 + h_3 + g_1^2)\right]\right\} \;,\\
  B^{a_0}_3 &= -\frac{1}{2} Z_{K^\pm}^2 {\mathds{O}_{S}}^{a_0}_{2} \left[\lambda_2(2\phi_N -\sqrt2\phi_S) + 2\phi_3 (2\lambda_1 + \lambda_2) \right]\nonumber\\
  &-\frac{\sqrt{2}}{6}Z_{K^\pm}^2 {\mathds{O}_{S}}^{a_0}_{NS0}\left[ \phi_N(4\lambda_1+\lambda_2)+\phi_3\lambda_2+\sqrt{2}\phi_S(2\lambda_1+\lambda_2)\right]\nonumber\\
  &-\frac{1}{6}Z_{K^\pm}^2 {\mathds{O}_{S}}^{a_0}_{NS8}\left[ 4\phi_N(\lambda_1+\lambda_2)+4\phi_3\lambda_2-\sqrt{2}\phi_S(4\lambda_1+5\lambda_2)\right]\;,\\
  B^{a_0}_4 &= -\frac{1}{2} Z_{K^\pm}^2 {\mathds{O}_{S}}^{a_0}_{2}w_{K_1^\pm} \left\{2g_1 - w_{K_1^\pm} \left[\phi_N(h_2 + g_1^2) + \phi_3(2h_1 + h_2 + g_1^2) + \sqrt{2}\phi_S (g_1^2 - h_3)\right]\right\}\nonumber \\
  &- \frac{\sqrt{2}}{6}Z_{K^\pm}^2 {\mathds{O}_{S}}^{a_0}_{NS0} w_{K_1^\pm} \left\{4g_1 - w_{K_1^\pm} \left[\phi_N(2h_1+ h_2 -h_3 + 2g_1^2) + \phi_3(h_2 - h_3 + 2g_1^2) \right.\right. \nonumber\\
  &+\left.\left. \sqrt{2}\phi_S (h_1 + h_2 - h_3 + 2g_1^2)\right]\right\} \nonumber \\
  &+ \frac{1}{6}Z_{K^\pm}^2 {\mathds{O}_{S}}^{a_0}_{NS8} w_{K_1^\pm} \left\{2g_1 + w_{K_1^\pm} \left[\phi_N(2h_1+ h_2 + 2h_3 - g_1^2) + \phi_3(h_2 + 2h_3 - g_1^2) \right.\right. \nonumber\\
  &-\left.\left. \sqrt{2}\phi_S (2h_1 + 2h_2 + h_3 + g_1^2)\right]\right\}\;,\\
  C^{a_0}_1 &= -\frac{g_1}{2} Z_{K^0}^2 {\mathds{O}_{S}}^{a_0}_{2} w_{K_1^0} + g_1\frac{\sqrt{2}}{3} Z_{K^0}^2 {\mathds{O}_{S}}^{a_0}_{NS0} w_{K_1^0} - \frac{g_1}{6} Z_{K^0}^2 {\mathds{O}_{S}}^{a_0}_{NS8} w_{K_1^0}
  \;, \quad C^{a_0}_2 = C^{a_0}_1 \;,\\
  C^{a_0}_3 &= \frac{g_1}{2} Z_{K^\pm}^2 {\mathds{O}_{S}}^{a_0}_{2} w_{K_1^\pm} + g_1\frac{\sqrt{2}}{3} Z_{K^\pm}^2 {\mathds{O}_{S}}^{a_0}_{NS0} w_{K_1^\pm} - \frac{g_1}{6} Z_{K^\pm}^2 {\mathds{O}_{S}}^{a_0}_{NS8} w_{K_1^\pm} \;, \quad C^{a_0}_4 = C^{a_0}_3 \;,\\
  D^{a_0}_1 &=	-\frac{1}{\sqrt2}Z_{K^0}Z_{K^\pm}Z_{a_0^\pm}  \lambda_2(2\phi_N -\sqrt2\phi_S)\;,  \\
  D^{a_0}_2 &= -\frac{1}{\sqrt2} Z_{K^0}Z_{K^\pm}Z_{a_0^\pm} \left\{g_1(w_{K_1^0} + w_{K_1^\pm}) - w_{K_1^0}w_{K_1^\pm} \left[\phi_N(h_2 + g_1^2) + \sqrt{2}\phi_S (g_1^2 - h_3)\right]\right\} \;,\\
  F^{a_0}_1 &= \frac{1}{2\sqrt2} Z_{K^0}Z_{K^\pm}Z_{a_0^\pm} \left\{ 2 g_1(w_{K_1^\pm} - \bar{w}_{\rho^{+}}) - w_{K_1^\pm} \bar{w}_{\rho^{+}} \Big[\phi_N(h_2 + 2h_3 - g_1^2) - \phi_3(h_2 -2h_3 + 3g_1^2)\right. \nonumber\\
   &-\left. \sqrt{2}\phi_S (g_1^2 + h_2)\Big]\right\}\;, \\
   F^{a_0}_2 &= \frac{1}{2\sqrt2} Z_{K^0}Z_{K^\pm}Z_{a_0^\pm} \left\{ 2 g_1(w_{K_1^0} + \bar{w}_{\rho^{+}}) + w_{K_1^0} \bar{w}_{\rho^{+}} \Big[\phi_N(h_2 + 2h_3 - g_1^2) + \phi_3(h_2 -2h_3 + 3g_1^2)\right. \nonumber\\
  &-\left. \sqrt{2}\phi_S (g_1^2 + h_2)\Big]\right\}\;,  
\end{align}
\end{subequations}
and
\begin{equation}
\bar{w}_{\rho^{+}} \equiv iw_{\rho^{+}} = -\frac{g_1\phi_3}{m_{\rho^{\pm}}^2}\;.
\end{equation}
The tree-level decay widths for the neutral and charged $a_0$ are given by
\begin{subequations}
\begin{align}
\Gamma_{a_0^0\to K^{0,+}K^{\bar{0},-}} &= \Gamma_{a_0^0\to K^{0}\bar K^{0}} + \Gamma_{a_0^0\to K^{+}K^{-}}= \nonumber\\
&= \frac{1}{8\pi M_{a_0^0}^2} \Bigg\lbrace k_{a_0^0\to K^{0}\bar K^{0}} \left| B^{a_0}_1 + \frac12 \left(C^{a_0}_1 + C^{a_0}_2 - B^{a_0}_2\right) M_{a_0^0}^2 + B^{a_0}_2 M_{K^{0}}^2 \right|^2  \nonumber\\
& + k_{a_0^0\to K^{+}K^{-}} \left| B^{a_0}_3 + \frac12 \left(C^{a_0}_3 + C^{a_0}_4 - B^{a_0}_4\right) M_{a_0^0}^2 + B^{a_0}_4 M_{K^{\pm}}^2 \right|^2 \Bigg\rbrace\;,\\
\Gamma_{a_0^{-}\to K^{0}K^{-}} &= \nonumber\\
=\frac{1}{8\pi M_{a_0^\pm}^2} &\Bigg\lbrace k_{a_0^{-}\to K^{0} K^{-}} \left| D^{a_0}_1 + \frac12 \left(F^{a_0}_1 + F^{a_0}_2 - D^{a_0}_2\right) (M_{a_0^\pm}^2 - M_{K^{\pm}}^2 - M_{K^{0}}^2) + F^{a_0}_1 M_{K^{\pm}}^2 + F^{a_0}_2 M_{K^{0}}^2 \right|^2 \Bigg\rbrace \;,
\end{align}
\end{subequations}
where $k_{A\to BC}$ is defined in \eref{Eq:decay_momABC}.

\subsubsection*{$a_0\to \pi\eta, \pi\eta^{\prime}$}
\label{SSSec:a0pieta}
We start from the following part of the Lagrangian
\begin{align}
\mathcal{L}_{a_0 \pi \eta_N/\eta_S} &= a_0^{0} \left[\mathbf{x}^T {\mathds{B}_{1}^{\eta}} \mathbf{x} + (\partial_{\mu}\mathbf{x})^T {\mathds{B}_{2}^{\eta}} \partial^{\mu}\mathbf{x} \right] + (\partial_{\mu}a_0^{0})\left[ (\partial^{\mu}\mathbf{x})^T {\mathds{C}^{\eta}} \mathbf{x} \right]\nonumber\\
&+ a_0^{+}\left[\pi^- {\mathbf{D}_{1}^{\eta}}^T \mathbf{x} + (\partial_{\mu}\pi^-) {\mathbf{D}_{2}^{\eta}}^T \partial^{\mu}\mathbf{x} \right] + (\partial_{\mu}a_0^{+})\left[ \pi^- {\mathbf{F}_{1}^{\eta}}^T \partial^{\mu}\mathbf{x} + (\partial^{\mu} \pi^-) {\mathbf{F}_{2}^{\eta}}^T \mathbf{x} \right] + \text{h.c.}\;,	
\end{align}
where $\mathbf{x}^T = (\tilde\eta_N,\tilde\pi^0, \tilde\eta_S)$ and the coefficient matrices and vectors are

\begin{subequations}
\begin{align}
{\mathds{B}_{1}^{\eta}}_{11} &= -\frac{1}{6}{\mathds{O}_{S}}^{a_0}_{NS0}\left[2c_1\phi_N\phi_S (\phi_N + \sqrt{2}\phi_S) +\sqrt{2}\phi_N(2\lambda_1+\lambda_2) + 2\lambda_1\phi_S\right]\nonumber\\
&-\frac{1}{2}{\mathds{O}_{S}}^{a_0}_{2}(2\lambda_1 + \lambda_2)\phi_3\nonumber\\
&+\frac{1}{6}{\mathds{O}_{S}}^{a_0}_{NS8}\left[2c_1\phi_N\phi_S (\sqrt{2}\phi_N - \phi_S) -\phi_N(2\lambda_1+\lambda_2) + 2\sqrt{2}\lambda_1\phi_S\right]\;,\\
{\mathds{B}_{1}^{\eta}}_{12} &= 
\frac{1}{6}{\mathds{O}_{S}}^{a_0}_{NS0}\left[c_1\phi_S (2\phi_N + \sqrt{2}\phi_S) - \sqrt{2}\lambda_2\right]\phi_3\nonumber\\
&+\frac{1}{2}{\mathds{O}_{S}}^{a_0}_{2}(c_1\phi_S^2 - \lambda_2)\phi_N\nonumber\\
&-\frac{1}{6}{\mathds{O}_{S}}^{a_0}_{NS8}\left[c_1\phi_S (2\sqrt{2}\phi_N - \phi_S) + \lambda_2\right]\phi_3\;,\\
{\mathds{B}_{1}^{\eta}}_{13} &= 
-\frac{1}{12}{\mathds{O}_{S}}^{a_0}_{NS0}c_1\left[\phi_N^2 (\phi_N + 3\sqrt{2}\phi_S) - \phi_3^2(\phi_N + \sqrt{2}\phi_S)\right]\nonumber\\
&+\frac{1}{2}{\mathds{O}_{S}}^{a_0}_{2}c_1\phi_N\phi_S\phi_3\nonumber\\
&+\frac{1}{12}{\mathds{O}_{S}}^{a_0}_{NS8}c_1\left[\phi_N^2 (\sqrt{2}\phi_N - 3\phi_S) - \phi_3^2(\sqrt{2}\phi_N - \phi_S)\right]\;,\\
{\mathds{B}_{1}^{\eta}}_{21} &= {\mathds{B}_{1}^{\eta}}_{12}\;\\
{\mathds{B}_{1}^{\eta}}_{22} &= 
-\frac{1}{6}{\mathds{O}_{S}}^{a_0}_{NS0}\left[2c_1 \phi_3^2 \phi_S +\sqrt{2}\phi_N(2\lambda_1+\lambda_2) + 2\lambda_1\phi_S\right]\nonumber\\
&-\frac{1}{2}{\mathds{O}_{S}}^{a_0}_{2}(2\lambda_1 + \lambda_2 + 2c_1\phi_S^2)\phi_3\nonumber\\
&+\frac{1}{6}{\mathds{O}_{S}}^{a_0}_{NS8}\left[2\sqrt{2} c_1\phi_3^2\phi_S -\phi_N(2\lambda_1+\lambda_2) + 2\sqrt{2}\lambda_1\phi_S\right]\;,\\
{\mathds{B}_{1}^{\eta}}_{23} &= 
\frac{1}{12}{\mathds{O}_{S}}^{a_0}_{NS0}c_1 \phi_3\left(\phi_N^2 - \phi_3^2 +2\sqrt{2}\phi_N \phi_S\right)\nonumber\\
&+\frac{1}{4}{\mathds{O}_{S}}^{a_0}_{2} c_1\phi_S(\phi_N^2 - 3\phi_3^2)\nonumber\\
&-\frac{1}{12}{\mathds{O}_{S}}^{a_0}_{NS8}c_1 \phi_3\left[\sqrt{2}(\phi_N^2 - \phi_3^2) -2\phi_N \phi_S\right]\;,\\
{\mathds{B}_{1}^{\eta}}_{31} &= {\mathds{B}_{1}^{\eta}}_{13}\;,\\
{\mathds{B}_{1}^{\eta}}_{32} &= {\mathds{B}_{1}^{\eta}}_{23}\;,\\
{\mathds{B}_{1}^{\eta}}_{33} &= 
-\frac{1}{6}{\mathds{O}_{S}}^{a_0}_{NS0}\left[\sqrt{2}c_1 \phi_N(\phi_N^2-\phi_3^2) +  2\sqrt{2} \lambda_1 \phi_N  + 2(\lambda_1 + \lambda_2)\phi_S\right]\nonumber\\
&+\frac{1}{2}{\mathds{O}_{S}}^{a_0}_{2}\left[-2\lambda_1 + c_1(\phi_N^2 - \phi_3^2)\right]\phi_3\nonumber\\
&+\frac{1}{6}{\mathds{O}_{S}}^{a_0}_{NS8}\left[-c_1 \phi_N(\phi_N^2-\phi_3^2) -  2 \lambda_1 \phi_N  + 2 \sqrt{2}(\lambda_1 + \lambda_2)\phi_S\right]\;,\\
{\mathds{B}_{2}^{\eta}}_{11} &= 
 \frac{\sqrt{2}}{6}{\mathds{O}_{S}}^{a_0}_{NS0}\left\{-2g_1 w_{\eta}^f + \left[({w_{\eta}^a})^2 + ({w_{\eta}^f})^2\right] (2g_1^2 + h_1 +h_2 - h_3)\phi_N + 2w_{\eta}^a w_{\eta}^f (2g_1^2 + h_2 - h_3)\phi_3 \right.\nonumber\\
 &\left.+ \frac{1}{\sqrt{2}}(({w_{\eta}^a})^2 + ({w_{\eta}^f})^2) h_1 \phi_S \right\}\nonumber\\
&+{\mathds{O}_{S}}^{a_0}_{2}\left\{ -g_1 w_{\eta}^a + w_{\eta}^a w_{\eta}^f (2g_1^2 + h_2 - h_3) \phi_N + \frac12 \left[({w_{\eta}^a})^2 + ({w_{\eta}^f})^2\right] (2g_1^2 + h_1 +h_2 - h_3)\phi_3\right\}\nonumber\\
&+\frac{1}{6}{\mathds{O}_{S}}^{a_0}_{NS8}\left\{-2g_1 w_{\eta}^f + \left[{w_{\eta}^a})^2 + ({w_{\eta}^f})^2\right] (2g_1^2 + h_1 +h_2 - h_3)\phi_N + 2w_{\eta}^a w_{\eta}^f (2g_1^2 + h_2 - h_3)\phi_3 \right.\nonumber\\
&\left.- \sqrt{2}(({w_{\eta}^a})^2 + ({w_{\eta}^f})^2) h_1 \phi_S \right\}\;,\\
 {\mathds{B}_{2}^{\eta}}_{12} &= 
 \frac{\sqrt{2}}{12}{\mathds{O}_{S}}^{a_0}_{NS0}\left\{-2g_1 (w_{\eta}^a + w_{\pi}^f) + 2(w_{\pi}^a w_{\eta}^a + w_{\pi}^f w_{\eta}^f ) \left[
 \vphantom{\frac{1}{\sqrt{2}}}(2g_1^2 + h_2 - h_3)(\phi_N + \phi_3)\right.\right. \nonumber \\ 
 & \left.\left.+ h_1 (\phi_N + \frac{1}{\sqrt{2}}\phi_S) \right]\right\}\nonumber\\
&+{\mathds{O}_{S}}^{a_0}_{2}\left[-\frac{g_1}{2} (w_{\pi}^a + w_{\eta}^f) + \frac12(w_{\eta}^a w_{\pi}^f + w_{\eta}^f w_{\pi}^a) (2g_1^2 + h_2 - h_3) \phi_N \right.\nonumber \\ 
 &+ \left.\frac12(w_{\eta}^a w_{\pi}^a + w_{\eta}^f w_{\pi}^f) (2g_1^2 + h_1 +h_2 - h_3)\phi_3\right] \nonumber\\
&+\frac{1}{12}{\mathds{O}_{S}}^{a_0}_{NS8}\left\{-2g_1 (w_{\eta}^a + w_{\pi}^f) + 2(w_{\pi}^a w_{\eta}^a + w_{\pi}^f w_{\eta}^f ) \left[ \vphantom{\frac{1}{\sqrt{2}}} (2g_1^2 + h_2 - h_3)(\phi_N + \phi_3)\right.\right. \nonumber\\ 
 & \left.\left.+ h_1 (\phi_N - \sqrt{2}\phi_S) \right]\right\}\;,\\
 {\mathds{B}_{2}^{\eta}}_{13} &= 0\;, \\
 {\mathds{B}_{2}^{\eta}}_{21} &= {\mathds{B}_{2}^{\eta}}_{12} \;, \\
 {\mathds{B}_{2}^{\eta}}_{22} &=
 \frac{\sqrt{2}}{6}{\mathds{O}_{S}}^{a_0}_{NS0}\left\{ \vphantom{\frac{1}{\sqrt{2}}}-2g_1 w_{\pi}^a + \left[({w_{\pi}^a})^2 + ({w_{\pi}^f})^2\right] (2g_1^2 + h_1 +h_2 - h_3)\phi_N + 2w_{\pi}^a w_{\pi}^f (2g_1^2 + h_2 - h_3)\phi_3 \right.\nonumber\\
 &\left.+ \frac{1}{\sqrt{2}}\left[({w_{\pi}^a})^2 + ({w_{\pi}^f})^2\right] h_1 \phi_S \right\} \nonumber\\
&+ {\mathds{O}_{S}}^{a_0}_{2}\left\{-g_1 w_{\pi}^f + w_{\pi}^a w_{\pi}^f (2g_1^2 + h_2 - h_3) \phi_N + \frac12 \left[({w_{\pi}^a})^2 + ({w_{\pi}^f})^2\right] (2g_1^2 + h_1 +h_2 - h_3)\phi_3\right\} \nonumber  \\
&+\frac{1}{6}{\mathds{O}_{S}}^{a_0}_{NS8}\left\{-2g_1 w_{\pi}^a + \left[({w_{\pi}^a})^2 + ({w_{\pi}^f})^2\right] (2g_1^2 + h_1 +h_2 - h_3)\phi_N + 2w_{\pi}^a w_{\pi}^f (2g_1^2 + h_2 - h_3)\phi_3 \right.\nonumber\\
 &\left.- \sqrt{2}\left[({w_{\pi}^a})^2 + ({w_{\pi}^f})^2\right] h_1 \phi_S \right\}\;, \\
 {\mathds{B}_{2}^{\eta}}_{23} &= 0 \;,\\
 {\mathds{B}_{2}^{\eta}}_{31} &= 0 \;,\\
 {\mathds{B}_{2}^{\eta}}_{32} &= 0 \;,\\
 {\mathds{B}_{2}^{\eta}}_{33} &=
 \frac{\sqrt{2}}{6}{\mathds{O}_{S}}^{a_0}_{NS0}\left\{-2g_1 w_{f_{1S}} + (w_{f_{1S}})^2 \left[h_1 \phi_N +  \frac{1}{\sqrt{2}}(4g_1^2 + h_1 + 2h_2 - 2h_3)\phi_S\right] \right\}\nonumber\\
&+\frac12{\mathds{O}_{S}}^{a_0}_{2} h_1 \phi_3 w_{f_{1S}}^2\nonumber\\
&+\frac{1}{6}{\mathds{O}_{S}}^{a_0}_{NS8}\left\{4g_1 w_{f_{1S}} + (w_{f_{1S}})^2 \left[h_1 \phi_N - \sqrt{2}(4g_1^2 + h_1 + 2h_2 - 2h_3)\phi_S\right] \right\}\;,\\
 {\mathds{C}^{\eta}} &= g_1\left(\begin{array}{@{}ccc@{}} 
 {\mathds{O}_{S}}_{21} w_{\eta}^f + {\mathds{O}_{S}}_{22} w_{\eta}^a & {\mathds{O}_{S}}_{21} w_{\eta}^a + {\mathds{O}_{S}}_{22} w_{\eta}^f & 0 \\ 
 {\mathds{O}_{S}}_{21} w_{\pi}^f + {\mathds{O}_{S}}_{22} w_{\pi}^a & {\mathds{O}_{S}}_{21} w_{\pi}^a + {\mathds{O}_{S}}_{22} w_{\pi}^f & 0 \\
 0 & 0 & \sqrt{2}{\mathds{O}_{S}}_{23} w_{f_{1S}}\\
 \end{array}\right)\;,\\
 {\mathbf{D}_1^{\eta}} &= Z_{a_0^{\pm}} Z_{\pi^{\pm}}
 \left(\begin{array}{@{}c@{}}
   (c_1\phi_S^2 - \lambda_2)\phi_N \\ 
   -(c_1\phi_S^2 - \lambda_2)\phi_3 \\
   \frac12 c_1\phi_S(\phi_N^2 - \phi_3^2) \\
 \end{array}\right)\;,\\
  {\mathbf{D}_2^{\eta}} &= Z_{a_0^{\pm}} Z_{\pi^{\pm}}
 \left(\begin{array}{@{}c@{}}
   -g_1w_{\eta}^f + w_{a_1^{\pm}}\left[w_{\eta}^f(2g_1^2 + h_2 - h_3)\phi_N + w_{\eta}^a(g_1^2 - h_3)\phi_3 - g_1\right] \\ 
   -g_1w_{\pi}^f + w_{a_1^{\pm}}\left[w_{\pi}^f(2g_1^2 + h_2 - h_3)\phi_N + w_{\pi}^a(g_1^2 - h_3)\phi_3 \right] \\ 
   0 \\
 \end{array}\right)\;,\\
  {\mathbf{F}_1^{\eta}} &= Z_{a_0^{\pm}} Z_{\pi^{\pm}}
 \left(\begin{array}{@{}c@{}}
   g_1w_{\eta}^f + \bar{w}_{\rho^{+}}\left[w_{\eta}^f(2g_1^2 + h_2 - h_3)\phi_3 + w_{\eta}^a(g_1^2 - h_3)\phi_N\right] \\  \\ 
   g_1w_{\pi}^f + \bar{w}_{\rho^{+}}\left[w_{\pi}^f(2g_1^2 + h_2 - h_3)\phi_3 + w_{\pi}^a(g_1^2 - h_3)\phi_N\right] \\  \\ 
   0 \\
 \end{array}\right)\\
   {\mathbf{F}_2^{\eta}} &= Z_{a_0^{\pm}} Z_{\pi^{\pm}}
 \left(\begin{array}{@{}c@{}}
   w_{a_1^{\pm}}\left[g_1 + \bar{w}_{\rho^{+}}(g_1^2 - h_3)\phi_3\right] \\ 
   \bar{w}_{\rho^{+}}\left[g_1 - w_{a_1^{\pm}}(g_1^2 - h_3)\phi_N\right] \\ 
   0 \\
 \end{array}\right)\;.
\end{align}
\end{subequations}
Then we apply for $\mathbf{x}$ the transformation in \eref{Eq:pseudoN3S_transf}  to obtain
\begin{align}
\mathcal{L}_{a_0 \pi \eta/\eta^{\prime}} &= a_0^{0} \left[{\mathbf{y}^{\text{ph}}}^T {\tilde{\mathds{B}}_{1}^{\eta}} \mathbf{y}^{\text{ph}} + (\partial_{\mu}\mathbf{y}^{\text{ph}})^T {\tilde{\mathds{B}}_{2}^{\eta}} \partial^{\mu}\mathbf{y}^{\text{ph}} \right] + (\partial_{\mu} a_0^{0})\left[ (\partial^{\mu}\mathbf{y}^{\text{ph}})^T \tilde{\mathds{C}}^{\eta} \mathbf{y}^{\text{ph}} \right]\nonumber\\
&+ a_0^{+} \left[\pi^- \left.\tilde{\mathbf{D}}_{1}^{\eta}\right.^T \mathbf{y}^{\text{ph}} + (\partial_{\mu}\pi^-) \left.\tilde{\mathbf{D}}_{2}^{\eta}\right.^T \partial^{\mu}\mathbf{y}^{\text{ph}} \right] + (\partial_{\mu}a_0^{+})\left[ \pi^- \left.\tilde{\mathbf{F}}_{1}^{\eta}\right.^T \partial^{\mu}\mathbf{y}^{\text{ph}} + (\partial^{\mu} \pi^-) \left.\tilde{\mathbf{F}}_{2}^{\eta}\right.^T \mathbf{y}^{\text{ph}} \right] + \text{h.c.}\;,	
\end{align} 
where
\begin{subequations}
\begin{align}
  \tilde{\mathds{B}}_{1}^{\eta} &\equiv \mathds{O}_{P}^T \mathds{B}_{1}^{\eta} \mathds{O}_{P}\;,\quad   \tilde{\mathds{B}}_{2}^{\eta}\equiv \mathds{O}_{P}^T \mathds{B}_{2}^{\eta} \mathds{O}_{P}\;,\quad \tilde{\mathds{C}}^{\eta} \equiv \mathds{O}_{P}^T {\mathds{C}}^{\eta}\mathds{O}_{P}\;, \\
  \left.\tilde{\mathbf{D}}_{1}^{\eta}\right.^T &\equiv {\mathbf{D}_{1}^{\eta}}^T \mathds{O}_{P}\;,\quad \left.\tilde{\mathbf{D}}_{2}^{\eta}\right.^T \equiv {\mathbf{D}_{2}^{\eta}}^T \mathds{O}_{P}\;, \quad
\left.\tilde{\mathbf{F}}_{1}^{\eta}\right.^T \equiv {\mathbf{F}_{1}^{\eta}}^T \mathds{O}_{P}\;,\quad \left.\tilde{\mathbf{F}}_{2}^{\eta}\right.^T \equiv {\mathbf{F}_{2}^{\eta}}^T \mathds{O}_{P}\;,  
\end{align}
\end{subequations}
which finally leads to
\begin{align}
  \mathcal{L}_{a_0 \pi \eta/\eta^{\prime}} &= a_0^{0} \left[ (\left.\tilde{\mathds{B}}_{1}^{\eta}\right._{12} + \left.\tilde{\mathds{B}}_{1}^{\eta}\right._{21}) \pi^0\eta +
    (\left.\tilde{\mathds{B}}_{2}^{\eta}\right._{12} + \left.\tilde{\mathds{B}}_{2}^{\eta}\right._{21}) (\partial_{\mu}\pi^0)(\partial^{\mu}\eta) \right] + (\partial_{\mu} a_0^{0}) \left[{\tilde{\mathds{C}}^{\eta}}_{21}\pi^0\partial^{\mu}\eta + {\tilde{\mathds{C}}^{\eta}}_{12}(\partial^{\mu}\pi^0) \eta \right]\nonumber	 \\
    &+ a_0^{0} \left[(\left.\tilde{\mathds{B}}_{1}^{\eta}\right._{13} + \left.\tilde{\mathds{B}}_{1}^{\eta}\right._{31}) \pi^0\eta^{\prime} +
    (\left.\tilde{\mathds{B}}_{2}^{\eta}\right._{13} + \left.\tilde{\mathds{B}}_{2}^{\eta}\right._{31}) (\partial_{\mu}\pi^0)(\partial^{\mu}\eta^{\prime}) \right]\nonumber	\\
  &+ a_0^{+} \left[ \left(\left.\tilde{\mathbf{D}}_{1}^{\eta}\right.^T\right)_{2} \pi^- \eta + \left(\left.\tilde{\mathbf{D}}_{2}^{\eta}\right.^T\right)_{2} (\partial_{\mu}\pi^-) (\partial^{\mu}\eta) \right] + (\partial_{\mu}a_0^{+})\left[ \left(\left.\tilde{\mathbf{F}}_{1}^{\eta}\right.^T\right)_{2} \pi^-\partial^{\mu}\eta +  \left(\left.\tilde{\mathbf{F}}_{2}^{\eta}\right.^T\right)_{2} (\partial^{\mu} \pi^-)\eta \right] \nonumber	\\
  &+ a_0^{+} \left(\left.\tilde{\mathbf{D}}_{1}^{\eta}\right.^T\right)_{3} \pi^-\eta^{\prime} + \text{h.c.}\;.
\end{align} 
The tree-level decay widths for the neutral and charged $a_0$ are given by
\begin{subequations}
\begin{align}
  \Gamma_{a_0^{0}\to \pi^{0}\eta} &=  \frac{k_{a_0^{0}\to \pi^{0}\eta}}{8\pi M_{a_0^{0}}^2}  \bigg| \left.\tilde{\mathds{B}}_{1}^{\eta}\right._{12} + \left.\tilde{\mathds{B}}_{1}^{\eta}\right._{21} + \frac12 \left({\tilde{\mathds{C}}^{\eta}}_{12} + {\tilde{\mathds{C}}^{\eta}}_{21} - \left.\tilde{\mathds{B}}_{2}^{\eta}\right._{12} - \left.\tilde{\mathds{B}}_{2}^{\eta}\right._{21}\right) \left(M_{a_0^{0}}^2 - M_{\eta}^2 - M_{\pi^{0}}^2\right) \nonumber \\
  &+ {\tilde{\mathds{C}}^{\eta}}_{21} M_{\eta}^2 + {\tilde{\mathds{C}}^{\eta}}_{12} M_{\pi^{0}}^2 \bigg|^2 \;,\\
\Gamma_{a_0^{0}\to \pi^{0}\eta^{\prime}} &=  \frac{k_{a_0^{0}\to \pi^{0}\eta^{\prime}}}{8\pi M_{a_0^{0}}^2} \bigg| \left.\tilde{\mathds{B}}_{1}^{\eta}\right._{13} + \left.\tilde{\mathds{B}}_{1}^{\eta}\right._{31} + \frac12 \left({\tilde{\mathds{C}}^{\eta}}_{13} + {\tilde{\mathds{C}}^{\eta}}_{31}- \left.\tilde{\mathds{B}}_{2}^{\eta}\right._{13} - \left.\tilde{\mathds{B}}_{2}^{\eta}\right._{31}\right) \left(M_{a_0^{0}}^2 - M_{\eta^{\prime}}^2 - M_{\pi^{0}}^2\right) \nonumber \\
 &+ {\tilde{\mathds{C}}^{\eta}}_{31} M_{\eta^{\prime}}^2 + {\tilde{\mathds{C}}^{\eta}}_{13} M_{\pi^{0}}^2 \bigg|^2 \;,\\
\Gamma_{a_0^{-}\to \pi^{-}\eta} &=  \frac{k_{a_0^{-}\to \pi^{-}\eta}}{8\pi M_{a_0^{\pm}}^2}  \bigg| \left(\left.\tilde{\mathbf{D}}_{1}^{\eta}\right.^T\right)_{2} + \frac12 \left[\left(\left.\tilde{\mathbf{F}}_{1}^{\eta}\right.^T\right)_{2} + \left(\left.\tilde{\mathbf{F}}_{2}^{\eta}\right.^T\right)_{2} - \left(\left.\tilde{\mathbf{D}}_{2}^{\eta}\right.^T\right)_{2}\right] \left(M_{a_0^{\pm}}^2 - M_{\eta}^2 - M_{\pi^{\pm}}^2\right) \nonumber \\
&+ \left(\left.\tilde{\mathbf{F}}_{1}^{\eta}\right.^T\right)_{2} M_{\eta}^2 + \left(\left.\tilde{\mathbf{F}}_{2}^{\eta}\right.^T\right)_{2} M_{\pi^{\pm}}^2 \bigg|^2\;, \\
\Gamma_{a_0^{-}\to \pi^{-}\eta^{\prime}} &=  \frac{k_{a_0^{-}\to \pi^{-}\eta^{\prime}}}{8\pi M_{a_0^{\pm}}^2} \bigg| \left(\left.\tilde{\mathbf{D}}_{1}^{\eta}\right.^T\right)_{3}  + \frac12 \left[\left(\left.\tilde{\mathbf{F}}_{1}^{\eta}\right.^T\right)_{3} + \left(\left.\tilde{\mathbf{F}}_{2}^{\eta}\right.^T\right)_{3} - \left(\left.\tilde{\mathbf{D}}_{2}^{\eta}\right.^T\right)_{3}\right] \left(M_{a_0^{\pm}}^2 - M_{\eta^{\prime}}^2 - M_{\pi^{\pm}}^2\right) \nonumber \\
&+\left(\left.\tilde{\mathbf{F}}_{1}^{\eta}\right.^T\right)_{3} M_{\eta^{\prime}}^2 + \left(\left.\tilde{\mathbf{F}}_{2}^{\eta}\right.^T\right)_{3} M_{\pi^{\pm}}^2 \bigg|^2 \;,
\end{align}
\end{subequations}
where $k_{A\to BC}$ is defined in \eref{Eq:decay_momABC}.



\subsubsection*{$f_0^{L/H}\to K K$}
\label{SSSec:f0KK}
The relevant part of the Lagrangian reads
\begin{align}
\mathcal{L}_{f_0 K K} &= f_0^{L/H} \left[ B^{L/H}_1  K^0 \bar{K^0} + B^{L/H}_2 (\partial_{\mu}K^0)\partial^{\mu}\bar{K^0} + B^{L/H}_3 K^+ K^- + B^{L/H}_4 (\partial_{\mu}K^+) \partial^{\mu}K^- \right]\nonumber\\
&+ (\partial_{\mu}f_0^{L/H})\left[ C^{L/H}_1 K^0\partial^{\mu}\bar{K^0} + C^{L/H}_2 (\partial^{\mu}K^0) \bar{K^0} + C^{L/H}_3 K^+\partial^{\mu}K^- + C^{L/H}_4 (\partial^{\mu}K^+) K^-\right]\;,	
\end{align}
where we introduced
\begin{subequations}
\begin{align}
 {\mathds{O}_{S}}^{L}_{2} &= {\mathds{O}_{S}}_{12}\;,\quad &{\mathds{O}_{S}}^{H}_{2} &= {\mathds{O}_{S}}_{32}\;,\\
  {\mathds{O}_{S}}^{L}_{NS0} &= \sqrt{2}{\mathds{O}_{S}}_{11}+{\mathds{O}_{S}}_{13}\;,\quad  &{\mathds{O}_{S}}^{H}_{NS0} &= \sqrt{2}{\mathds{O}_{S}}_{31}+{\mathds{O}_{S}}_{33}\;,\\
   {\mathds{O}_{S}}^{L}_{NS8} &= {\mathds{O}_{S}}_{11}-\sqrt{2}{\mathds{O}_{S}}_{13}\;,\quad  &{\mathds{O}_{S}}^{H}_{NS8} &= {\mathds{O}_{S}}_{31}-\sqrt{2}{\mathds{O}_{S}}_{33}\;,
\end{align}
\end{subequations}
\begin{subequations}
\begin{align}
  B^{L/H}_1 &= \frac{1}{2} Z_{K^0}^2 {\mathds{O}_{S}}^{L/H}_{2} \left[\lambda_2(2\phi_N -\sqrt2\phi_S) - 2\phi_3 (2\lambda_1 + \lambda_2) \right]\nonumber\\
  &-\frac{\sqrt{2}}{6}Z_{K^0}^2 {\mathds{O}_{S}}^{L/H}_{NS0}\left[ \phi_N(4\lambda_1+\lambda_2)-\phi_3\lambda_2+\sqrt{2}\phi_S(2\lambda_1+\lambda_2)\right]\nonumber\\
  &-\frac{1}{6}Z_{K^0}^2 {\mathds{O}_{S}}^{L/H}_{NS8}\left[ 4\phi_N(\lambda_1+\lambda_2)-4\phi_3\lambda_2-\sqrt{2}\phi_S(4\lambda_1+5\lambda_2)\right]\;,\\
  B^{L/H}_2 &= \frac{1}{2} Z_{K^0}^2 {\mathds{O}_{S}}^{L/H}_{2}w_{K_1^0} \left\{2g_1 - w_{K_1^0} \left[\phi_N(h_2 + g_1^2) - \phi_3(2h_1 + h_2 + g_1^2) + \sqrt{2}\phi_S (g_1^2 - h_3)\right]\right\} \nonumber\\
  &- \frac{\sqrt{2}}{6}Z_{K^0}^2 {\mathds{O}_{S}}^{L/H}_{NS0} w_{K_1^0} \left\{4g_1 - w_{K_1^0} \left[\phi_N(2h_1+ h_2 -h_3 + 2g_1^2) - \phi_3(h_2 - h_3 + 2g_1^2) \right.\right. \nonumber\\
  &+\left.\left. \sqrt{2}\phi_S (h_1 + h_2 - h_3 + 2g_1^2)\right]\right\} \nonumber \\
  &+ \frac{1}{6}Z_{K^0}^2 {\mathds{O}_{S}}^{L/H}_{NS8} w_{K_1^0} \left\{2g_1 + w_{K_1^0} \left[\phi_N(2h_1+ h_2 + 2h_3 - g_1^2) - \phi_3(h_2 + 2h_3 - g_1^2) \right.\right. \nonumber\\
  &-\left.\left. \sqrt{2}\phi_S (2h_1 + 2h_2 + h_3 + g_1^2)\right]\right\} \;, \\
  B^{L/H}_3 &= -\frac{1}{2} Z_{K^\pm}^2 {\mathds{O}_{S}}^{L/H}_{2} \left[\lambda_2(2\phi_N -\sqrt2\phi_S) + 2\phi_3 (2\lambda_1 + \lambda_2) \right]\nonumber\\
  &-\frac{\sqrt{2}}{6}Z_{K^\pm}^2 {\mathds{O}_{S}}^{L/H}_{NS0}\left[ \phi_N(4\lambda_1+\lambda_2)+\phi_3\lambda_2+\sqrt{2}\phi_S(2\lambda_1+\lambda_2)\right]\nonumber\\
  &-\frac{1}{6}Z_{K^\pm}^2 {\mathds{O}_{S}}^{L/H}_{NS8}\left[ 4\phi_N(\lambda_1+\lambda_2)+4\phi_3\lambda_2-\sqrt{2}\phi_S(4\lambda_1+5\lambda_2)\right]\;,\\
  B^{L/H}_4 &= -\frac{1}{2} Z_{K^\pm}^2 {\mathds{O}_{S}}^{L/H}_{2}w_{K_1^\pm} \left\{2g_1 - w_{K_1^\pm} \left[\phi_N(h_2 + g_1^2) + \phi_3(2h_1 + h_2 + g_1^2) + \sqrt{2}\phi_S (g_1^2 - h_3)\right]\right\}\nonumber \\
  &- \frac{\sqrt{2}}{6}Z_{K^\pm}^2 {\mathds{O}_{S}}^{L/H}_{NS0} w_{K_1^\pm} \left\{4g_1 - w_{K_1^\pm} \left[\phi_N(2h_1+ h_2 -h_3 + 2g_1^2) + \phi_3(h_2 - h_3 + 2g_1^2) \right.\right. \nonumber\\
  &+\left.\left. \sqrt{2}\phi_S (h_1 + h_2 - h_3 + 2g_1^2)\right]\right\} \nonumber\\
  &+ \frac{1}{6}Z_{K^\pm}^2 {\mathds{O}_{S}}^{L/H}_{NS8} w_{K_1^\pm} \left\{2g_1 + w_{K_1^\pm} \left[\phi_N(2h_1+ h_2 + 2h_3 - g_1^2) + \phi_3(h_2 + 2h_3 - g_1^2) \right.\right. \nonumber\\
  &-\left.\left. \sqrt{2}\phi_S (2h_1 + 2h_2 + h_3 + g_1^2)\right]\right\} \;,\\
  C^{L/H}_1 &= -\frac{g_1}{2} Z_{K^0}^2 {\mathds{O}_{S}}^{L/H}_{2} w_{K_1^0} + g_1\frac{\sqrt{2}}{3} Z_{K^0}^2 {\mathds{O}_{S}}^{L/H}_{NS0} w_{K_1^0} - \frac{g_1}{6} Z_{K^0}^2 {\mathds{O}_{S}}^{L/H}_{NS8} w_{K_1^0}
  \;, \quad C^{L/H}_2 = C^{L/H}_1\;, \\
  C^{L/H}_3 &= \frac{g_1}{2} Z_{K^\pm}^2 {\mathds{O}_{S}}^{L/H}_{2} w_{K_1^\pm} + g_1\frac{\sqrt{2}}{3} Z_{K^\pm}^2 {\mathds{O}_{S}}^{L/H}_{NS0} w_{K_1^\pm} - \frac{g_1}{6} Z_{K^\pm}^2 {\mathds{O}_{S}}^{L/H}_{NS8} w_{K_1^\pm}\; , \quad C^{L/H}_4 = C^{L/H}_3 \;.
\end{align}
\end{subequations}
The tree-level decay widths for $f_0^{L/H}$ are given by
\begin{align}
\Gamma_{f_0^{L/H}\to K^{0,+}K^{\bar{0},-}} &= \Gamma_{f_0^{L/H}\to K^{0}\bar K^{0}} + \Gamma_{f_0^{L/H}\to K^{+}K^{-}} \nonumber\\
&= \frac{1}{8\pi M_{f_0^{L/H}}^2} \Bigg\lbrace k_{f_0^{L/H}\to K^{0}\bar K^{0}} \left| B^{L/H}_1 + \frac12 \left(2C^{L/H}_1 - B^{L/H}_2\right) M_{f_0^{L/H}}^2 + B^{L/H}_2 M_{K^{0}}^2 \right|^2\nonumber \\
& + k_{f_0^{L/H}\to K^{+}K^{-}} \left| B^{L/H}_3 + \frac12 \left(2C^{L/H}_3 - B^{L/H}_4\right) M_{f_0^{L/H}}^2 + B^{L/H}_4 M_{K^{\pm}}^2 \right|^2 \Bigg\rbrace \;,
\end{align}
where $k_{A\to BC}$ is defined in \eref{Eq:decay_momABC}.



\subsubsection*{$f_0^{L/H}\to \pi\pi$}
\label{SSSec:f0pipi}
We start from the following part of the Lagrangian,
\begin{align}
\mathcal{L}_{f_0 \pi \pi } &= f_0^{L/H} \left[\mathbf{x}^T {\mathds{B}_{1}^{L/H}} \mathbf{x} + (\partial_{\mu}\mathbf{x})^T {\mathds{B}_{2}^{L/H}} \partial^{\mu}\mathbf{x} \right] + (\partial_{\mu}f_0^{L/H})\left[ (\partial^{\mu}\mathbf{x})^T {\mathds{C}^{L/H}} \mathbf{x} \right]\nonumber\\
&+ f_0^{L/H}\left[\pi^+ D_{1}^{L/H} \pi^- + (\partial_{\mu}\pi^+) D_{2}^{L/H} \partial^{\mu}\pi^- \right] + (\partial_{\mu}f_0^{L/H})\left[ \pi^+ F_{1}^{L/H} \partial^{\mu}\pi^{-} + (\partial^{\mu} \pi^+) F_{2}^{L/H} \pi^{-} \right]\;,	
\end{align}
where $\mathbf{x}^T = (\tilde\eta_N,\tilde\pi^0, \tilde\eta_S)$ and the coefficient matrices and vectors are,

\begin{subequations}
\begin{eqnarray}
{\mathds{B}_{1}^{L/H}}_{11} &=& -\frac{1}{6}{\mathds{O}_{S}}^{L/H}_{NS0}\left[2c_1\phi_N\phi_S (\phi_N + \sqrt{2}\phi_S) +\sqrt{2}\phi_N(2\lambda_1+\lambda_2) + 2\lambda_1\phi_S\right]\nonumber\\
&&-\frac{1}{2}{\mathds{O}_{S}}^{L/H}_{2}(2\lambda_1 + \lambda_2)\phi_3\nonumber\\
&&+\frac{1}{6}{\mathds{O}_{S}}^{L/H}_{NS8}\left[2c_1\phi_N\phi_S (\sqrt{2}\phi_N - \phi_S) -\phi_N(2\lambda_1+\lambda_2) + 2\sqrt{2}\lambda_1\phi_S\right]\;,\\
{\mathds{B}_{1}^{L/H}}_{12} &=& 
\frac{1}{6}{\mathds{O}_{S}}^{L/H}_{NS0}\left[c_1\phi_S (2\phi_N + \sqrt{2}\phi_S) - \sqrt{2}\lambda_2\right]\phi_3\nonumber\\
&&+\frac{1}{2}{\mathds{O}_{S}}^{L/H}_{2}(c_1\phi_S^2 - \lambda_2)\phi_N\nonumber\\
&&-\frac{1}{6}{\mathds{O}_{S}}^{L/H}_{NS8}\left[c_1\phi_S (2\sqrt{2}\phi_N - \phi_S) + \lambda_2\right]\phi_3\;,\\
{\mathds{B}_{1}^{L/H}}_{13} &=& 
-\frac{1}{12}{\mathds{O}_{S}}^{L/H}_{NS0}c_1\left[\phi_N^2 (\phi_N + 3\sqrt{2}\phi_S) - \phi_3^2(\phi_N + \sqrt{2}\phi_S)\right]\nonumber\\
&&+\frac{1}{2}{\mathds{O}_{S}}^{L/H}_{2}c_1\phi_N\phi_S\phi_3\nonumber\\
&&+\frac{1}{12}{\mathds{O}_{S}}^{L/H}_{NS8}c_1\left[\phi_N^2 (\sqrt{2}\phi_N - 3\phi_S) - \phi_3^2(\sqrt{2}\phi_N - \phi_S)\right]\;,\\
{\mathds{B}_{1}^{L/H}}_{21} &=& {\mathds{B}_{1}^{L/H}}_{12}\;,\\
{\mathds{B}_{1}^{L/H}}_{22} &=& 
-\frac{1}{6}{\mathds{O}_{S}}^{L/H}_{NS0}\left[2c_1 \phi_3^2 \phi_S +\sqrt{2}\phi_N(2\lambda_1+\lambda_2) + 2\lambda_1\phi_S\right]\nonumber\\
&&-\frac{1}{2}{\mathds{O}_{S}}^{L/H}_{2}(2\lambda_1 + \lambda_2 + 2c_1\phi_S^2)\phi_3\nonumber\\
&&+\frac{1}{6}{\mathds{O}_{S}}^{L/H}_{NS8}\left[2\sqrt{2} c_1\phi_3^2\phi_S -\phi_N(2\lambda_1+\lambda_2) + 2\sqrt{2}\lambda_1\phi_S\right]\;,\\
{\mathds{B}_{1}^{L/H}}_{23} &=& 
\frac{1}{12}{\mathds{O}_{S}}^{L/H}_{NS0}c_1 \phi_3\left[\phi_N^2 - \phi_3^2 +2\sqrt{2}\phi_N \phi_S\right]\nonumber\\
&&+\frac{1}{4}{\mathds{O}_{S}}^{L/H}_{2} c_1\phi_S(\phi_N^2 - 3\phi_3^2)\nonumber\\
&&-\frac{1}{12}{\mathds{O}_{S}}^{L/H}_{NS8}c_1 \phi_3\left[\sqrt{2}(\phi_N^2 - \phi_3^2) -2\phi_N \phi_S\right]\;,\\
{\mathds{B}_{1}^{L/H}}_{31} &=& {\mathds{B}_{1}^{L/H}}_{13}\;,\\
{\mathds{B}_{1}^{L/H}}_{32} &=& {\mathds{B}_{1}^{L/H}}_{23}\;,\\
{\mathds{B}_{1}^{L/H}}_{33} &=& 
-\frac{1}{6}{\mathds{O}_{S}}^{L/H}_{NS0}\left[\sqrt{2}c_1 \phi_N(\phi_N^2-\phi_3^2) +  2\sqrt{2} \lambda_1 \phi_N  + 2(\lambda_1 + \lambda_2)\phi_S\right]\nonumber\\
&&+\frac{1}{2}{\mathds{O}_{S}}^{L/H}_{2}(-2\lambda_1 + c_1(\phi_N^2 - \phi_3^2))\phi_3\nonumber \\
&&+\frac{1}{6}{\mathds{O}_{S}}^{L/H}_{NS8}\left[-c_1 \phi_N(\phi_N^2-\phi_3^2) -  2 \lambda_1 \phi_N  + 2 \sqrt{2}(\lambda_1 + \lambda_2)\phi_S\right]\;,\\
{\mathds{B}_{2}^{L/H}}_{11} &=& 
 \frac{\sqrt{2}}{6}{\mathds{O}_{S}}^{L/H}_{NS0}\left\{-2g_1 w_{\eta}^f + \left[({w_{\eta}^a})^2 + ({w_{\eta}^f})^2\right]e (2g_1^2 + h_1 +h_2 - h_3)\phi_N + 2w_{\eta}^a w_{\eta}^f (2g_1^2 + h_2 - h_3)\phi_3 \right.\nonumber\\
 &&\left.+ \frac{1}{\sqrt{2}}\left[({w_{\eta}^a})^2 + ({w_{\eta}^f})^2\right] h_1 \phi_S \right\}\nonumber\\
&+&{\mathds{O}_{S}}^{L/H}_{2}\left\{ -g_1 w_{\eta}^a + w_{\eta}^a w_{\eta}^f (2g_1^2 + h_2 - h_3) \phi_N + \frac12 \left[({w_{\eta}^a})^2 + ({w_{\eta}^f})^2\right] (2g_1^2 + h_1 +h_2 - h_3)\phi_3\right\}\nonumber\\
&+&\frac{1}{6}{\mathds{O}_{S}}^{L/H}_{NS8}\left[-2g_1 w_{\eta}^f + (({w_{\eta}^a})^2 + ({w_{\eta}^f})^2) (2g_1^2 + h_1 +h_2 - h_3)\phi_N + 2w_{\eta}^a w_{\eta}^f (2g_1^2 + h_2 - h_3)\phi_3 \right.\nonumber\\
&&\left.- \sqrt{2}(({w_{\eta}^a})^2 + ({w_{\eta}^f})^2) h_1 \phi_S \right]\;,\\
 {\mathds{B}_{2}^{L/H}}_{12} &=& 
 \frac{\sqrt{2}}{12}{\mathds{O}_{S}}^{L/H}_{NS0}\left\{-2g_1 (w_{\eta}^a + w_{\pi}^f) + 2(w_{\pi}^a w_{\eta}^a + w_{\pi}^f w_{\eta}^f ) \left[
 \vphantom{\frac{1}{\sqrt{2}}}(2g_1^2 + h_2 - h_3)(\phi_N + \phi_3)\right.\right. \\ 
 && \left.\left.+ h_1 (\phi_N + \frac{1}{\sqrt{2}}\phi_S) \right]\right\}\nonumber\\
&+&{\mathds{O}_{S}}^{L/H}_{2}\left[-\frac{g_1}{2} (w_{\pi}^a + w_{\eta}^f) + \frac12(w_{\eta}^a w_{\pi}^f + w_{\eta}^f w_{\pi}^a) (2g_1^2 + h_2 - h_3) \phi_N \right.\nonumber \\ 
 &+& \left.\frac12(w_{\eta}^a w_{\pi}^a + w_{\eta}^f w_{\pi}^f) (2g_1^2 + h_1 +h_2 - h_3)\phi_3\right] \nonumber\\
&+&\frac{1}{12}{\mathds{O}_{S}}^{L/H}_{NS8}\left\{ -2g_1 (w_{\eta}^a + w_{\pi}^f) + 2(w_{\pi}^a w_{\eta}^a + w_{\pi}^f w_{\eta}^f ) \left[  \vphantom{\frac{1}{\sqrt{2}}} (2g_1^2 + h_2 - h_3)(\phi_N + \phi_3)\right.\right. \nonumber\\ 
 && \left.\left.+ h_1 (\phi_N - \sqrt{2}\phi_S) \right]\right\}\;,\\
 {\mathds{B}_{2}^{L/H}}_{13} &=& 0\;, \\
 {\mathds{B}_{2}^{L/H}}_{21} &=& {\mathds{B}_{2}^{L/H}}_{12}\;,\\
 {\mathds{B}_{2}^{L/H}}_{22} &=& \frac{\sqrt{2}}{6} {\mathds{O}_{S}}^{L/H}_{NS0}\left\{\vphantom{\frac{1}{\sqrt{2}}}-2g_1 w_{\pi}^a + \left[({w_{\pi}^a})^2 + ({w_{\pi}^f})^2\right] (2g_1^2 + h_1 +h_2 - h_3)\phi_N + 2w_{\pi}^a w_{\pi}^f (2g_1^2 + h_2 - h_3)\phi_3 \right.\nonumber\\
 &&\left.+ \frac{1}{\sqrt{2}}\left[({w_{\pi}^a})^2 + ({w_{\pi}^f})^2\right] h_1 \phi_S \right\}\nonumber\\
&+& {\mathds{O}_{S}}^{L/H}_{2}\left[-g_1 w_{\pi}^f + w_{\pi}^a w_{\pi}^f (2g_1^2 + h_2 - h_3) \phi_N + \frac12 \left[({w_{\pi}^a})^2 + ({w_{\pi}^f})^2\right] (2g_1^2 + h_1 +h_2 - h_3)\phi_3\right]\nonumber \\
&+&\frac{1}{6}{\mathds{O}_{S}}^{L/H}_{NS8}\left\{ -2g_1 w_{\pi}^a + \left[({w_{\pi}^a})^2 + ({w_{\pi}^f})^2\right] (2g_1^2 + h_1 +h_2 - h_3)\phi_N + 2w_{\pi}^a w_{\pi}^f (2g_1^2 + h_2 - h_3)\phi_3 \right.\nonumber\\
 &&\left.- \sqrt{2}\left[({w_{\pi}^a})^2 + ({w_{\pi}^f})^2\right] h_1 \phi_S \right\}\;,\\
 {\mathds{B}_{2}^{L/H}}_{23} &=& 0 \;,\\
 {\mathds{B}_{2}^{L/H}}_{31} &=& 0 \;,\\
 {\mathds{B}_{2}^{L/H}}_{32} &=& 0 \;,\\
 {\mathds{B}_{2}^{L/H}}_{33} &=&
 \frac{\sqrt{2}}{6}{\mathds{O}_{S}}^{L/H}_{NS0}\left\{-2g_1 w_{f_{1S}} + (w_{f_{1S}})^2 \left[h_1 \phi_N +  \frac{1}{\sqrt{2}}(4g_1^2 + h_1 + 2h_2 - 2h_3)\phi_S\right] \right\}\nonumber\\
&+&\frac12{\mathds{O}_{S}}^{L/H}_{2} h_1 \phi_3 w_{f_{1S}}^2\nonumber\\
&+&\frac{1}{6}{\mathds{O}_{S}}^{L/H}_{NS8}\left\{4g_1 w_{f_{1S}} + (w_{f_{1S}})^2 \left[h_1 \phi_N - \sqrt{2}(4g_1^2 + h_1 + 2h_2 - 2h_3)\phi_S\right] \right\}\;,\\
 {\mathds{C}^{L/H}} &=& 
 \begin{cases}
 g_1\left(\begin{array}{@{}ccc@{}} 
 {\mathds{O}_{S}}_{11} w_{\eta}^f + {\mathds{O}_{S}}_{12} w_{\eta}^a & {\mathds{O}_{S}}_{11} w_{\eta}^a + {\mathds{O}_{S}}_{12} w_{\eta}^f & 0 \\ 
 {\mathds{O}_{S}}_{11} w_{\pi}^f + {\mathds{O}_{S}}_{12} w_{\pi}^a & {\mathds{O}_{S}}_{11} w_{\pi}^a + {\mathds{O}_{S}}_{12} w_{\pi}^f & 0 \\
 0 & 0 & \sqrt{2}{\mathds{O}_{S}}_{13} w_{f_{1S}}\\
 \end{array}\right),\; \text{for}\; L\\
 g_1\left(\begin{array}{@{}ccc@{}} 
 {\mathds{O}_{S}}_{31} w_{\eta}^f + {\mathds{O}_{S}}_{32} w_{\eta}^a & {\mathds{O}_{S}}_{31} w_{\eta}^a + {\mathds{O}_{S}}_{32} w_{\eta}^f & 0 \\ 
 {\mathds{O}_{S}}_{31} w_{\pi}^f + {\mathds{O}_{S}}_{32} w_{\pi}^a & {\mathds{O}_{S}}_{31} w_{\pi}^a + {\mathds{O}_{S}}_{32} w_{\pi}^f & 0 \\
 0 & 0 & \sqrt{2}{\mathds{O}_{S}}_{33} w_{f_{1S}}\\
 \end{array}\right),\; \text{for}\; H
 \end{cases} \\
D_1^{L/H} &=& -\frac{\sqrt{2}}{3}Z_{\pi^{\pm}}^2{\mathds{O}_{S}}^{L/H}_{NS0}\left[\phi_N (2\lambda_1+\lambda_2) + \sqrt{2}\lambda_1\phi_S\right]\nonumber\\
&&-Z_{\pi^{\pm}}^2{\mathds{O}_{S}}^{L/H}_{2}(2\lambda_1 + 3\lambda_2)\phi_3\nonumber\\
&&-\frac{1}{3}Z_{\pi^{\pm}}^2{\mathds{O}_{S}}^{L/H}_{NS8}\left[\phi_N (2\lambda_1+\lambda_2) - 2\sqrt{2}\lambda_1\phi_S\right]\;,\\
D_2^{L/H} &=& \frac{\sqrt{2}}{3} Z_{\pi^{\pm}}^2{\mathds{O}_{S}}^{L/H}_{NS0}\left\{-2g_1 w_{a_1^{\pm}} + w_{a_1^{\pm}}^2\left[ (2g_1^2 + h_1 +h_2 - h_3)\phi_N + \frac{1}{\sqrt{2}} h_1 \phi_S \right]\right\} \nonumber\\
&+& Z_{\pi^{\pm}}^2{\mathds{O}_{S}}^{L/H}_{2}w_{a_1^{\pm}}^2 (h_1 +h_2 + h_3) \phi_3\nonumber\\
&+&\frac{1}{3}Z_{\pi^{\pm}}^2{\mathds{O}_{S}}^{L/H}_{NS8}\left\{-2g_1 w_{a_1^{\pm}} + w_{a_1^{\pm}}^2\left[ (2g_1^2 + h_1 +h_2 - h_3)\phi_N - \sqrt{2} h_1 \phi_S \right]\right\}\;,\\
F_1^{L/H} &=&  Z_{\pi^{\pm}}^2 g_1 w_{a_1^{\pm}} \begin{cases}
{\mathds{O}_{S}}_{11} ,\; \text{for}\; L\\
{\mathds{O}_{S}}_{13} ,\; \text{for}\; H
\end{cases} \\
F_2^{L/H} &=& F_1^{L/H}\;.
\end{eqnarray}
\end{subequations}
Then we apply for $\mathbf{x}$ the transformation in \eref{Eq:pseudoN3S_transf}  to obtain
\begin{align}
\mathcal{L}_{f_0 \pi \pi }  &= f_0^{L/H} \left[{\mathbf{y}^{\text{ph}}}^T {\tilde{\mathds{B}}_{1}^{L/H}} \mathbf{y}^{\text{ph}} + (\partial_{\mu}\mathbf{y}^{\text{ph}})^T {\tilde{\mathds{B}}_{2}^{L/H}} \partial^{\mu}\mathbf{y}^{\text{ph}} \right] + (\partial_{\mu} f_0^{L/H})\left[ (\partial^{\mu}\mathbf{y}^{\text{ph}})^T \tilde{\mathds{C}}^{L/H} \mathbf{y}^{\text{ph}} \right] \nonumber \\
&+ f_0^{L/H}\left[\pi^+ D_{1}^{L/H} \pi^- + (\partial_{\mu}\pi^+) D_{2}^{L/H} \partial^{\mu}\pi^- \right] + (\partial_{\mu}f_0^{L/H})\left[ \pi^+ F_{1}^{L/H} \partial^{\mu}\pi^{-} + (\partial^{\mu} \pi^+) F_{2}^{L/H} \pi^{-} \right]\;,	
\end{align} 
where
\begin{equation}
  \tilde{\mathds{B}}_{1}^{L/H} \equiv \mathds{O}_{P}^T \mathds{B}_{1}^{L/H} \mathds{O}_{P}\;,\quad   \tilde{\mathds{B}}_{2}^{L/H}\equiv \mathds{O}_{P}^T \mathds{B}_{2}^{L/H} \mathds{O}_{P}\;,\quad \tilde{\mathds{C}}^{L/H} \equiv \mathds{O}_{P}^T {\mathds{C}}^{L/H}\mathds{O}_{P}\;,
\end{equation}
which finally leads to
\begin{align}
  \mathcal{L}_{f_0 \pi \pi } &= f_0^{L/H} \left[ \left(\tilde{\mathds{B}}_{1}^{L/H}\right)_{11} \pi^0 \pi^0 +
    \left(\tilde{\mathds{B}}_{2}^{L/H}\right)_{11} (\partial_{\mu}\pi^{0})(\partial^{\mu}\pi^{0}) \right] + (\partial_{\mu} f_0^{L/H}) \left[{\tilde{\mathds{C}}^{L/H}}_{11}\pi^0\partial^{\mu}\pi^{0} \right] \nonumber\\
  &+ f_0^{L/H}\left[\pi^+ D_{1}^{L/H} \pi^- + (\partial_{\mu}\pi^+) D_{2}^{L/H} \partial^{\mu}\pi^- \right] + (\partial_{\mu}f_0^{L/H})\left[ \pi^+ F_{1}^{L/H} \partial^{\mu}\pi^{-} + (\partial^{\mu} \pi^+) F_{2}^{L/H} \pi^{-} \right]\;.
\end{align} 
The tree-level decay widths for $f_0^{L/H}$ are given by
\begin{align}
  \Gamma_{f_0^{L/H}\to \pi^{0,+}\pi^{0,-}} &=  \Gamma_{f_0^{L/H}\to \pi^{0} \pi^{0}} + \Gamma_{f_0^{L/H}\to \pi^{+}\pi^{-}} \nonumber\\
  &= \frac{1}{8\pi M_{f_0^{L/H}}^2}  \Bigg\lbrace 2 k_{f_0^{L/H}\to \pi^{0}\pi^{0}} \bigg| \left(\tilde{\mathds{B}}_{1}^{L/H}\right)_{11} + \frac12 \left({\tilde{\mathds{C}}^{L/H}}_{11} -  \left(\tilde{\mathds{B}}_{2}^{L/H}\right)_{11}\right) M_{f_0^{L/H}}^2 + \left(\tilde{\mathds{B}}_{2}^{L/H}\right)_{11} M_{\pi^0}^2 \bigg|^2\nonumber\\
  &+  k_{f_0^{L/H}\to \pi^{\pm}\pi^{\mp}} \bigg|  D_{1}^{L/H} + \frac12 \left(F_{1}^{L/H} + F_{2}^{L/H} - D_{2}^{L/H}\right) M_{f_0^{L/H}}^2 +  D_{2}^{L/H} M_{\pi^{\pm}}^2 \bigg|^2 \Bigg\rbrace \;,
\end{align}
where $k_{A\to BC}$ is defined in \eref{Eq:decay_momABC}.
\end{widetext}

\section{Detailed fit results and parameter values}
\label{App:fit_res}

\begin{widetext}
In this appendix, the best-fit results with and without the $\omega$ decay are presented (Table~\ref{Tab:fit_res}) for the DS and for the DVS-I (see \secref{Sec:phys_mass} for more details). 
For comparison, we have also listed the experimental values, which we have already discussed in detail in \secref{Sec:results}. Note that the decays of the scalar-isoscalar $f_0$'s are not fitted; the values in the tables are just calculated with the parameters obtained.
\begin{table}
  \caption{Detailed fit results. Second column contains the experimental values, third, fifth, seventh and ninth columns hold our fit results in the DS and in the DVS-I, and cases where the $\omega \to \pi \pi$ decay is fitted ($\omega$) and where it is not (no-$\omega$), while in the fourth, sixth, eighth, and tenth columns the $\chi^2$ values for the given quantities are listed. \label{Tab:fit_res}}
  \centering
\begin{tabular} [c]{|c||c|c|c|c|c|c|c|c|c|}
\hline 
Observable &  Exp. val. [MeV]  &  $\text{Fit}_{\text{DS},\omega}$ [MeV] & $\chi^2$ &         $\text{Fit}_{\text{DS},\text{no-}\omega}$ [MeV]  & $\chi^2$ &  $\text{Fit}_{\text{DVS-I},\omega}$ [MeV] & $\chi^2$ & $\text{Fit}_{\text{DVS-I},\text{no-}\omega}$ [MeV]  & $\chi^2$\\\hline             
$f_{\pi^+}$                                  &  $92.06 \pm 4.60$     &   $96.78$    & $1.1$  & $96.72$   & $1.0$ & $96.13$   & $0.8$  & $96.61$   & $1.0$ \\\hline  
$f_{K^+}$                                    &  $110.10 \pm 5.51$    &   $109.20$   & $0.0$  & $110.45$  & $0.0$ & $108.73$  & $0.1$  & $109.54$  & $0.0$ \\\hline	
$\bar M_{\pi}$                               &  $138.04 \pm 6.90$    &   $140.56$   & $0.1$  & $140.20$  & $0.1$ & $140.32$  & $0.1$  & $140.61$  & $0.1$ \\\hline
$\Delta M_{\pi}$                             &  $-4.59 \pm 0.92$     &   $-4.54$    & $0.0$  & $-4.56$   & $0.0$ & $-4.57$   & $0.0$  & $-4.55$   & $0.0$ \\\hline
$M_{\eta}$                                   &  $547.86 \pm 27.39$   &   $550.38$   & $0.0$  & $547.39$  & $0.0$ & $546.90$  & $0.0$  & $548.07$  & $0.0$ \\\hline	
$M_{\eta^{\prime}}$                          &  $957.78 \pm 47.89$   &   $949.69$   & $0.0$  & $952.44$  & $0.0$ & $957.45$  & $0.0$  & $958.13$  & $0.0$ \\\hline
$\bar M_{K}$                                 &  $495.64 \pm 24.78$   &   $476.73$   & $0.6$  & $482.47$  & $0.3$ & $484.04$  & $0.2$  & $480.23$  & $0.4$ \\\hline
$\Delta M_{K}$                               &  $3.93 \pm 0.79$      &   $3.89$     & $0.0$  & $3.93$    & $0.0$ & $3.92$    & $0.0$  & $3.90$    & $0.0$ \\\hline
$\bar M_{\rho}$                              &  $775.16 \pm 38.76$   &   $762.00$   & $0.1$  & $761.61$  & $0.1$ & $744.48$  & $0.6$  & $762.31$  & $0.1$ \\\hline
$\Delta M_{\rho}$                            &  $0.15 \pm 0.57$      &   $0.13$     & $0.0$  & $0.13$    & $0.0$ & $0.14$    & $0.0$  & $0.10$    & $0.0$ \\\hline
$M_{\omega}$                                 &  $782.66 \pm 39.13$   &   $762.13$   & $0.3$  & $761.86$  & $0.3$ & $755.64$  & $0.5$  & $763.22$  & $0.2$ \\\hline
$M_{\phi}$                                   &  $1019.46 \pm 50.97$  &   $979.66$   & $0.6$  & $986.41$  & $0.4$ & $998.68$  & $0.2$  & $982.42$  & $0.5$ \\\hline
$\bar M_{K^{\star}}$                         &  $895.50 \pm 44.78$   &   $878.64$   & $0.1$  & $882.52$  & $0.1$ & $882.97$  & $0.1$  & $880.57$  & $0.1$ \\\hline
$\Delta M_{K^{\star}}$                       &  $0.08 \pm 0.90$      &   $0.14$     & $0.0$  & $0.19$    & $0.0$ & $0.15$    & $0.0$  & $0.22$    & $0.0$ \\\hline
$\bar M_{a_{1}}$                             &  $1230.00 \pm 246.00$ &   $1109.54$  & $0.2$  & $1115.71$ & $0.2$ & $1050.84$ & $0.5$  & $1115.80$ & $0.2$ \\\hline 
$M_{f_{1}^L}$                                &  $1281.90 \pm 256.38$ &   $1246.50$  & $0.0$  & $1222.40$ & $0.1$ & $1334.34$ & $0.0$  & $1233.96$ & $0.0$ \\\hline  	
$M_{f_{1}^H}$                                &  $1426.30 \pm 285.26$ &   $1357.51$  & $0.1$  & $1367.72$ & $0.0$ & $1366.92$ & $0.0$  & $1363.18$ & $0.0$ \\\hline  	
$\bar M_{K_{1}}$                             &  $1253.00 \pm 250.60$ &   $1256.58$  & $0.0$  & $1260.84$ & $0.0$ & $1256.35$ & $0.0$  & $1260.05$ & $0.0$ \\\hline  	
$\bar M_{a_{0}}$                             &  $1474.00 \pm 294.80$ &   $1251.40$  & $0.5$  & $1140.38$ & $1.0$ & $1308.90$ & $0.3$  & $1187.00$ & $0.8$ \\\hline  	
$\bar M_{K_{0}^{\star}}$                     &  $1425.00 \pm 285.00$ &   $1321.76$  & $0.1$  & $1237.23$ & $0.4$ & $1411.77$ & $0.0$  & $1282.16$ & $0.2$ \\\hline  	
$M_{f_0^L}$                                  &  $1350.00 \pm 675.00$ &   $1229.39$  & $0.0$  & $1136.72$ & $0.1$ & $1295.92$ & $0.0$  & $1187.92$ & $0.0$ \\\hline  	
$M_{f_0^H}$                                  &  $1733.00 \pm 866.50$ &   $1515.87$  & $0.1$  & $1326.58$ & $0.2$ & $1499.63$ & $0.1$  & $1375.34$ & $0.1$ \\\hline 	
$\bar \Gamma_{\rho\rightarrow\pi\pi}$        &  $148.53 \pm 7.43$    &   $154.81$   & $0.7$  & $154.85$  & $0.7$ & $153.63$  & $0.5$  & $155.85$  & $1.0$ \\\hline  	
$\Delta \Gamma_{\rho\rightarrow\pi\pi}$      &  $-1.70 \pm 1.60$     &   $-1.90$    & $0.0$  & $-1.89$   & $0.0$ & $-1.87$   & $0.0$  & $-1.78$   & $0.0$ \\\hline  	
$\bar \Gamma_{\omega\rightarrow\pi\pi}$      &  $0.13 \pm 0.03$      &   $0.00$     & $25.0$ & $0.0$     & $-$   & $0.00$    & $25.0$ & $0.0$     & $-$   \\\hline  	
$\bar \Gamma_{\phi\rightarrow \bar{K}K}$     &  $1.76 \pm 0.09$      &   $1.44$     & $0.4$  & $1.10$    & $0.3$ & $1.87$    & $0.3$  & $1.12$    & $0.4$ \\\hline  	      
$\Delta \Gamma_{\phi\rightarrow \bar{K}K}$   &  $-0.65 \pm 0.13$     &   $-0.63$    & $0.1$  & $-0.58$   & $0.2$ & $-0.70$   & $0.1$  & $-0.58$   & $0.1$ \\\hline  	      
$\bar \Gamma_{K^{\star}\rightarrow K\pi}$    &  $46.75 \pm 2.34$     &   $46.27$    & $0.0$  & $46.20$   & $0.0$ & $46.63$   & $0.0$  & $46.02$   & $0.1$ \\\hline  	      
$\Delta \Gamma_{K^{\star}\rightarrow K\pi}$  &  $1.10 \pm 1.80$      &   $0.66$     & $0.1$  & $0.40$    & $0.2$ & $1.09$    & $0.0$  & $0.31$    & $0.2$ \\\hline  	      
$\bar \Gamma_{a_{1}\rightarrow\rho\pi}$      &  $425.00 \pm 175.00$  &   $533.23$   & $0.4$  & $428.34$  & $0.0$ & $566.85$  & $0.7$  & $489.59$  & $0.1$ \\\hline     
$\Gamma_{a_{1}\rightarrow\pi\gamma}$         &  $0.64 \pm 0.25$      &   $0.62$     & $0.0$  & $0.68$    & $0.0$ & $0.50$    & $0.3$  & $0.65$    & $0.0$ \\\hline
$\Gamma_{f_{1}^H\rightarrow K^{\star}K}$     &  $43.60 \pm 8.72$     &   $43.53$    & $0.0$  & $43.84$   & $0.0$ & $43.42$   & $0.0$  & $43.72$   & $0.0$ \\\hline
$\Gamma_{a_{0}}$                             &  $265.00 \pm 53.00$   &   $238.25$   & $0.3$  & $239.03$  & $0.2$ & $263.73$  & $0.0$  & $259.45$  & $0.0$ \\\hline    
$\Gamma_{K_{0}^{\star}\rightarrow K\pi}$     &  $270.00 \pm 80.00$   &   $336.80$   & $0.7$  & $333.96$  & $0.6$ & $256.33$  & $0.0$  & $308.70$  & $0.2$ \\\hline
$\Gamma_{f_0^L\rightarrow\pi\pi}$  (no fit)  &  $250.00 \pm 125.00$  &   $0.004$    & $-$    & $0.96$    & $-$   & $152.68$  & $-$    & $1.74$    & $-$   \\\hline  
$\Gamma_{f_0^L\rightarrow K K}$   (no fit)   &  $150.00 \pm 100.00$  &   $114.17$   & $-$    & $86.74$   & $-$   & $0.40$    & $-$    & $95.34$   & $-$   \\\hline    
$\Gamma_{f_0^H\rightarrow\pi\pi}$  (no fit)  &  $20.20 \pm 10.10$    &   $1000.0$   & $-$    & $443.19$  & $-$   & $720.07$  & $-$    & $493.93$  & $-$   \\\hline
$\Gamma_{f_0^H\rightarrow K K}$    (no fit)  &  $87.70 \pm 43.85$    &   $594.43$   & $-$    & $984.60$  & $-$   & $60.15$   & $-$    & $999.64$  & $-$   \\\hline
\end{tabular}
\end{table}
The parameter sets for the fits are given in \tabref{Tab:par_sets}.
\begin{table}[h!]
    \centering
    \begin{tabular}{|c||c|c|c|c|}
        \hline 
        Parameter & DS, $\omega$ & DS, no-$\omega$  & DVS-I, $\omega$ & DVS-I, no-$\omega$ \\\hline\hline
        $\phi_{N}$ [MeV]               & $163.95$       & $163.93$       & $153.49$       & $166.37$      \\\hline
        $\phi_{S}$ [MeV]               & $127.65$       & $133.40$       & $119.63$       & $130.17$      \\\hline
        $\phi_{3}$ [MeV]               & $2.62\e{-2}$   & $-4.72\e{-3}$  & $-3.25\e{-3}$  & $-1.50\e{-2}$ \\\hline
        $m_{0}^2$ [MeV$^2$]            & $-9.91\e{+5}$  & $-6.39\e{+5}$  & $-8.13\e{+5}$  & $-7.04\e{+5}$ \\\hline 
        $\tilde{m}_1^2$ [MeV$^2$]      & $8.00\e{+5}$   & $8.00\e{+5}$   & $8.00\e{+5}$   & $8.00\e{+5}$  \\\hline 
        $\lambda_{1}$                  & $5.73$         & $0.09$         & $-0.82$        & $0.23$        \\\hline 
        $\lambda_{2}$                  & $55.91$        & $44.79$        & $72.55$        & $50.68$       \\\hline 
        $h_{1}$                        & $-72.31$       & $26.24$        & $-30.98$       & $26.23$       \\\hline
        $h_{2}$                        & $16.76$        & $23.82$        & $-1.55$        & $18.01$       \\\hline
        $h_{3}$                        & $4.64$         & $5.41$         & $2.58$         & $5.05$        \\\hline
        $g_{1}$                        & $5.63$         & $5.53$         & $5.75$         & $5.60$        \\\hline
        $g_{2}$                        & $2.42$         & $3.01$         & $0.38$         & $2.72$        \\\hline
        $c_{1}$ [$\text{MeV}^{-2}$]    & $2.98\e{-4}$   & $2.68\e{-4}$   & $-2.76\e{-5}$  & $2.95\e{-4}$  \\\hline 
        $\tilde{\delta}_{S}$ [MeV$^2$] & $1.59\e{+5}$   & $1.46\e{+5}$   & $2.16\e{+5}$   & $1.56\e{+5}$  \\\hline 
        $\delta_{3}$ [MeV$^2$]         & $91.47$        & $3.75$         & $4.025\e{+3}$  & $10.02$       \\\hline 
        $\delta m^2_V$ [MeV$^2$]       & $30.77$        & $-1.26\e{+2}$  & $7.13\e{+3}$   & $-6.41\e{+2}$ \\\hline
        $\delta m^2_A$ [MeV$^2$]       & $-2.45\e{+5}$  & $-1.91\e{+5}$  & $-5.01\e{+5}$  & $-2.11\e{+5}$ \\\hline
        $m^2_{\text{em},S}$ [MeV$^2$]  & $-9.95\e{+3}$  & $9.93\e{+3}$   & $-9.96\e{+3}$  & $-9.65\e{+3}$ \\\hline
        $m^2_{\text{em},P}$ [MeV$^2$]  & $-4.30\e{+3}$  & $-3.69\e{+3}$  & $-6.94\e{+3}$  & $-3.89\e{+3}$ \\\hline
        $m^2_{\text{em},V}$ [MeV$^2$]  & $-2.33\e{+2}$  & $-3.20\e{+2}$  & $-8.51\e{+3}$  & $-7.95\e{+2}$ \\\hline
        $m^2_{\text{em},A}$ [MeV$^2$]  & $9.70\e{+3}$   & $9.91\e{+3}$   & $8.42\e{+3}$   & $9.59\e{+3}$ \\\hline
        $m^2_{\text{em},K}$ [MeV$^2$]  & $-$            & $-$            & $3.99\e{+3}$   & $4.54\e{+2}$ \\\hline
    \end{tabular}
    \caption{Parameter sets for four different cases. From left to right, DS with fitting the $\omega$ decay, DS without fitting the $\omega$ decay, DVS-I with fitting the $\omega$ decay, and DVS-I without fitting the $\omega$ decay. The \emph{scientific E notation} is used here, where $m\e{\pm n}$ is equal to $m\times 10^{\pm n}.$ }
    \label{Tab:par_sets}
\end{table}
\end{widetext}

\FloatBarrier

\bibliography{Isospin_ELSM}

\end{document}